\documentclass[a4paper,11pt]{article}
\pdfoutput=1 % if your are submitting a pdflatex (i.e. if you have
\usepackage{jheppub} % for details on the use of the package, please
\usepackage[T1]{fontenc} % if needed
\usepackage{lmodern}
\usepackage{newtxtext,amssymb,amsmath,extarrows}
\usepackage{graphicx,microtype,physics,framed,empheq}
\usepackage{booktabs} 
\usepackage{multirow}
\newcommand*\widefbox[1]{\fbox{\hspace{2em}#1\hspace{2em}}}

\graphicspath{{fig/}}

\title{\boldmath Large Spin-2 Signals at the Cosmological Collider}

\author[a,b]{Xi Tong}
\author[c]{and Zhong-Zhi Xianyu}

\affiliation[a]{Department of Physics, The Hong Kong University of Science and Technology,\\Clear Water Bay, Kowloon, Hong Kong, P.R. China}
\affiliation[b]{The HKUST Jockey Club Institute for Advanced Study,\\The Hong Kong University of Science and Technology,\\Clear Water Bay, Kowloon, Hong Kong, P.R. China}
\affiliation[c]{Department of Physics, Tsinghua University,\\Beijing 100084, China}

\emailAdd{xtongac@connect.ust.hk}
\emailAdd{zxianyu@tsinghua.edu.cn}

\abstract{We study the theory and phenomenology of massive spin-2 fields during the inflation with nonzero background chemical potential, and extend the cosmological collider physics to tensor modes. We identify a unique dimension-5 and parity-violating chemical potential operator for massive spin-2 fields, which leads to a ghost-free linear theory propagating one scalar mode and two tensor modes. The chemical potential greatly boosts the production of one tensor mode even for very heavy spin-2 particles, and thereby leads to large and distinct cosmological collider signals for massive spin-2 particles. The large signals show up at the tree-level in both the curvature trispectrum and the tensor-curvature mixed bispectrum.}

\begin{document}
	\maketitle
	\flushbottom
	
	\section{Introduction}\label{Intro}
	It is widely believed that the physics of our world can be described by a quantum theory of fields at energy scales well below the Planck scale \cite{Weinberg:1995mt,Weinberg:1996kw}. The notion of particles with integer or half-integer spins then emerges from quantizing the fluctuating modes of fields \cite{Fierz1939,Pauli:1940zz}. Although only lower-spin particles with spin-$0,1/2,1$ are found as fundamental particles in the Standard Model, massive higher-spin particles do exist as composite particles. Also, massive higher-spin particles are expected to play an important role in many Beyond Standard Model (BSM) models, with massive spin-2 particle as a notable example. Models involving massive spin-2 states include large-$N$ gauge theories \cite{tHooft:1973alw,Witten:1979kh}, massive gravity \cite{Fierz:1939ix,Hinterbichler:2011tt,deRham:2014zqa}, extra-dimension models \cite{Kaluza:1921tu,Klein:1926tv,Aragone:1987dtt,Arkani-Hamed:1998jmv,Randall:1999ee,Randall:1999vf}, and, perhaps most notably, string theory \cite{Rahman:2012thy}. The mass scale of such BSM spin-2 particles covers a broad range, leaving us with a vast landscape for phenomenological exploration. 
	
	In this work, focusing on the high-energy side of the massive spin-2 landscape, we initiate a study of massive spin-2 particles in the context of Cosmological Collider (CC) physics. The correlators of curvature fluctuations and tensor fluctuations encode valuable information about the particle spectrum during the inflation. The program of CC physics aims at utilizing the inflationary spacetime as a test ground of particle physics \cite{Chen:2009zp,Baumann:2011nk,Noumi:2012vr,Gong:2013sma,Arkani-Hamed:2015bza,Lee:2016vti,Chen:2016nrs,Chen:2016uwp,Chen:2016hrz}. The soft limits of the inflationary correlators could exhibit an oscillatory behavior known as the CC signal or clock signal \cite{Chen:2015lza}. The CC signal reflects the mass, spin, coupling and CP properties of the interacting particle. With an inflationary energy scale reaching up to $H\lesssim 10^{14}$ GeV, the CC is an ideal probe of heavy spin-2 modes such as that from KK compactification \cite{Kumar:2018jxz} and string theory \cite{Kim:2019wjo,Noumi:2019ohm}. 
	
	It has been shown that one of the most salient feature of the corresponding CC signal in the curvature bispectrum is a characteristic $P_2(\cos\theta)$ angular dependence \cite{Arkani-Hamed:2015bza,Lee:2016vti,MoradinezhadDizgah:2018ssw}, due to the mixing between the longitudinal component of the spinning particle and the inflaton. However, the overall amplitude of the CC signal is exponentially suppressed by the particle mass $m$. This is because the non-local type CC signal comes from on-shell particle production \cite{Tong:2021wai}. But a pure gravitational production is subjected to a Boltzmann suppression $e^{-m/T_{dS}}$ with the intrinsic de Sitter (dS) temperature $T_{dS}=\frac{H}{2\pi}$. The Boltzmann suppression makes it difficult to probe particles with a large mass. This deficiency necessitates a mechanism to amplify the CC signal strength of massive spin-2 particles, or in other words, a mechanism to enhance their production.
	
	Although the inflationary geometry resembles that of a dS spacetime with ten isometries, the rolling speed of the inflaton $\dot\phi_0(t)$ spontaneously breaks four of them \cite{Cheung:2007st}, including three dS boost transformations ($i.e.$, the special conformal transformations in the boundary CFT). The rolling inflaton automatically selects a preferred frame in which the background is uniform. Any dS boost would change this uniformity and bring in space dependence in the inflaton background. In this special frame, particle distribution needs not to take the dS-invariant Boltzmann form. Instead, the symmetry-breaking scale $|\dot{\phi}_0|^{1/2}$ may bring dramatic enhancement to particle production. Such is the case if the inflaton background plays the role of an external chemical potential \cite{Chen:2018xck,Hook:2019zxa,Wang:2019gbi,Liu:2019fag,Wang:2020ioa,Bodas:2020yho,Sou:2021juh,Wang:2021qez}. It has been shown that in the presence of this symmetry-breaking chemical potential, the production rate of massive particles with lower spins ($S=0,\frac{1}{2},1$) deviate from the dS-invariant Boltzmann form $e^{-m_S/T_{dS}}$ with exponential enhancement, leading to sizable observational effects in the CC signals \cite{Chen:2018xck,Hook:2019zxa,Wang:2019gbi,Liu:2019fag,Wang:2020ioa,Bodas:2020yho}. It is therefore tantalizing to generalize the chemical potential to particles with higher spins, among which $S=2$ is a first step. We will show that the lowest dimensional chemical potential operator capable of enhancing particle production is uniquely fixed by the consistency of the theory. An inspection of the linear theory in dS shows that the massive spin-2 field is ghost-free, but only propagates three degrees of freedom (DoF) instead of five if there is a non-zero chemical potential. Among the three propagating degrees, the two tensor modes receive helicity-dependent modulation of particle production whereas the longitudinal mode is not affected. Once coupled to the ``visible'' inflaton and graviton, the enhanced mode can lead to large CC signals in the mixed bispectra and curvature trispectrum, both at the tree level. The resulting non-Gaussian signal strengths can be as large as $f_{NL,\zeta\zeta\gamma}\sim \mathcal{O}(10)$ and $g_{NL,\zeta^4}\sim\mathcal{O}(10^4)$, respectively. Therefore, they are promising targets for the next generation of Cosmic Microwave Background (CMB) measurements such as LiteBIRD, as well as future 21-cm interferometry experiments. 
	
	In the existing literature, works on such large CC signals \cite{Flauger:2016idt,Chen:2018xck,Tong:2018tqf,Chua:2018dqh,Lu:2019tjj,Kumar:2019ebj,Wang:2019gbi} mostly focuses on scalar non-Gaussianities, while those on large mixed/tensor non-Gaussianities \cite{Agrawal:2017awz,Goon:2018fyu,Dimastrogiovanni:2018uqy,Fujita:2021flu,Dimastrogiovanni:2021cif,Ozsoy:2021onx,Cabass:2021iii,Cabass:2021fnw} rarely discuss the non-analytic oscillatory signatures from on-shell effects. As a result, our model serves as a first extension of cosmological collider phenomenology to external tensor modes with potentially large and observable signals.

	This paper is structured as follows. In Sect.~\ref{ChemPtlReview}, we start with a brief review of the chemical potential, and then move on to the introduction of a chemical potential to a massive spin-2 field in dS in Sect.~\ref{FreeSpin2ChemPtlSect}. After working out the linear theory, we turn on interactions between the spin-2 field and the ``visible'' sector fields, and study its impact on the inflationary observables in Sect.~\ref{CCObservablesSect}. Finally, we conclude in Sect.~\ref{ConclusionsSect}.
	
	\section{Chemical potential briefly reviewed}\label{ChemPtlReview}
	In this section, we briefly review the chemical potential effects for lower-spin states previously studied in the context of cosmological collider physics. Such chemical potential has its origin from statistical mechanics, where the chemical potential $\kappa$ is introduced as a Lagrange multiplier to control the total particle number in a grand canonical ensemble. The partition function $Z$ then picks up a dependence on the particle number $N$ in the system,
	\begin{equation}
	Z=e^{-(H-\kappa N)/T}~.
	\end{equation}
	This partition function can be directly adapted to a field theory, in the form of a path integral over the phase space weighted by a phase factor containing the Hamiltonian and the particle number,
	\begin{equation}
		Z=\int\mathcal{D}\Phi\mathcal{D}\Pi e^{i\int dtd^3x \Pi\dot{\Phi}}e^{-i\int dt\left(H[\Phi,\Pi]-\kappa N[\Phi]\right)}~,
	\end{equation}
	where $\Phi$ represents a general matter field and $\Pi$ is its canonical conjugate. For simplicity, we have assumed a particle number independent of the canonical momentum $\Pi$, which turns out be the case in most of the scenarios we are interested in. We then perform the integration over $\Pi$ to obtain the Lagrangian $L=\int\Pi\dot{\Phi}-H$, since it is more convenient to develop a perturbation theory with a Lagrangian formalism. Evoking locality and (local) Lorentz covariance, we can promote the chemical potential to a local vector field $\kappa_\mu=\kappa_\mu (x)$. Then the partition function reads
	\begin{equation}
	\label{eq_ZLagPI}
		Z=\int\mathcal{D}\Phi e^{i\int d^4x\sqrt{-g}\left[\mathcal{L}(\Phi,\partial\Phi)+\kappa_\mu J^\mu(\Phi,\partial\Phi)\right]}~,
	\end{equation}
	where $J^\mu(\Phi,\partial\Phi)$ is the current density whose integral over a spatial slice $\Sigma$ gives the particle number, $N=\int_\Sigma d^3x_\mu J^\mu$. The second term $\kappa_\mu J^\mu$ in (\ref{eq_ZLagPI}) represents the effect of an external field $\kappa$ on the matter, in the form of a chemical potential. We note that the matter current does not have to be conserved. In fact, it can be shown that, for the chemical potential term to nontrivially impact on the particle production, a necessary condition is that either the one-form field $\kappa=\kappa_\mu(x)dx^\mu$ is not closed, or the current is not conserved \cite{Sou:2021juh}: 
	\begin{equation}
		d\kappa\neq 0~~~\text{or}~~~\nabla\cdot J\neq 0~.
	\end{equation}
	For spin-0 fields, the $U(1)$ current is conserved and we need a non-closed chemical potential with a non-zero field strength, $\nabla_\mu \kappa_\nu-\nabla_\nu \kappa_\mu\neq 0$. This essentially provides a background electromagnetic field whose Schwinger effect gives an anisotropic enhancement of particle production.
	
	For spin-$1/2$ and spin-1 fields, there are natural choices of non-conserved currents quadratic in the field, namely, the axial current and the Chern-Simons current. In such cases, one can introduce a closed 1-form field as the chemical potential, without breaking the background rotation symmetry. This 1-form field is most naturally realized by the gradient of a scalar field, usually the inflaton itself,
	\begin{equation}
		\kappa_\mu=\frac{\nabla_\mu\phi(t)}{\Lambda_{c}}~,
	\end{equation}
	where $\Lambda_{c}$ is a cutoff scale for the chemical potential operator $\kappa_\mu J^\mu$.
	
	More explicitly, we take a massive spin-1 Proca field $A_\mu$ in the dS spacetime ($g_{\mu\nu}=a^2(\tau)\eta_{\mu\nu}$, $a(\tau)=-1/H\tau$) as an example. In addition to the conventional kinetic term and mass term,
	\begin{equation}
		S_\text{Proca}[A]=\int d^4x\sqrt{-g}\left[-\frac{1}{4}F_{\mu\nu}F^{\mu\nu}-\frac{1}{2}m^2A_\mu A^\mu\right]~,
	\end{equation}
	we turn on the chemical potential by adding a dimension-5 operator
	\begin{equation}
		S_{c}[A,\phi]=\frac{1}{2}\int d^4x \sqrt{-g} \kappa_\mu J^\mu_{CS}=-\int d^4x\sqrt{-g}\frac{\phi}{4\Lambda_{c}}\mathcal{E}^{\mu\nu\rho\sigma}F_{\mu\nu}F_{\rho\sigma}~,
	\end{equation}
	where $\phi$ is the inflaton field that depends only on time $t$ at the background level, $J_{CS}^\mu=\mathcal{E}^{\mu\nu\rho\sigma}A_\nu F_{\rho\sigma}$ is the Abelian Chern-Simons current, and $\mathcal{E}^{\mu\nu\rho\sigma}\equiv\frac{\epsilon^{\mu\nu\rho\sigma}}{\sqrt{-g}}$ is the Levi-Civita tensor density. Notice that the shift symmetry of the inflaton $\phi\to\phi+C$ is respected, since the constant inflaton background contributes to a total derivative. However, with a rolling inflaton background, the chemical potential term $S_{c}[A,\phi(t)]$ gives non-trivial effects on the dynamics of the spin-1 field. Varying the action with respect to $A$, we obtain the Equation-of-Motion (EoM) for the Proca field,
	\begin{equation}
		0=\frac{\delta(S_\text{Proca}+S_c)}{\sqrt{-g}\delta A_\mu}=\Box A^\nu-\nabla_\mu\nabla^\nu A^\mu-m^2 A^\nu-\frac{2\nabla_\mu\phi}{\Lambda_{c}}\mathcal{E}^{\mu\nu\rho\sigma}\nabla_\rho A_\sigma~.
	\end{equation}
	Taking the divergence and utilizing a few useful identities given in Appendix~\ref{HelpfulIds}, we arrive at the constraint $\nabla^\mu A_\mu=0$. Insert this back into the EoM, and we obtain a modified Klein-Gordon equation with a constraint
	\begin{align}
		\text{EoM:}\quad&\left(\Box-(m^2+3H^2)\right) A_\nu-\frac{\nabla_\alpha\phi}{\Lambda_{c}} \mathcal{E}^{\alpha\kappa\rho\sigma} g_{\kappa\nu}\nabla_\rho A_\sigma=0\label{Spin1ChemLinearEoMSimp}~,\\
		\text{Constraint:}\quad&\nabla^\mu A_\mu=0~.
	\end{align}
	The constraint relates the temporal mode to the longitudinal mode and reduces the DoF to $4-1=3$. This can be seen more clearly under a helicity decomposition,
	\begin{equation}
		A_\mu=\sum_{\lambda=-1}^{1}A_\mu^{(\lambda)}~.
	\end{equation}
	Without loss of generality, we go to momentum space and focus on a mode propagating in the $+z$ direction,
	\begin{align}
		&A_0^{(0)}=v_0~,~&&A_0^{(\pm1)}=0~,\\
		&A_i^{(0)}=f_0e_i^0~,&&A_i^{(\pm1)}=v_{\pm1}e_i^{\pm1}~.
	\end{align}
	The polarization vectors are chosen as
	\begin{equation}
		e_i^0=\left(\begin{array}{c}
			0\\
			0\\
			1
		\end{array}\right)~~~,~~~e_i^{\pm 1}=\left(\begin{array}{c}
			1\\
			\pm i\\
			0\end{array}\right)~.
	\end{equation}
	Denoting the chemical potential by $\kappa\equiv\frac{\dot{\phi}_0}{\Lambda_{c}}$, the mode functions satisfy
	\begin{align}
		v_{0}''+2aHv_0'+\left(k^2+(m^2+2H^2) a^2\right)v_0&=0~,\\
		v_{\pm1}''+\left(k^2\mp 2k\kappa a+m^2 a^2\right)v_{\pm1}&=0~.\label{VecModeEoM}
	\end{align}
	The longitudinal component is solved from the temporal component via
	\begin{equation}
		f_0=-\frac{i}{k}(v_0'+2aH v_0)~.
	\end{equation}
	As a result, the temporal mode $v_0$ is not affected by the chemical potential, whereas the transverse modes $v_{\pm1}$ receive helicity-dependent corrections. This dS-symmetry-breaking correction is further reflected in the detailed particle production history. Using the Stokes-line method \cite{Berry:1989zz,Sou:2021juh}, one can show that mode functions satisfying the EoM (\ref{VecModeEoM}) develop a negative frequency part at the conformal time\footnote{(\ref{Spin1ChemProdTime}) only applies when $|\tilde{\kappa}|<\tilde{m}$. For a chemical potential greater than the mass, (\ref{VecModeEoM}) acquires a period of tachyonic instability in the region $-k\tau\in\left(\tilde{\kappa}-\sqrt{\tilde{\kappa}^2-\tilde{m}^2},\tilde{\kappa}+\sqrt{\tilde{\kappa}^2-\tilde{m}^2}\right)$, indicating copious classical production with $\langle n_{\pm 1}(\mathbf{k})\rangle'>1$.}
	\begin{equation}
		\tau_{*,\pm1}\simeq -\frac{0.66\tilde{m}\mp0.34\tilde{\kappa}}{k}~,\label{Spin1ChemProdTime}
	\end{equation}
	where $\tilde{m}\equiv\frac{m}{H}$ and $\tilde{\kappa}\equiv\frac{\kappa}{H}$. This emergent negative frequency is physically interpreted as real particle production through the Bogoliubov transformation. The production amount is given by the amplitude of the negative frequency mode as 
	\begin{equation}
		\langle n_{\lambda}(\mathbf{k})\rangle'=\frac{e^{2 \pi  (\tilde{m} -\lambda\tilde{\kappa} )}+1}{e^{4 \pi  \tilde{m} }-1}\xrightarrow{\tilde{m}\gg 1}e^{-2\pi(\tilde{m}+\lambda\tilde{\kappa})}~.~~~~~(\lambda=0,\pm1)
	\label{Spin1ChemProdAmount}
	\end{equation}
	In this way, we see that the dS-invariant Boltzmann factor is altered in the presence of the chemical potential. Assuming $\tilde\kappa>0$, the production of $\lambda=-1$ mode is exponentially enhanced while the $\lambda=+1$ mode is exponentially suppressed. Such is a general feature of a chemical potential, that it gives a linear and biased shift in the effective mass of a bosonic particle. Particles in the amplified mode, once produced, will subsequently decay into massless inflatons and gravitons, producing sizable CC signals in the non-Gaussian correlators of curvature fluctuations and tensor fluctuations.
	
	\section{Linear theory of a massive spin-2 field in dS with chemical potential}\label{FreeSpin2ChemPtlSect}
	\subsection{Introducing the chemical potential}\label{IntroducingSin2ChemPtl}
	Now, we move on to the discussion of a massive spin-2 field. A consistent theory of a free massive spin-2 field $h_{\mu\nu}$ in dS is described by the quadratic part of the Einstein-Hilbert action,
	\begin{align}
		\nonumber S_{EH}=\int d^4x \sqrt{-g}\Bigg[&\frac{1}{2}\nabla_\mu h^{\mu\lambda}\nabla^\nu h_{\nu\lambda}-\frac{1}{4}\nabla_\mu h_{\nu\lambda}\nabla^\mu h^{\nu\lambda}+\frac{1}{4}\nabla^\mu h \nabla_\mu h-\frac{1}{2}\nabla^\nu h^\mu_{~~\nu}\nabla_\mu h\\
		&-\frac{1}{2}H^2\left(h_{\mu\nu}h^{\mu\nu}+\frac{1}{2}h^2\right)\Bigg]~,\label{EHQuadAction}
	\end{align}
	and a Fierz-Pauli mass term to ensure the absence of ghosts \cite{Fierz:1939ix},
	\begin{equation}
	\label{eq_SFP}
		S_{FP}=\int d^4x \sqrt{-g}\left[\frac{1}{4}m^2\left(h_{\mu\nu}h^{\mu\nu}-h^2\right)\right]~.
	\end{equation}
	where $h\equiv h^\mu_{~\mu}$. Here the covariant derivatives and index contractions are computed under the dS metric $g$. In dS, the absence of ghost states further require that the mass $m$ appeared in (\ref{eq_SFP}) satisfies the Higuchi bound $m^2>2H^2$ \cite{Higuchi:1986py}. Then, the combination $S_{EH}+S_{FP}$ gives rise to five DoFs without negative-norm states. In this work, we are more interested in the states in the principal series which could give rise to oscillatory signals. For spin-2, they correspond to the mass range $m^2>9H^2/4$.
	
	Inspired by the chemical potential for the spin-1 field $A_\mu$, we wish to introduce a chemical potential for massive spin-2 field $h_{\mu\nu}$ through its coupling with a rolling inflaton background $\phi(t)$. \emph{A priori} there could be multiple choices for introducing this coupling operator. Thus we further introduce a set of conditions that the operator should meet:
	\begin{enumerate} \itemsep =-5pt
		\item[(i)] The chemical potential should be from a local operator with the lowest possible mass dimension.
		\item[(ii)] The operator must take the form of $\kappa_\mu J^\mu(h)$, where $\kappa_\mu\propto \partial_\mu\phi$ and $J$ must be non-conserved, to have nontrivial effects of the particle production.
		\item[(iii)] The operator must be quadratic in $h_{\mu\nu}$ so that it contributes to the linear theory.
		\item[(iv)] The operator should (spontaneously) break dS boosts while respecting spatial translations, spatial rotations and dilation.
		\item[(v)] The operator must give rise to a consistent linear theory without ghosts.
	\end{enumerate}
	These conditions uniquely fix the chemical potential operator to be
	\begin{empheq}[box=\widefbox]{align}
		S_{c}=\int d^4x\sqrt{-g}\left[\frac{\phi}{2\Lambda_{c}}\mathcal{E}^{\mu\nu\rho\sigma}\nabla_\mu h_{\nu\lambda}\nabla_\rho h_\sigma^{~\lambda}\right]~.\label{ChemOp}
	\end{empheq}
	This is the lowest dimensional parity-odd operator one can write down that is quadratic in $h_{\mu\nu}$. To see that it takes the form of $\kappa_\mu J^\mu(h)$, one can perform an integration-by-part,
	\begin{equation}
		S_{c}=\int d^4x\sqrt{-g}\kappa_\mu(\phi)J^\mu(h)=-\int d^4x\sqrt{-g}\frac{\nabla_\mu\phi}{2\Lambda_{c}}\mathcal{E}^{\mu\nu\rho\sigma}h_{\nu\lambda}\nabla_\rho h_\sigma^{~\lambda}~.
	\end{equation} 
	This form also makes clear the symmetry-breaking pattern. Namely, the rolling inflaton background defines a time-like vector $\kappa_\mu=\frac{\nabla_\mu\phi(t)}{\Lambda_{c}}$ normal to the spatial hypersurface $t=$ const, breaking dS boosts while preserving spatial isometries. In the slow-roll phase, $\dot{\phi}_0\approx$ const, and dilation is also preserved. Note also that (\ref{ChemOp}) is shift-symmetric, $\phi\to\phi+C$, hence it does not back-react to the inflation potential. The P-odd current $J$ only has one spatial derivative and no time derivative on $h$, which ensures the absence of Ostrogradski ghosts or singularities \cite{Ostrogradsky:1850fid,Woodard:2015zca}.

	A notable feature of the chemical potential term (\ref{ChemOp}) is that it is not invariant under the linear gauge transformation for the spin-2 field $\delta h_{\mu\nu}=-(\nabla_\mu\xi_\nu+\nabla_\nu\xi_\mu)$, which would be problematic in massless or partially massless cases. For massive spin-2 fields satisfying the Higuchi bound, we know that gauge invariance must be realized non-linearly, and (\ref{ChemOp}) can be understood as the unitary gauge result. At high enough energy scales, the spin-2 field becomes strongly-coupled and new DoFs must be introduced to maintain unitarity. Thus we need to make sure our relevant physical processes are under perturbative control. We will elaborate more on this in Sect.~\ref{ModelSetupSubSect}.
	
	Before moving on to further developments, we point out that experienced readers might have in mind a more familiar alternative to the operator (\ref{ChemOp}). Namely, we could have assigned a gravitational Chern-Simons term $\phi R\tilde{R}(h)$ constructed from the massive spin-2 field. This term is gauge-invariant and quadratic in $h$ at leading order. Upon integration-by-parts, it takes the required form $\kappa_\mu J_{CS}^\mu(h)$, with $J_{CS}^\mu(h)$ being the gravitational Chern-Simons current. It also satisfies the required symmetry-breaking patterns. Its only deficiency compared to our chemical potential operator seems to be having higher dimensionality, due to two more derivatives. However, this turns out to be the crucial reason why we favor (\ref{ChemOp}) instead. We will show later in Sect.~\ref{UVcompletionSect} that this deficiency of the Chern-Simons term leads to pathologies that are non-negligible for the parameter regime that we are interested in.

	\subsection{EoMs and DoF analysis}
	Now, let us take a closer look at the linear theory with chemical potential. A variation of the quadratic action gives
	\begin{equation}
		0=\frac{\delta\left(S_{EH}+S_{FP}+S_c\right)}{\sqrt{-g}\delta h^{\mu\nu}}\equiv X_{\mu\nu}-\frac{1}{2}m^2(h_{\mu\nu}-g_{\mu\nu}h)-Y_{\mu\nu}~,\label{Spin2ChemLinearEoM}
	\end{equation}
	where $X_{\mu\nu}$ comes from the linearized Einstein-Hilbert action, and $Y_{\mu\nu}$ represents the terms from the chemical potential $S_c$,
	\begin{align}
		\nonumber X_{\mu\nu}=&\frac{1}{2}\left(\Box h_{\mu\nu}-\nabla_\mu\nabla_\alpha h^\alpha_{~\nu}-\nabla_\nu\nabla_\alpha h^\alpha_{~\mu}+\nabla_\mu \nabla_\nu h\right)+\frac{1}{2}g_{\mu\nu}\left(\nabla_\alpha\nabla_\beta h^{\alpha\beta}-\Box h\right)\\
		&-H^2 \left(h_{\mu\nu}+\frac{1}{2}g_{\mu\nu}h\right)~,\\
		Y_{\mu\nu}=&\frac{\nabla_\alpha \phi}{2\Lambda_{c}} \mathcal{E}^{\alpha\kappa\rho\sigma}\left(g_{\kappa\nu}\nabla_\rho h_{\mu\sigma}+g_{\kappa\mu}\nabla_\rho h_{\nu\sigma}\right)~.
	\end{align}
	Take the divergence of the EoM (\ref{Spin2ChemLinearEoM}) and use the background Bianchi identity $\nabla^\mu X_{\mu\nu}=0$, we obtain
	\begin{equation}
		\nabla^\mu Y_{\mu\nu}=-\frac{1}{2}m^2(\nabla^\mu h_{\mu\nu}-\nabla_\nu h)~.\label{firstDiv}
	\end{equation}
	With a useful identity (\ref{epsilonIdentity}) due to the antisymmetric property of $\mathcal{E}^{\alpha\kappa\rho\sigma}$ and the conformally flat dS metric, (\ref{firstDiv}) can be simplified to
	\begin{equation}
		\frac{1}{\Lambda_{c}}\mathcal{E}^{\alpha\kappa\rho\sigma}\nabla_\alpha\phi g_{\kappa\nu}\nabla_\rho\nabla^\mu h_{\mu\sigma}=-m^2(\nabla^\mu h_{\mu\nu}-\nabla_\nu h)~,\label{constraint1}
	\end{equation}
	where we have assumed the spatial homogeneity of the inflaton background $\phi=\phi(t)$. Take a further divergence and use the identity (\ref{epsilonIdentity}) again, we arrive at a constraint on $h$ alone:
	\begin{equation}
		\nabla^\mu\nabla^\nu h_{\mu\nu}=\Box h~.\label{constraint2}
	\end{equation}
	This eliminates one out of ten DoFs in $h$. However, there are more constraints hidden in the EoM. This can be seen from taking the trace of (\ref{Spin2ChemLinearEoM}),
	\begin{equation}
		(m^2-2H^2)h=0~,
	\end{equation}
	where (\ref{constraint2}) is applied for simplification. Massive spin-2 particles in the principal series satisfy $m^2>2H^2$ and is therefore traceless, $h=0$. Inserting this zero trace back into (\ref{constraint1}) and (\ref{constraint2}), and denoting $C_\nu\equiv\nabla^\mu h_{\mu\nu}$, we have
	\begin{eqnarray}
		C_\nu+\frac{\nabla_\alpha\phi}{m^2\Lambda_{c}}\mathcal{E}^{\alpha\kappa\rho\sigma} g_{\kappa\nu}\nabla_\rho C_\sigma&=&0~,\label{constraintA1}\\
		\nabla^\nu C_\nu&=&0~.
	\end{eqnarray}
	Taking $\nu=0$ in (\ref{constraintA1}), it is easy to check that the temporal component vanishes by the spatial homogeneity of $\phi$ and the antisymmetry of the Levi-Civita tensor density:
	\begin{equation}
		C_0=0~.
	\end{equation}
	The spatial components satisfy the constraint
	\begin{equation}
		\epsilon_{ijk}\partial_j C_k=-\frac{m^2\Lambda_{c}a^2(\tau)}{\phi'(\tau)}C_i~.\label{constraintA2}
	\end{equation}
	Thus the 3-vector $C_i$ is the eigenfunction of the curl operator with a time-dependent eigenvalue. In plasma physics, this equation describes a force-free magnetic field and is well-studied in the context of planetary and stellar quasistatic magnetic fields \cite{Chandrasekhar1956,Chandrasekhar1957}.
	The general solution to $C_i$ consists of the toroidal and poloidal components:
	\begin{equation}
		C_i=T_i+P_i~,~~\text{with}~~T_i=\epsilon_{ijk}\partial_j \psi \hat{n}_k~,~~~P_i=-\frac{\phi'}{m^2\Lambda_{c}a^2} \epsilon_{ijk}\partial_j T_k~,
	\end{equation}
	where $\hat{n}$ is an arbitrary unit vector and $\psi$ satisfies the Helmholtz equation
	\begin{equation}
		\left(\partial_i^2+\frac{m^4\Lambda_{c}^2a^4}{\phi'^2}\right)\psi=0~.
	\end{equation}
	Thus we see that the vector $C_i$ does not propagate any dynamical degree of freedom (specifically, its momentum spectrum is of measure zero in $\mathbf{k}\in\mathbb{R}^3$). One trivial solution to (\ref{constraintA2}) is simply $C_i=0$. In fact, there is a uniqueness theorem \cite{1534-0392_2003_2_251}, which, when accompanied by suitable boundary conditions, guarantees that this trivial solution is the only one. We therefore adopt the boundary condition $C_1=C_2=0$ on the $x$-$y$ plane, then the uniqueness theorem dictates 
	\begin{equation}
		C_i=0
	\end{equation}
	throughout spacetime. Thus setting $C_\mu=0$, we arrive at the same set of five constraints as the case without chemical potential. These eliminates five out of ten DoFs in the massive spin-2 field. We summarize the simplified EoM and the constraints as follows:
	\begin{empheq}[box=\fbox]{align}
		\text{EoM:}\quad&\left(\Box-(m^2+2H^2)\right)h_{\mu\nu}-\frac{\nabla_\alpha\phi}{\Lambda_{c}} \mathcal{E}^{\alpha\kappa\rho\sigma} \left(g_{\kappa\nu}\nabla_\rho h_{\mu\sigma}+g_{\kappa\mu}\nabla_\rho h_{\nu\sigma}\right)=0~,\label{Spin2ChemLinearEoMSimp}\\
		~~\text{Constraints:}\quad&\nabla^\mu h_{\mu\nu}=0~,~~~h=0~.\label{Spin2ChemConstraints}
	\end{empheq}
	The last term in (\ref{Spin2ChemLinearEoMSimp}) further generalizes the linear EoM for spinning fields in dS considered in \cite{Kehagias:2017cym} by incorporating the parity-violating Levi-Civita tensor density. As we shall see below, just like the case of spin-1, this chemical potential term introduces a biased shift of mass for the spin-2 particle, thereby enhancing the production rate for one helicity while suppressing the other.
	
	It is tempting to conclude that $h_{\mu\nu}$ contains $10-5=5$ DoFs on-shell at this stage, just as what one would intuitively imagine for a massive spin-2 particle. However, we will demonstrate below that there are two more (non-local) constraints hidden in (\ref{Spin2ChemLinearEoMSimp}) and (\ref{Spin2ChemConstraints}), further reducing the number of on-shell DoF to $10-5-2=3$.
	
	Since the dS boosts are now broken, particles with different helicities do not linearly mix into each other, we can label them by their momentum $\mathbf{k}$ and their helicity $\lambda=\mathbf{\hat{k}}\cdot \mathbf{S}$. More explicitly, we go to the momentum space and perform the helicity decomposition,
	\begin{equation}
		h_{\mu\nu}(\tau,\mathbf{k})=\sum_{\lambda=-2}^{2}h^{(\lambda)}_{\mu\nu}(\tau,\mathbf{k})~.
	\end{equation}
	In component form, we have 
	\begin{align}
		\nonumber&h_{00}^{(0)}=h_0, &&h_{00}^{(\pm 1)}=0, &&h_{00}^{(\pm 2)}=0, \\
		\nonumber&h_{0i}^{(0)}=f_0 e_i^0, &&h_{0i}^{(\pm 1)}=h_{\pm 1} e^{\pm 1}_i, &&h_{0i}^{(\pm 2)}=0,\\
		&h_{ij}^{(0)}=g_0 e_{ij}^0+\frac{1}{3}h_0 \delta_{ij}, &&h_{ij}^{(\pm 1)}=v_{\pm 1} e^{\pm 1}_{ij}, &&h_{ij}^{(\pm 2)}=h_{\pm 2}e^{\pm 2}_{ij}~.
	\end{align}
	The polarization tensor is chosen to satisfy
	\begin{align}
		\nonumber e_i^0&=\hat{k}_i, &&k_i e_i^{\pm 1}=0, &&e_i^{\pm 1} e_i^{\pm 1 *}=2,&& &&\\
		\hat{k}_i e_{ij}^0&=e_j^0, &&\hat{k}_i e_{ij}^{\pm 1}=\frac{3}{2}e_{j}^{\pm 1}, &&~~~~\hat{k}_i e_{ij}^{\pm 2}=0, &&e_{ij}^{\pm 2}=e_{ij}^{\mp 2*}, &&e_{ij}^{\pm 2}e_{ij}^{\pm 2*}=4~.
	\end{align}
	For a plane wave propagating in the $+z$-direction, the component form of the polarization tensor is
	\begin{align}
		\nonumber e_i^0&=\left(\begin{array}{c}
			0\\
			0\\
			1
		\end{array}\right), &&e_i^{\pm 1}=\left(\begin{array}{c}
			1\\
			\pm i\\
			0
		\end{array}\right), &&\\
		e_{ij}^0&=\left(\begin{array}{ccc}
			-\frac{1}{2} & 0 & 0\\
			0 & -\frac{1}{2} & 0\\
			0 & 0 & 1
		\end{array}\right), &&e_{ij}^{\pm 1}=\frac{3}{2}\left(\begin{array}{ccc}
			0 & 0 & 1\\
			0 & 0 & \pm i\\
			1 & \pm i & 0
		\end{array}\right), &&e_{ij}^{\pm 2}=\left(\begin{array}{ccc}
			1 & \pm i & 0\\
			\pm i & -1 & 0\\
			0 & 0 & 0
		\end{array}\right)~.
	\end{align}
	Inserting them into (\ref{Spin2ChemLinearEoMSimp}), we obtain the EoMs for the five mode functions $h_\lambda$,
	\begin{subequations}
		\begin{eqnarray}
			h_0''+2a H h_0'+\left[k^2+m^2a^2\right]h_0&=&0~,\label{helicity0hEoM}\\
			h_{\pm 1}''+\left[k^2\pm k \kappa a +\left(m^2-2H^2\right)a^2\right]h_{\pm 1}&=&0~,\label{helicity1hEoM}\\
			h_{\pm 2}''-2a H h_{\pm 2}'+\left[k^2\pm 2k \kappa a +\left(m^2-2H^2\right)a^2\right]h_{\pm 2}&=&0~,\label{helicity2hEoM}
		\end{eqnarray}
	\end{subequations}
	as well as those for other mode functions,
	\begin{subequations}
		\begin{eqnarray}
			f_0''-2aH f_0'+ \left[k^2+(m^2-6H^2)a^2\right]f_0&=&-2 i k a H\left(
				g_0+\frac{8}{3}h_0\right)~,\\
			g_0''-2aH g_0'+\left[k^2+(m^2-2H^2)a^2\right]g_0&=&-\frac{8 i k a H}{3}f_0~,\\
			v_{\pm1}''-2aH v_{\pm1}'+\left[k^2\pm k\kappa a+(m^2-2H^2)a^2\right]v_{\pm1}&=&-\frac{4 i k a H}{3}h_{\pm1}~,\label{helicity1vEoM}
		\end{eqnarray}
	\end{subequations}
	where we have again denoted $\kappa=\frac{\dot{\phi}_0}{\Lambda_{c}}\sim \text{const}$. The constraints (\ref{Spin2ChemConstraints}) give relations among mode functions and their first-order derivatives,
	\begin{subequations}
		\begin{eqnarray}
			f_0&=&-\frac{i}{k}(h_0'+2a H h_0)~~~,~~~g_0=-\frac{i}{k}(f_0'+2a H f_0)-\frac{1}{3}h_0~,\\
			v_{\pm 1}&=&-\frac{2i}{3k}(h_{\pm 1}'+2a H h_{\pm 1})~.\label{helicity1Constraint}
		\end{eqnarray}
	\end{subequations}
	By differentiating twice on the constraint (\ref{helicity1Constraint}) and the repeated application of the EoM (\ref{helicity1hEoM}), one can arrive at an expression of $v_{\pm 1}''$ in terms of $h_{\pm 1}$ and $h_{\pm 1}'$. Alternatively, one can repeatedly apply the constraint (\ref{helicity1Constraint}) and its derivative on the EoM (\ref{helicity1vEoM}). This leads to another expression of $v_{\pm 1}''$ in terms of $h_{\pm 1}$ and $h_{\pm 1}'$. Interestingly, a comparison of these two expressions gives an \textit{algebraic} equation on $h_{\pm 1}$:
	\begin{equation}
		i\kappa a^2 h_{\pm 1}=0~.\label{helicity1algebraicConstraint}
	\end{equation}
	Therefore, with a non-zero $\kappa$, the vector modes are required to vanish by theory consistency. In contrast, there is no such requirement on the $\lambda=0,\pm2$ modes: Their EoMs and constraints allow non-trivial solutions. As a result, we conclude that our theory (\ref{Spin2ChemLinearEoM}) propagates only one scalar DoF and two tensor DoFs.
	
	Of course, physically speaking, in the limit of small chemical potential $\kappa\to 0$ (for example, by setting $\Lambda_{c}\to\infty$), the effect of the inflaton background must be negligible, and the theory must return to that of a freely propagating massive spin-2 particle with five DoFs. This apparent discontinuity is somewhat similar to the vDVZ discontinuity in massive gravity \cite{vanDam:1970vg,Zakharov:1970cc}. The vDVZ discontinuity is resolved by considering the non-linear theory and resumming the higher-order contributions \cite{Vainshtein:1972sx,Deffayet:2001uk}. This leads to a suppressed scalar mode at a distance scale smaller than the Vainshtein radius, effectively recovering massless gravity. In our case, although the full non-linear theory is unknown, we can see from the EoM (\ref{helicity1hEoM}) that in the IR limit $\tau\to 0$, the chemical potential term $\pm k\kappa a$ is negligible compared to the mass term $m^2a^2$. This somehow suggests an explanation similar to the Vainshtein mechanism, namely, that the two vector DoFs are frozen at distances smaller than $\mathcal{O}\left(\frac{\kappa}{m^2}\right)$. But in the late-time limit, with a longer physical wavelength, the vector modes start to emerge. We leave a more detailed study of this question for future.

	\subsection{Mode functions and particle production}\label{ModeFunSect}
	From (\ref{helicity0hEoM}) and (\ref{helicity2hEoM}), we see that similar to the case of spin-1, the chemical potential influences the transverse modes by providing a linear and biased shift of the effective mass. For $\kappa>0$, the effective mass of particles with $\lambda=+2$ is raised up, thereby increasing adiabaticity and leading to suppressed particle production. Particles with $\lambda=-2$, in contrast, enjoy a smaller effective mass and enhanced particle production.
	
	We can solve the EoMs under the Bunch-Davies initial condition. The longitudinal mode function is in a Hankel form
	\begin{equation}
		h_{0}(\tau,k)=\mathcal{N}_0(-\tau)^{3/2}H_{i\mu}^{(1)}(-k\tau)~,
	\end{equation}
	while the tensor mode function is given by the Whittaker W function,
	\begin{equation}
	h_{\pm 2}(\tau,k)=-\frac{\mathcal{N}_{\pm2}}{\tau}W_{\pm i\tilde{\kappa},i\mu}(2ik\tau)~,~~~\mu\equiv\sqrt{\tilde{m}^2-\frac{9}{4}}~,\label{MassiveSpin2ModeFun}
	\end{equation}
	with $\mu\equiv\sqrt{\tilde{m}^2-\frac{9}{4}}$ and $\tilde{\kappa}\equiv\frac{\kappa}{H}$. The normalization constants are given by
	\begin{align}
		\mathcal{N}_0&=\sqrt{\frac{\pi }{3}}\frac{k^2}{H}\frac{e^{-\pi \mu/2}}{H \sqrt{\left(\mu ^2+1/4\right) \left(\mu
				^2+9/4\right)}}~,\\
		\mathcal{N}_{\pm2}&=\frac{e^{\mp\pi\tilde{\kappa }/2}}{2 H \sqrt{k}}~.\label{normConsts}
	\end{align}
	 Traditionally, these normalization constants are fixed by imposing canonical commutation relations for different components of field operators (see \cite{Higuchi:1986py} for instance). Here, for simplicity, we follow a cleaner shortcut using the symplectic inner product method, which is also easier to generalize to higher spins. First notice that from any two solutions $A_{\mu\nu}$ and $B_{\mu\nu}$ of the simplified EoM (\ref{Spin2ChemLinearEoMSimp}), one can construct a current
	\begin{equation}
		K^\rho(A,B)\equiv A_{\mu\nu}^*\nabla^\rho B^{\mu\nu}-B_{\mu\nu}\nabla^\rho A^{*\mu\nu}+\frac{\phi}{\Lambda_{c}}\mathcal{E}^{\alpha\nu\rho\sigma}\nabla_\alpha\left(A^*_{\mu\nu} B^\mu_{~\sigma}-B_{\mu\nu} A^{*\mu}_{~~~\sigma}\right)~.
	\end{equation}
	Using the EoM (\ref{Spin2ChemLinearEoMSimp}), it can be shown that this current is conserved on-shell:
	\begin{equation}
		\nabla_\rho K^\rho=0~.
	\end{equation}
	As a result, we can define an inner product by integrating $K$ over a spatial hyperspace $\Sigma(\tau)$ with induced metric $\hat{g}$,
	\begin{equation}
		\Big(A_{\mu\nu},B_{\rho\sigma}\Big)_\tau\equiv \int_{\Sigma(\tau)} \sqrt{\hat{g}}d^3 x_\rho K^\rho(A,B)~.
	\end{equation}
	The inner product is time-independent by virtue of the conservation of $K$:
	\begin{equation}
		\Big(A_{\mu\nu},B_{\rho\sigma}\Big)_{\tau}-\Big(A_{\mu\nu},B_{\rho\sigma}\Big)_{\tau_0}=\int_{\Sigma(\tau)\cup \Sigma(\tau_0)} \sqrt{\hat{g}}d^3 x_\rho K^\rho=\int_{\mathcal{V}(\tau,\tau_0)}\sqrt{-g}d^4 x \nabla_\rho K^\rho=0~.
	\end{equation}
	Henceforth, we define the normalization condition as\footnote{Notice that this definition differs from that in the literature \cite{Lee:2016vti,Kehagias:2017cym} by a factor of $2$. This is because of our different convention of fundamental field $h_{\mu\nu}$ in the action (\ref{EHQuadAction}).}
	\begin{equation}
		\Bigg(h_{\mu\nu}^{(\lambda)}(\tau,\mathbf{k})e^{i\mathbf{k}\cdot\mathbf{x}},h_{\rho\sigma}^{(\lambda')}(\tau,\mathbf{k}')e^{i\mathbf{k}'\cdot\mathbf{x}}\Bigg)=2\delta^{\lambda\lambda'}\delta^3({\mathbf{k}-\mathbf{k}'})~.
	\end{equation}
	In dS, this condition reduces to
	\begin{equation}
		i a^{-2}\eta^{\mu\rho} \eta^{\nu\sigma} \mathcal{W}\left(h_{\mu\nu}^{(\lambda)*},h_{\rho\sigma}^{(\lambda)}\right)=2~,\label{NormCondExplicit}
	\end{equation}
	where $\mathcal{W}$ stands for the Wronskian. Inserting the mode functions into (\ref{NormCondExplicit}), we obtain the results of the normalization constants\footnote{Actually, an alternative way to see the absence of vector modes is that after solving (\ref{helicity1hEoM}) and (\ref{helicity1Constraint}) up to a normalization constant $\mathcal{N}_{\pm1}$, the inner product turns out to be time-dependent, $\Big(h_{\mu\nu}^{(\pm1)},h_{\rho\sigma}^{(\pm1)}\Big)\propto i a^{-2}\eta^{\mu\rho} \eta^{\nu\sigma} \mathcal{W}\left(h_{\mu\nu}^{(\pm1)*},h_{\rho\sigma}^{(\pm1)}\right)=\frac{8H^2}{k}\left(\mu^2+\frac{9}{4}\pm\tilde{\kappa} k\tau\right)|\mathcal{N}_{\pm1}|^2$. The only way to maintain theory consistency is to have $\mathcal{N}_{\pm1}=0$.} (\ref{normConsts}).
	
	Focusing on the $\lambda=\pm2$ helicities, the mode functions exhibit Stokes phenomenon and grow a negative-frequency part after the Stokes-line crossing time
	\begin{equation}
		\tau_{*,\pm2}\simeq -\frac{0.66\mu\mp0.34\tilde{\kappa}}{k}~.\label{Spin2ChemProdTime}
	\end{equation}
	Due to the dilation symmetry of dS, the production time for modes with different momenta all corresponds to a particular energy scale, defined by the physical momentum at production,
	\begin{equation}
		\mathcal{M}_{*,\pm 2}\equiv\frac{k}{a(\tau_{*,\pm2})}\simeq 0.66\mu H\mp0.34\kappa~.\label{Spin2ChemProdScale}
	\end{equation}
	It is only \textit{below} this energy scale are particles produced and begin their subsequent interactions.	The IR behavior of the mode function reads
	\begin{equation}
		h_{\pm 2}(\tau,k)\xrightarrow{\tau\to 0}\alpha_{\pm 2} \frac{1}{2\sqrt{\mu}H}(-\tau)^{-\frac{1}{2}+i\mu}+\beta_{\pm 2} \frac{1}{2\sqrt{\mu}H}(-\tau)^{-\frac{1}{2}-i\mu}~,
	\end{equation}
	where
	\begin{eqnarray}
		\alpha_{\pm 2}&=&(1-i) (2k)^{i \mu } e^{\frac{1}{2} \pi  (\mu \mp\kappa )}\frac{\sqrt{\mu }~\Gamma (-2 i \mu )}{\Gamma
			\left(\frac{1}{2} -i \mu\mp i \tilde{\kappa} \right)}~,\\
		\beta_{\pm 2}&=&(1-i) (2k)^{-i \mu } e^{-\frac{1}{2} \pi  (\mu\pm\tilde{\kappa} )}\frac{ \sqrt{\mu }~\Gamma (2 i \mu )}{\Gamma
			\left(\frac{1}{2}+i \mu\mp i \tilde{\kappa}  \right)}~,
	\end{eqnarray}
	are the Bogoliubov coefficients satisfying the condition $|\alpha_{\pm 2}|^2-|\beta_{\pm 2}|^2=1$. The average particle number density is
	\begin{equation}
		\langle n_{\pm 2}(\mathbf{k})\rangle'=|\beta_{\pm 2}|^2=\frac{e^{2 \pi  (\mu \mp\tilde{\kappa} )}+1}{e^{4 \pi  \mu }-1}\xrightarrow{\mu\gg 1}e^{-2\pi(\mu\pm\tilde{\kappa})}~.\label{Spin2ChemProdAmount}
	\end{equation}
	Therefore, massive spin-2 particles with helicity $\lambda=-2$ receive an exponential enhancement from the external chemical potential (\ref{ChemOp}).
	
	\subsection{UV completions?}\label{UVcompletionSect}
	Before concluding this section, we make a few remarks on the UV completion regarding the spin-2 gauge symmetry of our chemical potential operator.
	\begin{enumerate}
		\item[$\bullet$] We have constrained the chemical potential operator to take the lowest mass dimension possible. On one hand, this follows from the effective field theory logic, that operators with lower dimensions is more relevant in the IR. On the other hand, this is because higher dimensionality implies more derivatives, and the resulting EoM may incite the appearance of ghosts or singularities. This is the case, for example, if one picks the dimension-7 operator derived from the Chern-Simons term instead of (\ref{ChemOp}),
		\begin{align}
			\nonumber S_{CS}=&\int d^4x\sqrt{-\bar{g}}\frac{\phi}{8\Lambda_{CS}} \mathcal{E}^{\mu\nu\rho\sigma}\bar{R}_{\alpha\beta\mu\nu}\bar{R}^{\alpha\beta}_{~~~\rho\sigma}\\
			=&\int\sqrt{-g}\frac{\phi}{2\Lambda_{CS}^3} \mathcal{E}^{\mu\nu\rho\sigma}\nabla_\mu\nabla_{[\alpha} h_{\beta]\nu}\nabla_\rho\nabla^{[\alpha}h^{\beta]}_{~\sigma}+\mathcal{O}(h^3)~,\label{CSterm}
		\end{align}
		where $\bar{R}[\bar{g}]$ is computed with the metric $\bar{g}\equiv g+\frac{h}{\Lambda_{CS}}$. The massless version of this term leads to the so-called Chern-Simons gravity \cite{Jackiw:2003pm,Alexander:2009tp}, and has been proposed as a source of gravitational birefringence \cite{Lue:1998mq} as well as leptogenesis \cite{Alexander:2004us}. This action is invariant under the linear gauge transformation $\delta h_{\mu\nu}=-(\nabla_\mu\xi_\nu+\nabla_\nu\xi_\mu)$, suggesting a possible completion when going from the unitary-gauge massive IR theory to the UV picture with linear symmetry realization. However, this operator contains higher-order derivatives, and leads to pathologies if we try to use this operator to assist particle production in the IR. This can be seen by inspecting the EoM of the helicity $\lambda=\pm2$ tensor modes,
		\begin{equation}
			\left(1\mp\frac{2\hat{\kappa}k}{a H}\right)h_{\pm2}''-2 (a H\mp4\hat{\kappa}k) h_{\pm2}'+\left[k^2+(m^2-2H^2)a^2\mp4\hat{\kappa} k aH\left(1+\frac{k^2}{2a^2 H^2}\right)\right]h_{\pm2}=0~,\label{CSEoM}
		\end{equation}
		where $\hat{\kappa}\equiv\frac{\dot{\phi}_0 H}{\Lambda_{CS}^3}$ is the equivalent of our chemical potential $\tilde{\kappa}$. In the late-time limit $\tau\to 0$, one can canonicalize this mode and find enhancement/suppression of particle production for the negative-/positive-helicity mode, in a way similar to our chemical potential $\tilde{\kappa}$. However, the coefficient before the $h_{\pm2}''$ term is not positive-definite. This means that, at an early time, the $\lambda=+2$ mode first encounters a singularity once the coefficient of the second-order time derivative term vanishes, and then becomes a ghost when its physical momentum exceeds the scale $k/a(\tau)\gtrsim H/\hat{\kappa}=\Lambda_{CS}^3/\dot{\phi}_0$. This is essentially the vacuum instability problem in Chern-Simons gravity \cite{Dyda:2012rj}. In contrast, our chemical potential operator avoids such a problem by leaving the second-order time derivative unchanged.
		
		Interestingly, however, we note that it is possible to deform the problematic equation (\ref{CSEoM}) into a healthy one such as (\ref{helicity2hEoM}). Generically in EFTs, higher-order derivative terms are present in the field EoM. One can then evoke the reduction-of-order procedure \cite{Solomon:2017nlh} to reduce these potentially pathological terms. In our case, we can reorganize (\ref{CSEoM}) into the form
		\begin{equation}
			h_{\pm2}''-2 a H h_{\pm2}'+\left[k^2+(m^2-2H^2)a^2\right]h_{\pm2}=\pm\frac{2\hat{\kappa}k}{a H}\left[h_{\pm2}''-4a H h_{\pm2}'+2 a^2H^2\left(1+\frac{k^2}{2a^2 H^2}\right)h_{\pm2}\right]~,
		\end{equation}
		and then replace the first term on the right-hand side by the free on-shell EoM,
		\begin{equation}
			h_{\pm2}''\to2 a H h_{\pm2}'-\left[k^2+(m^2-2H^2)a^2\right]h_{\pm2}~.
		\end{equation}
		This changes the form of the EoM to
		\begin{equation}
			\mathrm{h}_{\pm 2}''-2a H \mathrm{h}_{\pm 2}'+\left[k^2\pm 2k a\frac{\hat{\kappa}(m^2-2H^2)}{H} +\left(m^2-2H^2\right)a^2\right]\mathrm{h}_{\pm 2}+\mathcal{O}(\hat{\kappa}^2)=0~,\label{CSEoMReduced}
		\end{equation}
		where $\mathrm{h}_{\pm2}\equiv e^{\mp 2k\hat{\kappa}/aH}h_{\pm2}$. Comparing this EoM to (\ref{helicity2hEoM}), we find that they take the same form upon identification $\tilde{\kappa}\leftrightarrow (m^2/H^2-2)\hat{\kappa}$. Of course, one should be reminded that the reduction-of-order procedure is accurate only up to $\mathcal{O}(\hat{\kappa}^2)$ errors. The solution of (\ref{CSEoMReduced}) (analogous to our case) only represents a resummation of a selected piece of the full effect in the massive Chern-Simons gravity, which still has pathologies on its own. The question is whether this selected piece reflects the correct UV physics. After all, the Chern-Simons term (\ref{CSterm}) is only one of many non-renormalizable operators ($e.g.$, $\phi\bar{R}^2,\phi\bar{R}_{\mu\nu}^2$) that are also capable of influencing the linear EoM. This question remains unclear to us. On the other hand, even if we directly adopt (\ref{CSEoMReduced}), to have significant enhancement of particle production, we expect $\tilde{\kappa}\sim\mu^2\hat{\kappa}\sim \mu$ and thus $\hat{\kappa}\sim \mu^{-1}$. But then the problematic scale $\frac{H}{\hat{\kappa}}\sim m$ lies just near the particle production scale $\mathcal{M}_{*,\pm 2}\sim m$. Therefore, the parameter regime that we are interested in is exactly where the theory breaks down. In view of the above difficulties, our dimension-5 chemical potential operator seems still more reasonable than the Chern-Simons term. 
		\item[$\bullet$] On the other hand, it may be possible to embed the chemical potential operator into something similar to the axion-$SU(2)$ model \cite{Dimastrogiovanni:2016fuu,Agrawal:2017awz,Thorne:2017jft,Agrawal:2018mrg,Maleknejad:2018nxz,Dimastrogiovanni:2018xnn,Fujita:2018vmv,Fujita:2021flu}, where a gauge boson VEV introduces a mixing between spacetime indices and internal indices. This leads to the emergence of a spin-2 particle with similar dynamics as that under a chemical potential. We point out there are two key differences in our setups and objectives. First, the underlying symmetries and what we mean by spin-2 particles are different. We are focusing on spin-2 particles that transform under $SO(3,1)$ in the high-energy limit. In the axion-$SU(2)$ model, the spin-2 particle originates from the VEV of the gauge bosons, and transforms under the remaining $SO(3)$ subgroup of spacetime symmetries and internal symmetries. In the high-energy limit, the gauge boson VEV becomes negligible and Lorentz symmetry is approximately restored, then they should be distinguishable by having different Lorentz transformation properties. Second, in addition to the analytical non-Gaussianity shapes, we are targeting at the non-analytic CC signals which provide direct and clear information about the system. 
		\item[$\bullet$] Another possible UV completion may be similar to that of teleparallel gravity with a Nieh-Yan term \cite{Li:2020xjt,Li:2021wij}. By coupling the torsion tensors to an axion-like field, parity-violating production of chiral gravitational waves can be produced \cite{Cai:2021uup}. It is then interesting to see the effect of including a graviton mass in such theories, which would be the equivalent of massive spin-2 field with chemical potential.
	\end{enumerate}

	\section{Impact on observables}\label{CCObservablesSect}
	With a healthy theory of free massive spin-2 particle with chemical potential, In this section, we turn on interactions and couple the massive spin-2 field to the ``visible'' sector, namely the inflaton $\varphi$ and graviton $\gamma_{ij}$. We will show that with the help of the chemical potential, the enhanced helicity $\lambda=-2$ modes can leave sizable and distinctive CC signals on the mixed bispectra and curvature trispectrum even when its mass is much larger than the Hubble scale. Our model thus provides a first example of CC physics generalized to tensor modes. 
	
	\subsection{Model setup}\label{ModelSetupSubSect}
\begin{figure}[h!]
		\centering
		\includegraphics[width=15cm]{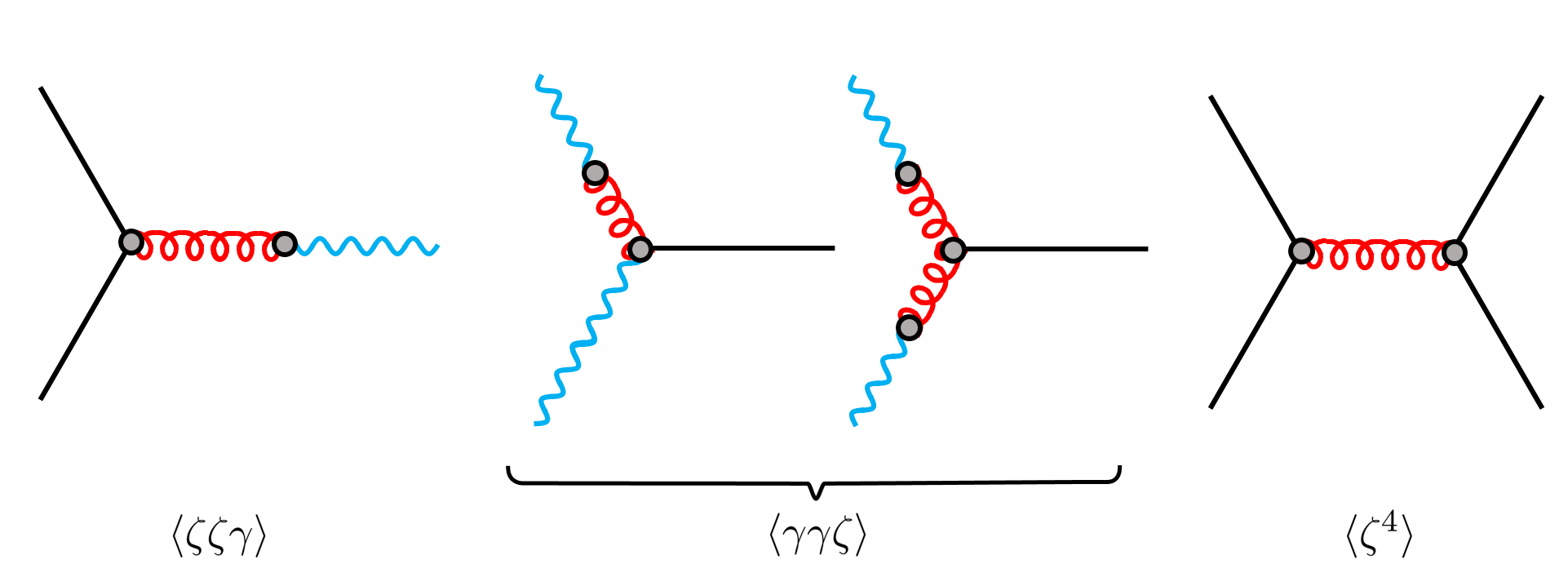}
		\caption{Non-Gaussian correlators induced at tree level from (\ref{ChemPtlNonCovariantForm}),~(\ref{MixingVertex}) and (\ref{hphiphiVertex}).}\label{NGdiags}
	\end{figure}

	Before turning on additional couplings, notice that the dimension-5 chemical potential operator $S_c$ itself brings a non-linear interaction between $h_{\mu\nu}$ and the inflaton fluctuation $\varphi\equiv\phi-\phi_0$. It induces 1-loop corrections to the correlators $\langle\zeta^n\rangle$, with $n=2,3,\cdots$. However, the loop integral introduces a loop factor $(4\pi)^{-2}$, and suppresses the signal strength. This issue can be avoided if we are able to observe the enhanced spin-2 particles at tree-level. We can form tree-level amplitudes with spin-2 signals by the following set of effective operators:
	\begin{empheq}[box=\widefbox]{align}
		S_{c}&=\int d\tau d^3x\frac{\phi'}{2\Lambda_{c}}a^{-2} \epsilon_{ijk}h_{il}\partial_j h_{kl}~,\label{ChemPtlNonCovariantForm}\\
		S_{h\gamma}&=\int d\tau d^3x \rho \phi'\gamma_{ij}'h_{ij}~,\label{MixingVertex}\\
		S_{h\phi^2}&=-\int d\tau d^3x\frac{1}{M} h_{ij}\partial_i\phi\partial_j\phi~.\label{hphiphiVertex}
	\end{empheq}
	Notice that these operators are all of the lowest possible dimension giving the field content and the symmetry requirement, and are all dimension-5. Therefore, our choice of these EFT operators is natural and does not require any tuning \textit{a priori}.  	The first term $S_{c}$ is just the chemical potential operator (\ref{ChemOp}) written in a non-covariant form. $S_{h\gamma}$ gives a linear mixing between the massive spin-2 particle and the graviton \cite{Lee:2016vti,Dimastrogiovanni:2018uqy,Dimastrogiovanni:2021cif,Dimastrogiovanni:2021mfs}, and $S_{h\phi^2}$ is the massive spin-2 field's coupling to the inflaton.
	These coupling can induce non-Gaussianities\footnote{Notice that the mixing $S_{h\gamma}$ also brings a parity-violating correction to the graviton power spectrum $\langle\gamma^2\rangle$ at Gaussian level, producing helical primordial gravitational waves. In this work, we will keep the mixing perturbative, thus the primordial gravitational wave is still dominated by vacuum fluctuations and is not maximally helical.\label{HelicalGWFootnote}} at tree level via the Feynman diagrams shown in Fig.~\ref{NGdiags}.
	
	As mentioned in Sect.~\ref{IntroducingSin2ChemPtl}, $S_c$ is not invariant under a linear gauge transformation $\delta h_{\mu\nu}=-(\nabla_\mu\xi_\nu+\nabla_\nu\xi_\mu)$. Consequently, it must be understood as an effective operator that is valid below the strong-coupling scale. And the gauge invariance of the unknown UV theory is realized non-linearly below this cutoff. Now we have introduced extra couplings of the massive spin-2 field as EFT operators that break both the massive spin-2 gauge symmetry and time diffeomorphism. As a result, for the sake of consistency, we should explore the CC phenomenology of chemical potential with the EFT under control. In particular, the energy scale $\mathcal{M}_{*,\pm 2}$ of chemical-potential-assisted particle production must be lower than both the strong-coupling scales $\Lambda_{sc}$ of the massive spin-2 field, and the non-linear scales $\Lambda_{nl}$ of time diffeomorphism,
	\begin{empheq}[box=\widefbox]{align}
		\mathcal{M}_{*,\pm 2}<\Lambda_{sc},\Lambda_{nl}~,\label{EFTConsistency}
	\end{empheq}
	where $\mathcal{M}_{*,\pm 2}$ is given by (\ref{Spin2ChemProdScale}) and is not far away from the mass scale $m$.
	
	Both $\Lambda_{sc}$ and $\Lambda_{nl}$ can be found by using the Naive Dimensional Analysis (NDM) method \cite{Weinberg:1978kz,Manohar:1983md}. Namely, we perform a Stueckelberg trick and enforce perturbative unitarity on the newly generated EFT operators to find their cutoffs. For example, introducing a Stueckelberg field $A_\mu$, we can decompose the spin-2 field as 
	\begin{equation}
		h_{\mu\nu}=\hat{h}_{\mu\nu}+\frac{\nabla_\mu A_\nu+\nabla_\nu A_\mu}{m}~.
	\end{equation}
	Then the chemical potential term changes into
	\begin{align}
		\nonumber S_{c}&=\frac{1}{2}\int d^4x \epsilon^{\mu\nu\rho\sigma}\left[\frac{\phi}{\Lambda_{c}}\nabla_\mu \hat h_{\nu\lambda}\nabla_\rho \hat h_\sigma^{~\lambda}+\frac{2\phi}{m\Lambda_{c}}\nabla_\mu\hat{h}_{\nu\lambda}\nabla_\rho\nabla^\lambda A_\sigma+\frac{\phi}{m^2\Lambda_{c}}\nabla_\mu\nabla_\lambda A_\nu\nabla_\rho \nabla^\lambda A_\sigma\right]~.~~~
	\end{align}
	These are dimension-5,6,7 operators, respectively, if we take the inflaton field $\phi$ on its quantum fluctuations $\varphi$. Perturbative unitarity then requires the cutoff energy scale being the smallest derived from them,
	\begin{equation}
		\Lambda_{sc}^{(c)}\sim\min \left\{4\pi\Lambda_{c},(4\pi m\Lambda_{c})^{1/2},(4\pi m^2\Lambda_{c})^{1/3}\right\}~,
	\end{equation}
	where $4\pi$ comes from the loop factor. For $\Lambda_{c}> m$, the constraint from the dimension-7 operator is the most strict one, giving $\Lambda_{sc}^{(c)}=(4\pi m^2\Lambda_{c})^{1/3}$. Thus if the chemical potential is not significantly larger than the mass scale, $\kappa\sim m$, the weak-coupling constraint $\mathcal{M}_{*,\pm 2}\sim m< (4\pi m^2\Lambda_{c})^{1/3}=\Lambda_{sc}^{(c)}$ is always satisfied. Therefore, the consistency of the weakly coupled picture of a massive spin-2 particle with enhanced particle production set up in Sect.~\ref{FreeSpin2ChemPtlSect} is reassured from an EFT point of view. Repeating the procedure for the other two operators, we obtain their strong coupling scales:
	\begin{equation}
		\Lambda_{sc}^{(h\gamma)}=\sqrt{\frac{4\pi m M_p}{\rho}}~,~\Lambda_{sc}^{(h\phi^2)}=\sqrt{4\pi m M}~.
	\end{equation}
	\begin{figure}[h!]
		\centering
		\includegraphics[width=13cm]{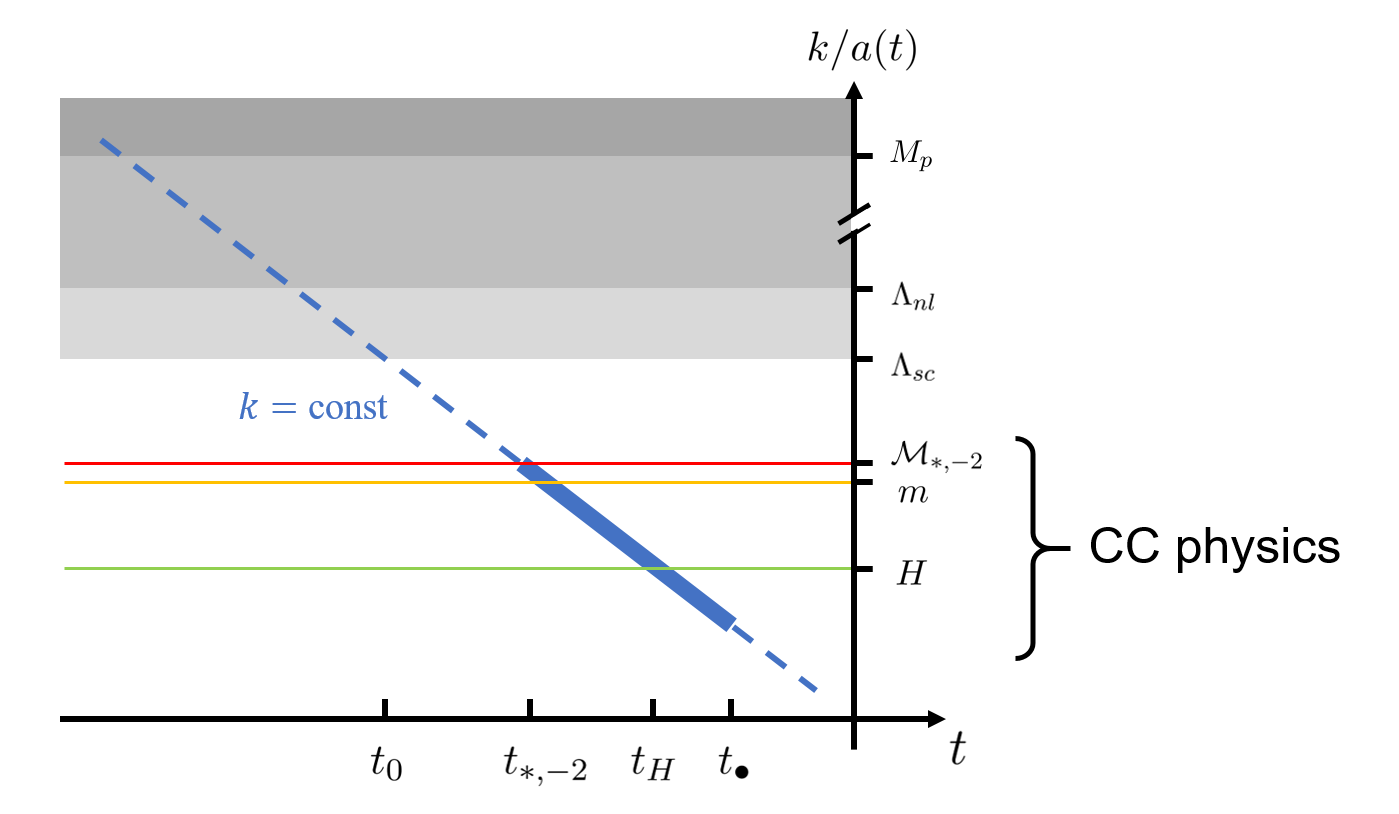}
		\caption{A schematic illustration of the life of a helicity $\lambda=-2$ mode with constant comoving momentum $k$. The horizontal axis is time and the vertical axis represents the redshifted physical momentum of this mode. The blue line is dashed when the mode is at the vacuum and is solid when it is occupied by particles. The characteristic events are $t_0$: the mode emerges out of the cutoffs as the BD vacuum, $t_{*,-2}$: particle production with chemical potential, $t_H$: horizon-crossing, $t_\bullet$: the particle decays into inflatons and gravitons, with the mode going back to vacuum. The whole CC process occurs well below the strong-coupling scale and the non-linear scale of time diffeomorphism, and is under perturbative control.}\label{PhysicalPicture}
	\end{figure}
	For the time diffeomorphism, we perform the Stueckelberg trick $t\to t+\varphi/\dot{\phi}_0$ and apply NDA to the operators above. The resulting non-linear scales for time diffeomorphism are
	\begin{equation}
		\Lambda_{nl}^{(h\gamma)}=\left(\frac{4\pi M_p\dot{\phi}_0}{\rho}\right)^{1/3}~,~\Lambda_{nl}^{(h\phi^2)}=\left(4\pi M\dot{\phi}_0\right)^{1/3}~.
	\end{equation}
	We do not expect any non-linear scale due to broken time diffeomorphism for chemical potential operator (\ref{ChemPtlNonCovariantForm}), because it is already completed to a covariant-form (\ref{ChemOp}) in which the spacetime diffeomorphisms are made explicit\footnote{One can also start with covariant completions, such as for example, $S_{h\gamma}\subset \int d^4x \sqrt{-g}\frac{\nabla_\rho \phi}{\Lambda_{mix}^2} \nabla^\rho R^{\mu\nu}h_{\mu\nu}$ and $S_{h\phi^2}\subset\int d^4x \sqrt{-g}\frac{1}{M}h^{\mu\nu}\nabla_\mu\phi \nabla_\nu\phi$. However, the size of the couplings may be small, limited by naive choices of higher-dimensional operators as covariant completions. In this work, we would like to keep the discussion general and adopt a bottom-up approach, while paying attention to the EFT validity.}. As a summary, we will work within the energy scale below the lowest cutoff found above,
	\begin{align}
		\Lambda_{sc}&=\min\left\{\Lambda_{sc}^{(c)},\Lambda_{sc}^{(h\gamma)},\Lambda_{sc}^{(h\phi^2)}\right\}~,\label{StrongCouplingScale}\\
		\Lambda_{nl}&=\min\left\{\Lambda_{nl}^{(h\gamma)},\Lambda_{nl}^{(h\phi^2)}\right\}~.\label{TimeDiffNonlinearScale}
	\end{align}
	
	Physically speaking, the modes of the massive spin-2 field are assumed to stay at the Bunch-Davies (BD) vacuum when they exit the cutoffs $\Lambda_{sc},\Lambda_{nl}$. Then as they are redshifted to a lower energy scale $\mathcal{M}_{*,\pm2}$, chemical-potential-assisted particle production happens, and the modes are occupied abundantly with pairs of spin-2 particles. At a lower scale, these particle pairs decay into inflatons and gravitons via the corresponding couplings, producing observable CC signals. This physical picture is illustrated in Fig.~\ref{PhysicalPicture}.

	\subsection{Cosmological collider signals}
	With the interactions in hand, we can now perform a detailed computation of the non-Gaussian correlators,
	\begin{equation}
		\langle\mathcal{O}(\{\mathbf{k}_i\})\rangle\equiv(2\pi)^3\delta^3\left(\sum_i\mathbf{k}_i\right)\langle\mathcal{O}(\{\mathbf{k}_i\})\rangle'~,
	\end{equation}
	where $\mathcal{O}(\{\mathbf{k}_i\})$ represents products of curvature and tensor perturbations $\zeta(\mathbf{k}_i),\gamma(\mathbf{k}_j)|_{\tau\to 0}$ observed at the end of inflation. The Schwinger-Keldysh propagators of various fields and corresponding Feynman rules are given in Appendix~\ref{FeynmanRuleAppendix}. To avoid cluttering of unnecessary lengthy equations, we also put the technical expressions of various diagrams in Appendix~\ref{FeynmanRuleAppendix}. See \cite{Chen:2017ryl} for a review of our diagrammatic notations and Feynman rules. Our strategy is to apply the recently developed cutting rule \cite{Tong:2021wai} (see a brief review in Appendix~\ref{CRAppendix}) to obtain analytical results of the leading-order CC signals. They are then checked against the results from direct numerical integration.

	\subsubsection{$\langle\zeta\zeta\gamma\rangle$}
\begin{figure}[h!]
		\centering
		\includegraphics[width=14cm]{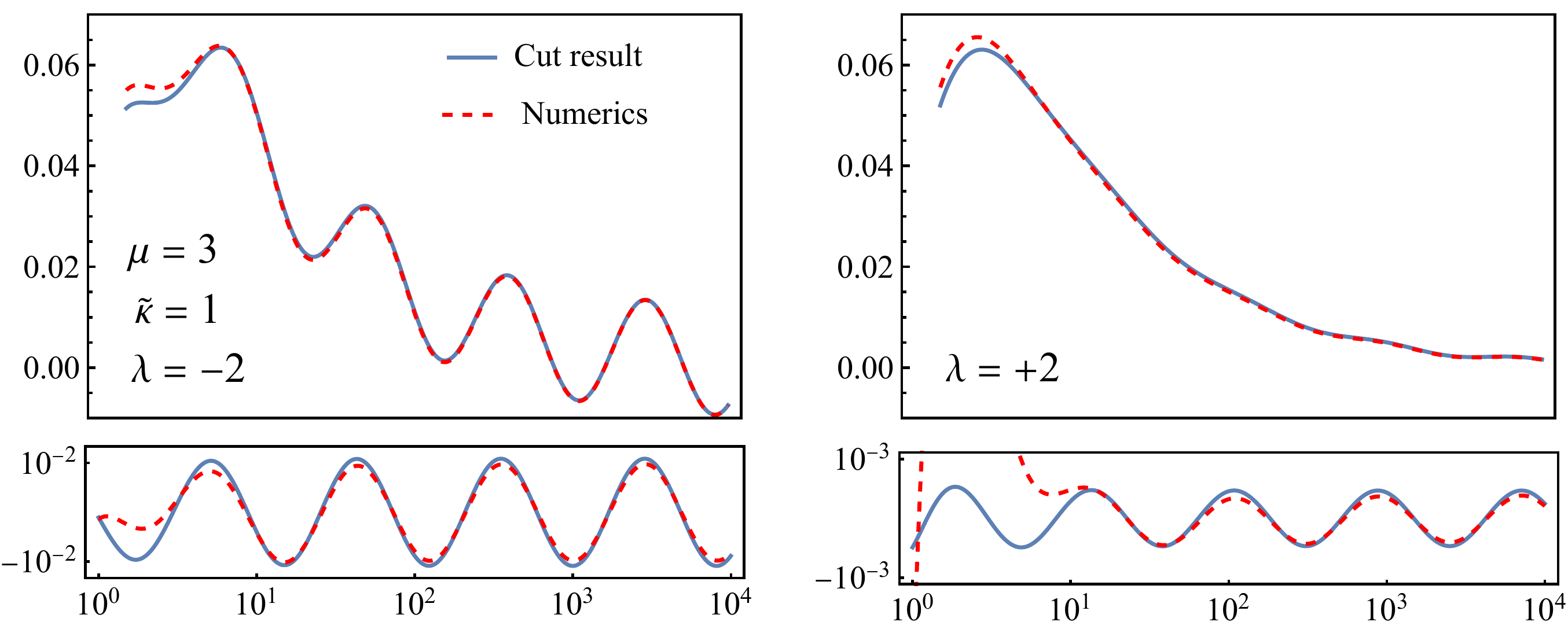}\\
		\includegraphics[width=14cm]{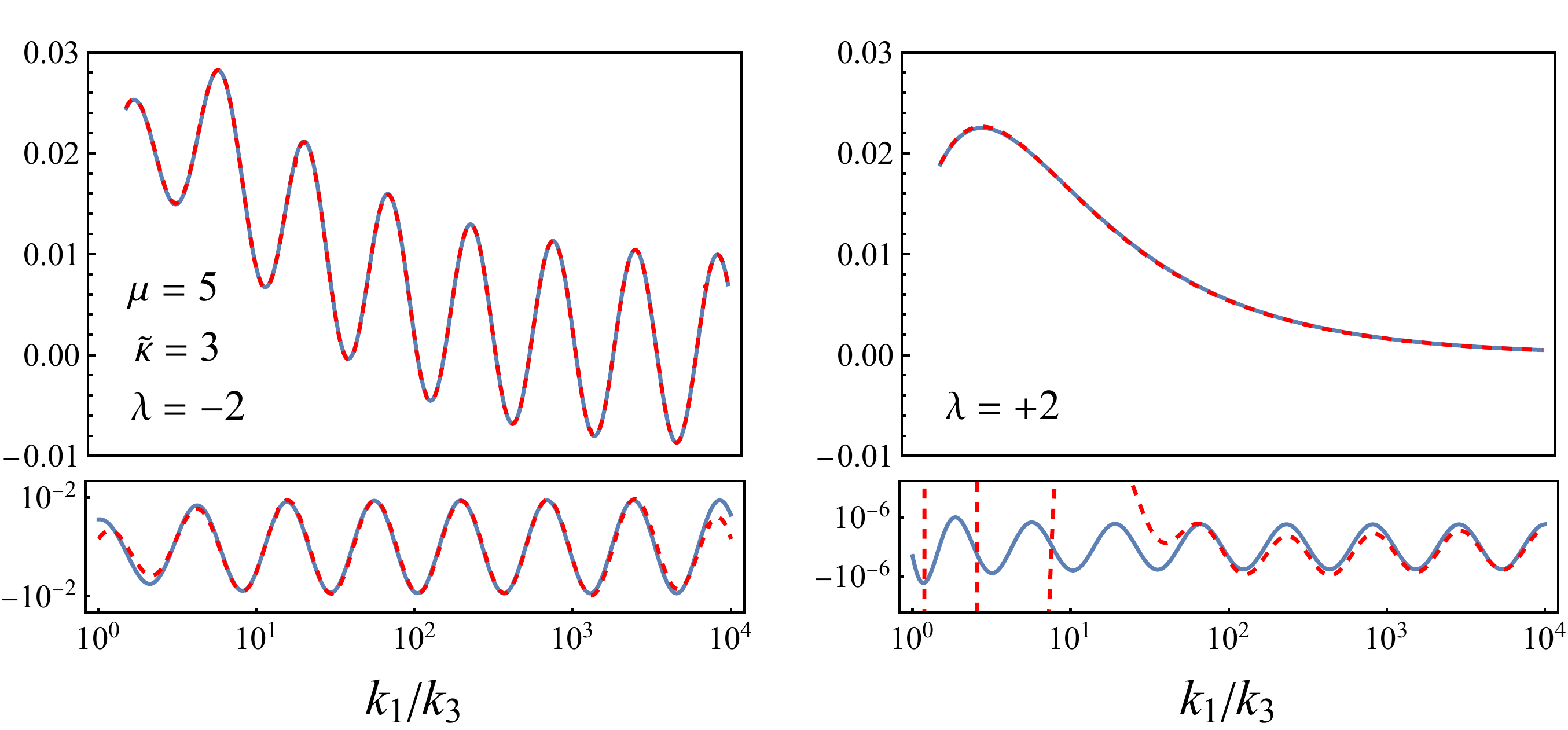}
		\caption{The CC signals in the $\langle\zeta\zeta\gamma_{\lambda}\rangle$ correlator for different mass $\mu$ and chemical potential $\tilde{\kappa}$. The correlator is normalized by $\left(k_1/k_3\right)^{9/2}k_3^6\langle\zeta(k_1)\zeta(k_1)\gamma_\lambda(k_3)\rangle'/\frac{\rho H^5}{M M_p^2\dot{\phi}_0}$, with the upper panels showing the full result and the lower panels showing the oscillatory CC signal only. The blue solid lines correspond to the analytical cut result while the red dashed lines correspond to numerical results (the signals are filtered from the total numeric integration result using high-pass filters with a Blackman window function).}\label{BphiphigammaCCS}
	\end{figure}

	The mixed correlator $\langle\zeta\zeta\gamma\rangle$ can be decomposed into different helicity components,
	\begin{equation}
		\langle\zeta(\mathbf{k}_1)\zeta(\mathbf{k}_2)\gamma_{ij}(\mathbf{k}_3)\rangle'=\left(\frac{H}{\dot{\phi}_0}\right)^2\sum_{\lambda=\pm 2} e^{\lambda*}_{ij} \langle\zeta(\mathbf{k}_1)\zeta(\mathbf{k}_2)\gamma_\lambda(\mathbf{k}_3)\rangle'~.
	\end{equation}
	More explicitly, the helicity-basis correlator can be written as the product of the kinematic factor and the dynamical factor,
	\begin{equation}
		\langle\zeta(\mathbf{k}_1)\zeta(\mathbf{k}_2)\gamma_\lambda(\mathbf{k}_3)\rangle'=\Pi^\lambda \mathcal{I}_\lambda~.
	\end{equation}
	The kinematic factor reduces to that of an associated Legendre polynomial,
	\begin{equation}
		\Pi^\lambda\equiv(\mathbf{k}_1)_i(\mathbf{k}_2)_j e^\lambda_{ij}(\mathbf{\hat{k}}_3)=-\frac{k_1^2}{3}P_2^2(\cos\theta_{13})~,
	\end{equation}
	reflecting the spin-2 nature of the field $h_{\mu\nu}$.
	\begin{figure}[h!]
		\centering
		\includegraphics[width=12cm]{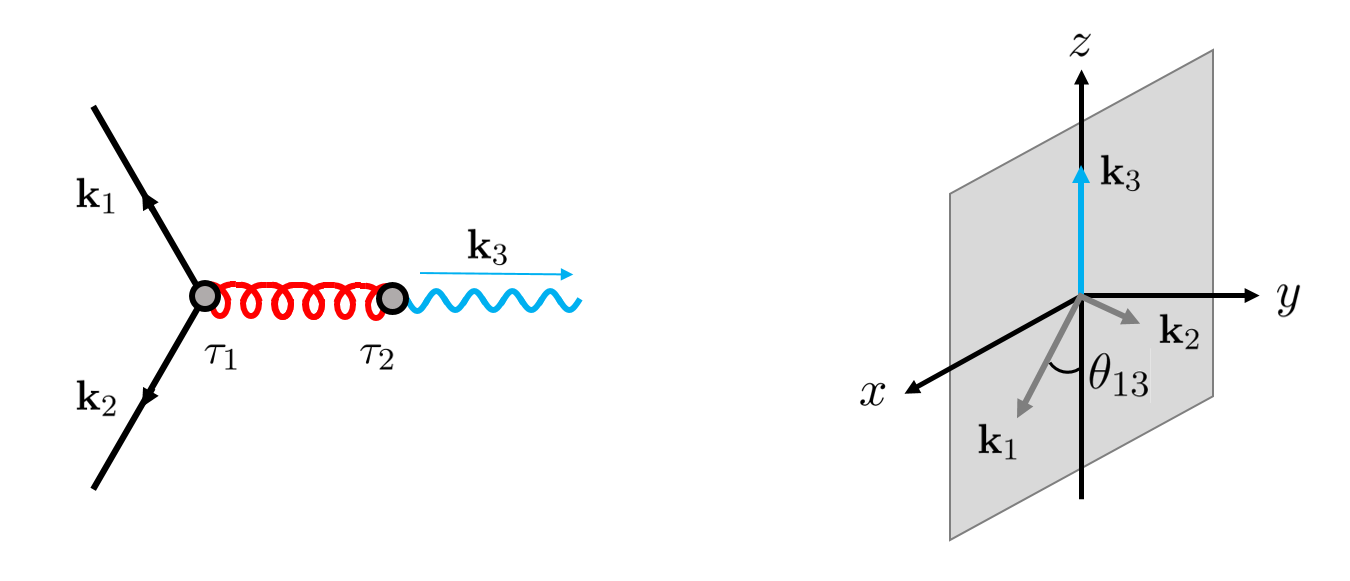}
		\caption{Left: The leading-order contribution of the amplified massive spin-2 field to $\langle\zeta\zeta\gamma\rangle$. Right: The kinematic configuration.}\label{phigammagammaDiag}
	\end{figure}
	The technical expression for the dynamical factor $\mathcal{I}_\lambda$ is given by (\ref{zetazetagammaI}). It can be evaluated analytically to leading order via the cutting rule. The total bispectrum is mimicked by the oscillatory signal part $S_\lambda$ and the analytic EFT\footnote{Note that this EFT stands for the effective theory of inflatons and gravitons with the massive spin-2 field integrated out, which is further in the IR than the EFT operators we are considering. The characteristic energy scales for them are $H<m$ and $\mathcal{M}_{*,\pm2}<\Lambda_{sc},\Lambda_{nl}$, respectively. As a matter of fact, much interesting physics can also be extracted from studying the graviton non-Gaussianities from this pure massless EFT point-of-view \cite{Bordin:2020eui,Cabass:2021fnw}.} background $B$.
	\begin{equation}
		\langle\zeta(\mathbf{k}_1)\zeta(\mathbf{k}_2)\gamma_\lambda(\mathbf{k}_3)\rangle'\simeq\Pi^\lambda\left( S_\lambda+B_\lambda\right)~.\label{zetazetagammaSplusB}
	\end{equation}
	The CC signal part can be analytically computed as
	\begin{align}
		\nonumber S_{\pm 2}=\frac{\rho H^5}{M M_p^2\dot{\phi}_0}&\frac{\pi  e^{\mp\pi  \tilde{\kappa }} \sinh \left(\pi  \tilde{\kappa }\right) \Gamma \left(2\mp i \tilde{\kappa }\right)\text{sech}^2(\pi  \mu ) }{4 \tilde{\kappa } \left(1\mp i\tilde{\kappa }\right)}\frac{1}{k_1^3 k_2^3 k_3^2}\\
		\nonumber&\times\Bigg\{\, {}_2\mathbf {F}_1\Bigg[\begin{array}{c} \frac{1}{2}-i \mu , \frac{1}{2}+i \mu\\[2pt] 1\mp i \tilde{\kappa }\end{array}\Bigg|\,\frac{k_{123}}{2 k_3}\Bigg]\\
		\nonumber&~~~~~-\frac{k_{12} }{2 k_3}\left(\mu ^2+\frac{1}{4}\right)\, {}_2\mathbf {F}_1\Bigg[\begin{array}{c} \frac{3}{2}-i \mu ,\frac{3}{2}+i \mu\\[2pt] 2\mp i \tilde{\kappa }\end{array}\Bigg|\,\frac{k_{123}}{2 k_3}\Bigg]\\
		\nonumber&~~~~~+\frac{k_1 k_2}{4 k_3^2}\left(\mu ^2+\frac{1}{4}\right) \left(\mu ^2+\frac{9}{4}\right)\, {}_2\mathbf {F}_1\Bigg[\begin{array}{c} \frac{5}{2}-i \mu ,\frac{5}{2}+i \mu\\[2pt] 3\mp i \tilde{\kappa }\end{array}\Bigg|\,\frac{k_{123}}{2 k_3}\Bigg]\Bigg\}+\text{c.c.}\\
		+\mathcal{O}\Big(|\beta_{\pm2}&|^2\Big)~,
	\end{align}
	Here we have used the abbreviation $k_{i_1\cdots i_m}\equiv k_{i_1}+\cdots+k_{i_m}$, and the regularized hypergeometric function ${}_2\mathbf {F}_1$ is defined from the usual hypergeometric function ${}_2{F}_1$ as
	\begin{align}
		{}_2\mathbf {F}_1\Bigg[\begin{array}{c} a,b\\[2pt] c \end{array}\Bigg|\,z\Bigg]\equiv
		{}_2{F}_1\Bigg[\begin{array}{c} a,b\\[2pt] c \end{array}\Bigg|\,z\Bigg]/\Gamma(c)~.
	\end{align}
	The leading order EFT background is generated from the effective vertex,
	\begin{equation}
		\int d\tau d^3x\frac{4\rho\dot{\phi}_0}{m^2 M}a\gamma'_{ij}\partial_i\varphi\partial_j\varphi~,
	\end{equation}
	which leads to
	\begin{equation}
		B_\lambda=-\frac{\rho H^7}{m^2 M M_p^2\dot{\phi}_0}\frac{2 k_{12}^2+2 k_1 k_2+3 k_{12} k_3+k_3^2}{k_1^3 k_2^3 k_3 k_{123}^3}~.\label{zetazetagammaEFTpart}
	\end{equation}
	In Fig.~\ref{BphiphigammaCCS}, we plot the total mixed bispectrum (\ref{zetazetagammaSplusB}) as well as the signals $S_\lambda$ for different parameter choices. The numerical result is also presented for comparison. It can be seen that they match very well in most parameter regions. The numerical CC signals for $\lambda=+2$ becomes less precise when the mass of the particle becomes large, because the heavy Boltzmann suppression demands an exponentially increasing numerical accuracy. This is not a problem for the enhanced mode $\lambda=-2$, where the oscillatory features are apparent.
	
\begin{figure}[h!]
		\centering
		\includegraphics[width=12cm]{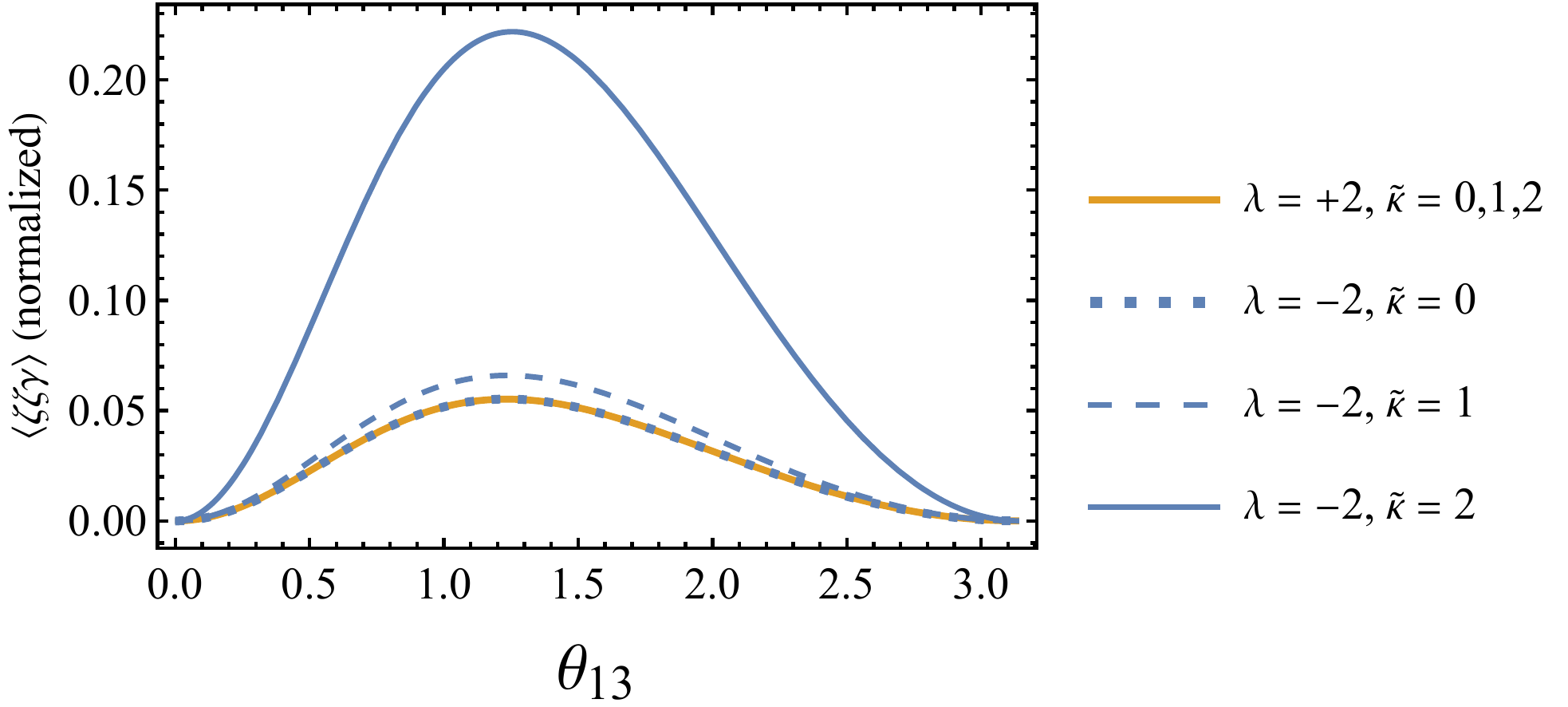}
		\caption{The angular dependence of the mixed bispectrum $\langle\zeta\zeta\gamma\rangle$ for different helicities and chemical potential choices. The horizontal axis is the angle $\theta_{13}$ between momenta $\mathbf{k}_1,\mathbf{k}_3$. The vertical axis is the mixed bispectrum normalized by $(k_1/k_3)^{9/2}k_3^6\langle\zeta(k_1)\zeta(k_2)\gamma_\lambda(k_3)\rangle'/\frac{\rho H^5}{M M_p^2\dot{\phi}_0}$. The momentum ratio is held fixed as $k_1/k_3=5$, with the mass of the spin-2 particle taken to be $\mu=3$. We observe the characteristic $P_2^2(\cos\theta_{13})$ dependence of the massive spin-2 particle. In addition, the chemical potential introduces a bias for the two helicities, and the $\lambda=-2$ mode is enhanced in amplitude.}\label{BphiphigammaAngleDep}
	\end{figure}

	The equilateral limit of the mixed bispectrum is dominated by the EFT contribution, with the signal strength characterized by the $f_{NL}$ parameter\footnote{An alternative definition is given by \cite{Lee:2016vti} $f_{NL,\zeta\zeta\lambda}=\frac{6}{17}\frac{\langle\zeta(\mathbf{k})\zeta(\mathbf{k})\gamma_\lambda(\mathbf{k})\rangle'}{\langle\zeta(\mathbf{k})^2\rangle'^{2}}\simeq \frac{\sqrt{r}\rho\dot{\phi}_0 H}{12 m^2 M}|\alpha_\lambda|^2$, with single field inflation predicting $\sum_{\lambda=\pm 2}f_{NL,\zeta\zeta\lambda}=\sqrt{r}/16$.}. 
	\begin{align}
		f_{NL,\zeta\zeta\lambda}&\equiv \frac{10}{9}\frac{\langle\zeta(k)\zeta(k)\gamma_\lambda(k)\rangle'}{\langle\zeta(k)^2\rangle'^{2}}\simeq 0.9\times|\alpha_\lambda|^2\left(\frac{\rho}{10}\right)\left(\frac{r}{0.05}\right)\left(\frac{m}{5H}\right)^{-2}\left(\frac{M}{20H}\right)^{-1}~,
	\end{align}
	where only the leading order EFT part (\ref{zetazetagammaEFTpart}) is used, and $r\lesssim 0.1$ is the tensor-to-scalar ratio. For a large chemical potential with $\tilde{\kappa}\gtrsim\mu$, the $f_{NL,\zeta\zeta-2}$ parameter receives exponential amplification from $|\alpha_{-2}|^2$ due to copious particle production. This can be much greater than that in the single field slow-roll inflation scenario, where $f_{NL,\zeta\zeta\lambda}\sim 0.1r\lesssim 0.01$. Furthermore, the squeezed-limit bispectrum is dominated by the signal part, and the helicity $\lambda=-2$ mode exhibits enhanced characteristic oscillations with the momentum ratio,
	\begin{align}
		\nonumber \lim_{k_3\to 0}\langle\zeta(\mathbf{k}_1)\zeta(\mathbf{k}_2)\gamma_\lambda(\mathbf{k}_3)\rangle'=&\frac{\pi\rho H^5\mathcal{A}(\mu,\tilde{\kappa})}{12\sqrt{2}M M_p^2\dot{\phi}_0} \frac{P_2^2(\cos\theta_{13})}{k_1^3 k_3^3}\left(\frac{k_3}{k_1}\right)^{3/2}|\alpha_{-2}^*\beta_{-2}|\sin\left(\mu\ln\frac{k_1}{k_3}+\chi(\mu,\tilde{\kappa})\right)\\
		&+\mathcal{O}\left(|\beta_{-2}|^2\right)~.
	\end{align}
	Here the dimensionless factor is
	\begin{equation}
		\mathcal{A}(\mu,\tilde{\kappa})\equiv \sqrt{\frac{\left(\mu ^2+\frac{9}{4}\right) \left(\mu ^2+\frac{81}{4}\right)}{\tilde\kappa  \mu }}~,
	\end{equation}
	and $\chi(\mu,\tilde{\kappa})$ is a phase dependent on the mass and chemical potential, whose detailed form is not important. Thus we see that the dS-invariant Boltzmann suppression of a massive spin-2 particle is largely alleviated, as a direct consequence of enhanced particle production $\beta_{-2}\sim e^{-\pi(\mu-\tilde{\kappa})}\gg e^{-\pi\mu}$. This is also reflected in Fig.~\ref{BphiphigammaCCS}, where large $\beta_{-2}$-suppressed oscillations are found for one helicity and smaller Boltzmann-suppressed oscillations are found for the other. One can also observe this in the angular dependence of the bispectrum (see Fig.~\ref{BphiphigammaAngleDep}). In the presence of chemical potential, the characteristic $P_2^2(\cos\theta_{13})$ behavior of the massive spin-2 mode acquires helicity-dependent modulations. In practice, an observation of such parity-violating angular dependence and CC signal oscillations will be a clear signature of spin-2 chemical potential.

	\subsubsection{$\langle\gamma\gamma\zeta\rangle$}
\begin{figure}[h!]
		\centering
		\includegraphics[width=13cm]{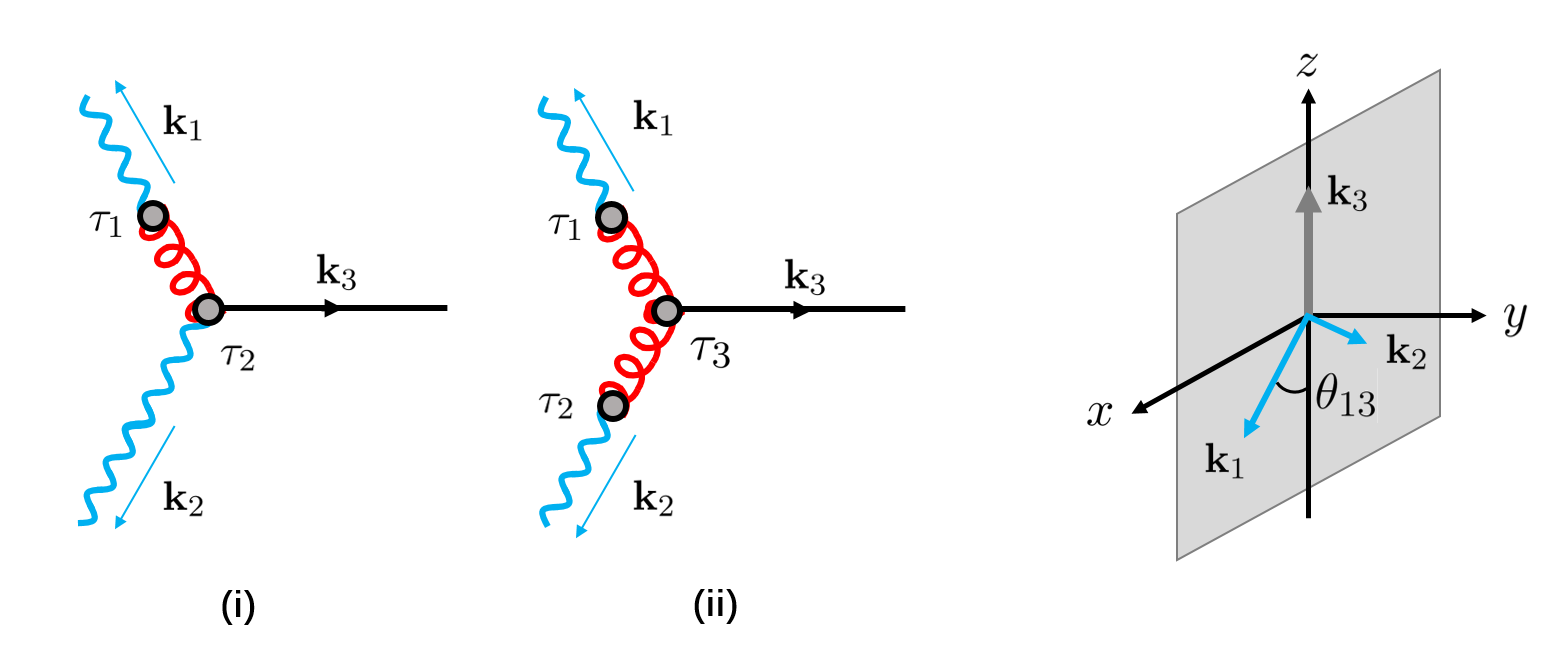}
		\caption{Left: The two tree diagrams contributing to $\langle\gamma\gamma\zeta\rangle$. Right: The kinematic configuration.}\label{gammagammaphiDiag}
	\end{figure}

	Another type of mixed bispectrum $\langle\gamma\gamma\zeta\rangle$ also arises in our model. It consists of two diagrams ((i) and (ii) in Fig.~\ref{gammagammaphiDiag}) whose kinematics and dynamics differ from each other, thus we consider them separately. We first decompose the correlator in terms of helicity components,
	\begin{align}
		\nonumber\langle\gamma_{ij}(\mathbf{k}_1)\gamma_{kl}(\mathbf{k}_2)\zeta(\mathbf{k}_3)\rangle'&=\sum_{\lambda_1,\lambda_2=\pm2}e^{\lambda_1*}_{ij}(\hat{\mathbf{k}}_1)e^{\lambda_2*}_{kl}(\hat{\mathbf{k}}_2)\langle\gamma_{\lambda_1}(\mathbf{k}_1)\gamma_{\lambda_2}(\mathbf{k}_2)\zeta(\mathbf{k}_3)\rangle'~.
	\end{align}
	Thus the helicity-basis 3-point function can be written as
	\begin{equation}
		\langle\gamma_{\lambda_1}(\mathbf{k}_1)\gamma_{\lambda_2}(\mathbf{k}_2)\zeta(\mathbf{k}_3)\rangle'=\Pi_{(\text{i})}^{\lambda_1\lambda_2}\mathcal{I}^{(\text{i})}_{\lambda_1\lambda_2}+\Pi_{(\text{ii})}^{\lambda_1\lambda_2}\mathcal{I}^{(\text{ii})}_{\lambda_1\lambda_2}~.
	\end{equation}
	The kinematic factors are
	\begin{align}
		\Pi_{(\text{i})}^{\lambda_1\lambda_2}&\equiv  e^{\lambda_1}_{ij}(\hat{\mathbf{k}}_1)e^{\lambda_2}_{ij}(\hat{\mathbf{k}}_2)~,\\
		\Pi_{(\text{ii})}^{\lambda_1\lambda_2}&\equiv\epsilon_{ijk}(\mathbf{k}_1-\mathbf{k}_2)_i e^{\lambda_1}_{jl}(\hat{\mathbf{k}}_1)e^{\lambda_2}_{kl}(\hat{\mathbf{k}}_2)/k_3~.
	\end{align}
	Here, four combinations of helicities are all allowed. In the soft limit $k_1\to 0$, the kinematic factors reduce to a characteristic angular dependence
	\begin{align}
		&\Pi_{(\text{i})}^{+2,+2}=\Pi_{(\text{i})}^{-2,-2}=i\Pi_{(\text{ii})}^{+2,+2}=-i\Pi_{(\text{ii})}^{-2,-2}=4\sin^4\frac{\theta_{13}}{2}~,\\
		&\Pi_{(\text{i})}^{-2,+2}=\Pi_{(\text{i})}^{+2,-2}=i\Pi_{(\text{ii})}^{-2,+2}=-i\Pi_{(\text{ii})}^{+2,-2}=4\cos^4\frac{\theta_{13}}{2}~.
	\end{align}
	The dynamical factors $\mathcal{I}^{(\text{i}),(\text{ii})}_{\lambda_1\lambda_2}$ are again given by (\ref{gammagammazetaIi}) and (\ref{gammagammazetaIii}) in the appendix. Evoking the cutting rule, we proceed in a diagram-by-diagram fashion. The first diagram is approximated as
	\begin{equation}
		\langle\gamma_{\lambda_1}(\mathbf{k}_1)\gamma_{\lambda_2}(\mathbf{k}_2)\zeta(\mathbf{k}_3)\rangle'_{(\text{i})}\simeq\Pi_{(\text{i})}^{\lambda_1\lambda_2}\left(S^{(\text{i})}_{\lambda_1\lambda_2}+B^{(\text{i})}_{\lambda_1\lambda_2}\right)~,\label{BgammagammazetaCutResulti}
	\end{equation}
	The CC signal in the $k_3/k_1$ channel is
	\begin{align}
		\nonumber S^{(\text{i})}_{\pm2,\lambda_2}=&\frac{i \pi \rho^2 H^4 }{32M_p^4}e^{\mp\pi  \tilde{\kappa }}  \sinh \left(\mp\pi  \tilde{\kappa }\right) \Gamma \left(\mp i \tilde{\kappa }\right)\text{sech}^2(\pi  \mu )\\
		&\times\frac{1}{k_1^4 k_2 k_3}\left(\mu ^2+\frac{1}{4}\right) \left(\mu ^2+\frac{9}{4}\right)\, {}_2\mathbf {F}_1\Bigg[\begin{array}{c} \frac{5}{2}-i \mu , \frac{5}{2}+i \mu\\[2pt] 3\mp i \tilde{\kappa }\end{array}\Bigg|\,\frac{k_{123}}{2 k_1}\Bigg]+\mathcal{O}\left(|\beta_{\pm2}|^2\right)+\text{c.c.}~,
	\end{align}
	and the CC signal in the $k_3/k_2$ channel is obtained via a simple replacement $k_1\leftrightarrow k_2$. The background $B^{(\text{i})}_{\lambda_1\lambda_2}$ is mimicked by the operator
	\begin{equation}
		\int d\tau d^3x \frac{\rho^2 H^2}{m^2} a\varphi'\gamma_{ij}'^2~,
	\end{equation}
	yielding
	\begin{equation}
		B^{(\text{i})}_{\lambda_1\lambda_2}=\frac{\rho^2 H^6}{4M_p^4m^2}\frac{1}{k_1 k_2 k_3 k_{123}^3}~.
	\end{equation}

	The second diagram is slightly more complicated. According to the cutting algorithm, to obtain the leading order CC signal, we can cut one massive propagator at a time, and approximate the others by local EFT operators. For instance, in the soft limit $k_1\ll k_2,k_3$, the massive spin-2 propagator with momentum $\mathbf{k}_2$ can be integrated out to give an effective operator
	\begin{equation}
		\int d\tau d^3x\frac{\rho\dot{\phi}_0}{2m^2\Lambda_{c}}a^{-1}\varphi'\epsilon_{ijk}\left(h_{il}\partial_j \gamma_{kl}'+\gamma_{il}'\partial_j h_{kl}\right)~.
	\end{equation}
	Applying the cutting rule, we can write the correlator as
	\begin{equation}
		\langle\gamma_{\lambda_1}(\mathbf{k}_1)\gamma_{\lambda_2}(\mathbf{k}_2)\zeta(\mathbf{k}_3)\rangle'_{(\text{ii})}\simeq\Pi_{(\text{ii})}^{\lambda_1\lambda_2}\left(S^{(\text{ii})}_{\lambda_1\lambda_2}+B^{(\text{ii})}_{\lambda_1\lambda_2}\right)~.\label{BgammagammazetaCutResult}
	\end{equation}
	The CC signal in this limit is then approximated by
	\begin{align}
		\nonumber S^{(\text{ii})}_{\pm2,\lambda_2}\simeq\frac{i\rho^2\dot{\phi}_0 H^3}{64M_p^4 m^2\Lambda_{c}}\Bigg\{&\frac{\pi e^{\mp\pi  \tilde{\kappa }} \sinh \left(\mp\pi  \tilde{\kappa }\right) \Gamma \left(\mp i \tilde{\kappa }\right)\text{sech}^2(\pi  \mu )}{k_1^5 k_2}\\
		\nonumber&\times\left(\mu ^2+\frac{1}{4}\right) \left(\mu ^2+\frac{9}{4}\right) \left(\mu ^2+\frac{25}{4}\right)\, {}_2\mathbf {F}_1\Bigg[\begin{array}{c} \frac{7}{2}-i \mu ,\frac{7}{2}+i \mu\\[2pt] 4\mp i \tilde{\kappa }\end{array}\Bigg|\,\frac{k_{123}}{2 k_ 1}\Bigg]\\
		&+\text{c.c.}\Bigg\}+\mathcal{O}\left(|\beta_{\pm2}|^2\right)~.
	\end{align}
	Interestingly, the background $B^{(\text{ii})}_{\lambda_1\lambda_2}$ in this case receives no contribution from the large-mass EFT at all orders. One might naively come up with a leading order large-mass EFT operator from integrating out all massive spin-2 propagators as
	\begin{equation}
		\int d\tau d^3x\frac{\rho^2\dot{\phi}^2}{m^4\Lambda_{c}}\varphi'\epsilon_{ijk}\gamma_{il}'\partial_j\gamma_{kl}'~.\label{naiveEFTOp}
	\end{equation}
	However, although this operator does affect the dynamics of the massless $(\varphi,\gamma)$-EFT, it fails to enter the correlator $\langle\gamma\gamma\zeta\rangle$ at tree-level. One can easily verify this by either directly computing $B_{\lambda_1\lambda_2}$ from (\ref{naiveEFTOp}), or more generally, from inspecting the diagram using the power-counting method presented in \cite{Liu:2019fag}. As a result, we conclude that the mixed-type non-Gaussianity generated via diagram $(\text{ii})$ does not admit a perturbative expansion in $\mu^{-1}$. Its leading order is already non-perturbatively suppressed,
	\begin{equation}
	B^{(\text{ii})}_{\lambda_1\lambda_2}=0+\mathcal{O}\left(e^{-\pi\mu}\right)~.
	\end{equation}
	
	\begin{figure}[h!]
		\centering
		\includegraphics[width=14cm]{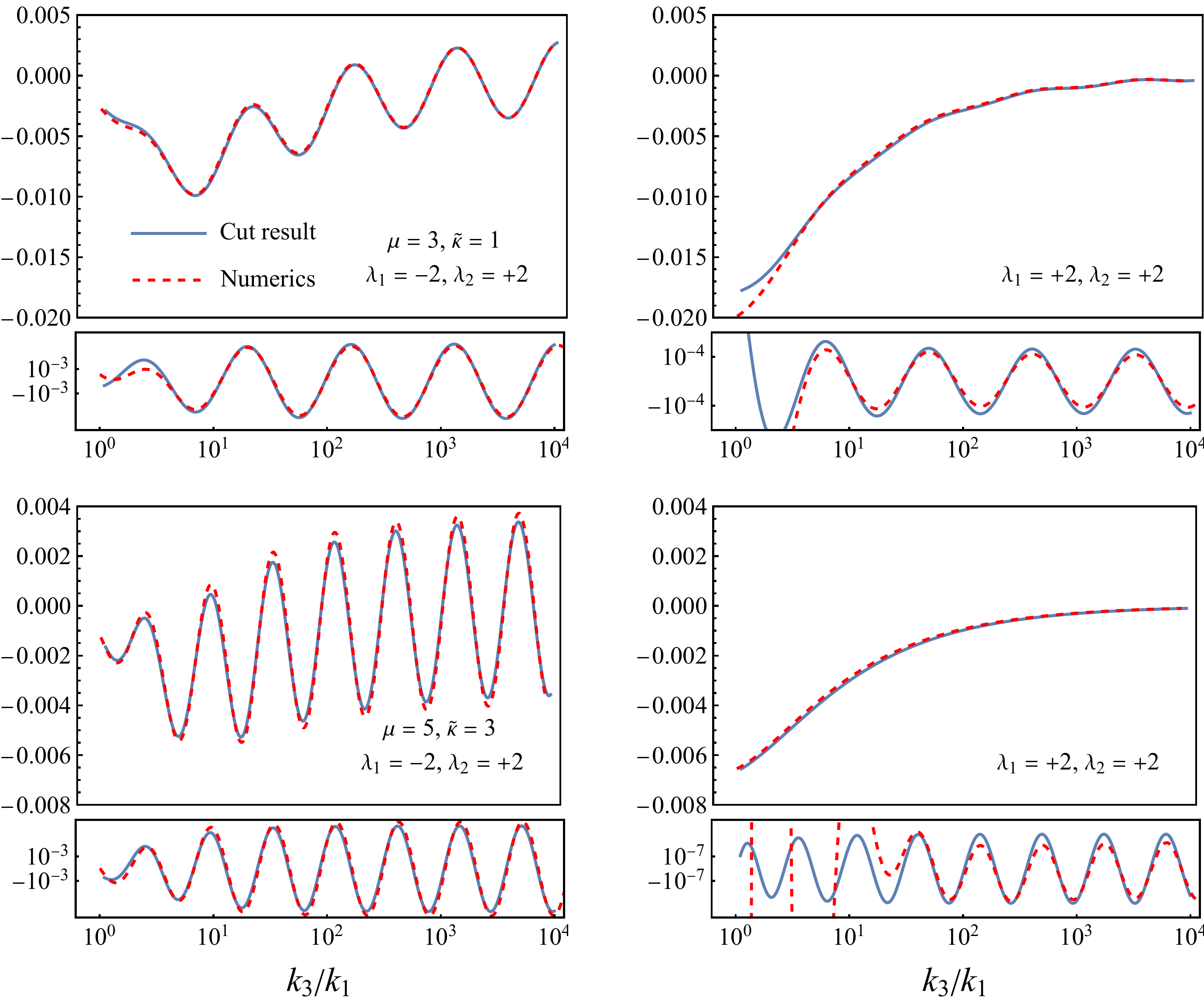}
		\caption{The CC signals in the $\langle\gamma_{\lambda_1}\gamma_{\lambda_2}\zeta\rangle_{(\text{i})}$ correlator for different choices of mass $\mu$ and chemical potential $\tilde{\kappa}$. For the purpose of demonstration, here we have only shown two helicity choices, namely $\lambda_1=-2,\lambda_2=+2$ (left) and $\lambda_1=\lambda_2=+2$ (right). The other helicities can also be calculated in a similar fashion. The correlator is normalized by $\left(k_3/k_1\right)^{9/2}k_1^6\langle\gamma_{\lambda_1}(k_1)\gamma_{\lambda_2}(k_1)\zeta(k_3)\rangle'_{(\text{i})}/\frac{\rho^2 H^4}{M_p^4}$, with the upper panels showing the full result and the lower panels showing the oscillatory CC signal only. The blue solid lines correspond to the analytical cut result while the red dashed lines correspond to numerical results (the signals are filtered from the total numeric integration result using high-pass filters with a Blackman window function).}\label{BgammagammaphidiagiCCS}
	\end{figure}
	
	\begin{figure}[h!]
		\centering
		~~~\includegraphics[width=14cm]{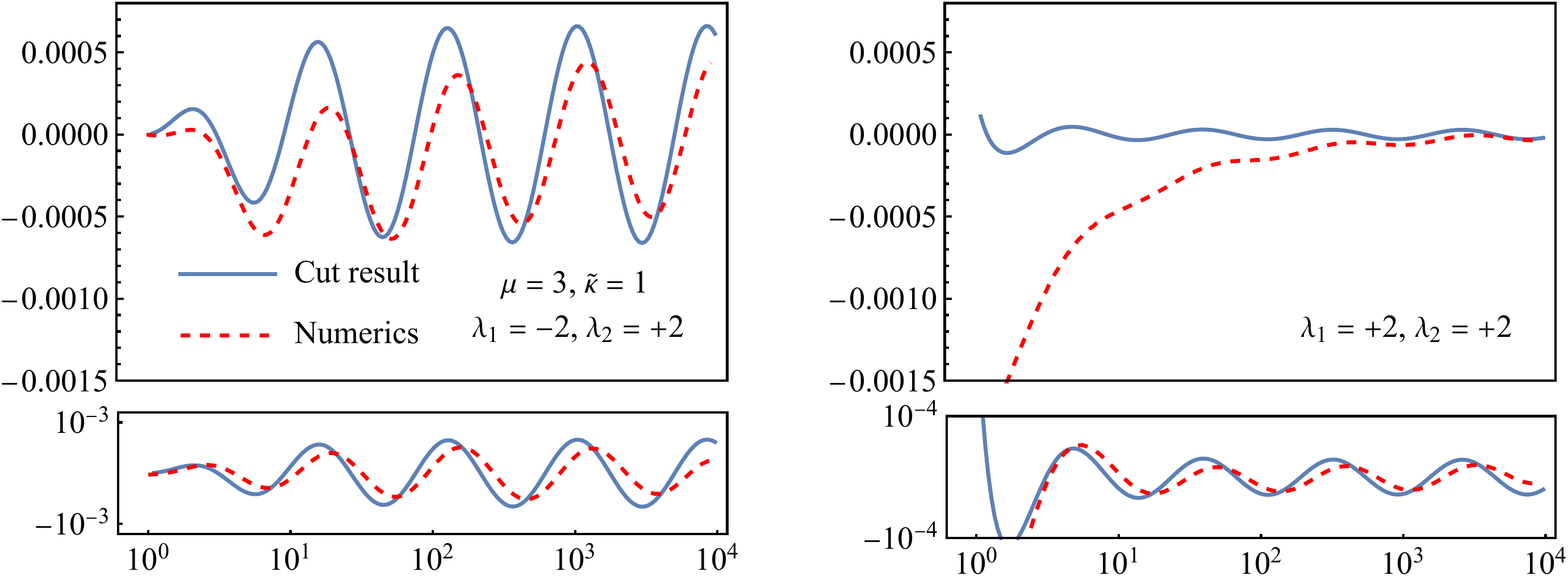}\\
		\includegraphics[width=14.2cm]{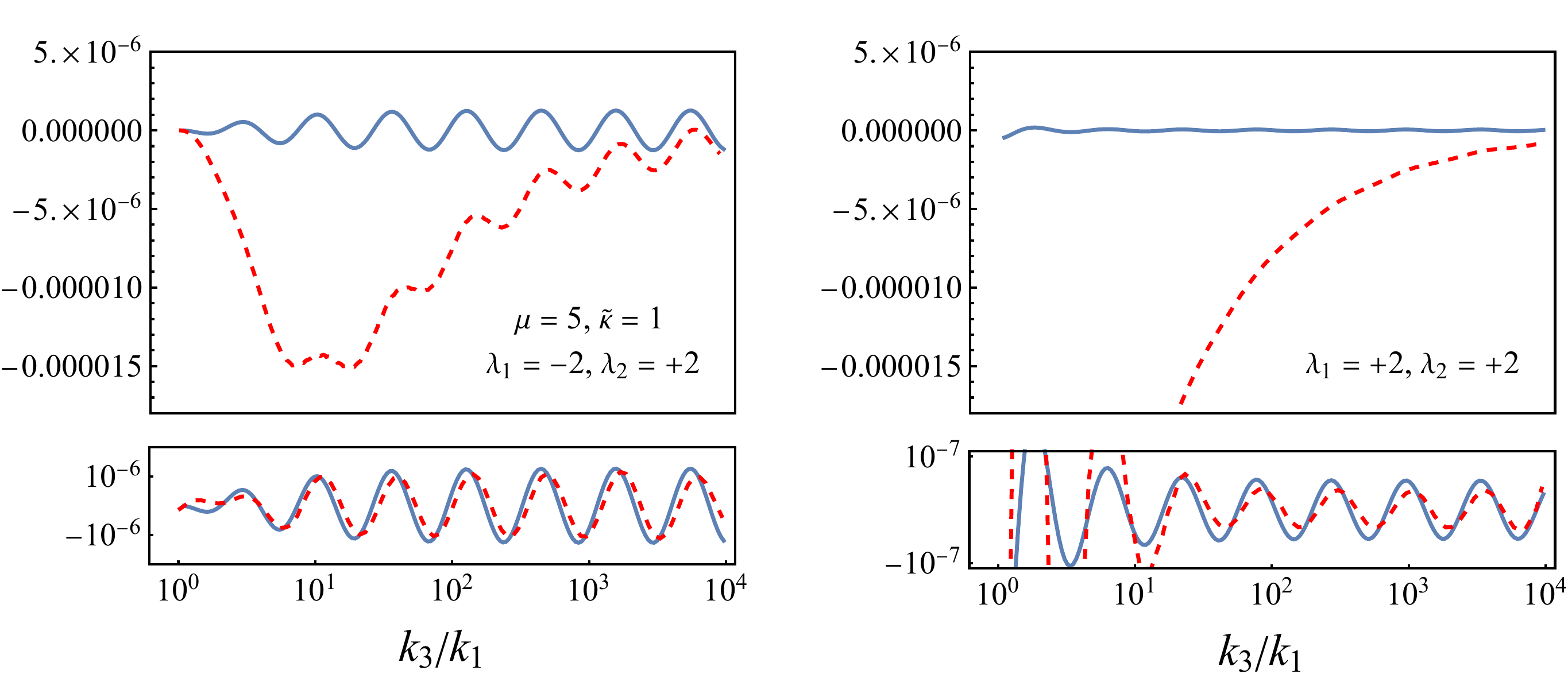}
		\caption{The CC signals in the $\langle\gamma_{\lambda_1}\gamma_{\lambda_2}\zeta\rangle_{(\text{ii})}$ correlator for different choices of mass $\mu$ and chemical potential $\tilde{\kappa}$. For the purpose of demonstration, here we have only shown two helicity choices, namely $\lambda_1=-2,\lambda_2=+2$ (left) and $\lambda_1=\lambda_2=+2$ (right). The other helicities can also be calculated in a similar fashion. The correlator is normalized by $\left(k_3/k_1\right)^{9/2}k_1^6\langle\gamma_{\lambda_1}(k_1)\gamma_{\lambda_2}(k_1)\zeta(k_3)\rangle'_{(\text{ii})}/\frac{\rho^2\dot{\phi} H^3}{M M_p^4\Lambda_{c}}$, with the upper panels showing the full result and the lower panels showing the oscillatory CC signal only. The blue solid lines correspond to the analytical cut result while the red dashed lines correspond to numerical results (the signals are filtered from the total numeric integration result using high-pass filters with a Blackman window function).}\label{BgammagammaphidiagiiCCS}
	\end{figure}

	We compare our cut results for both diagrams to numerical results in Fig.~\ref{BgammagammaphidiagiCCS} and Fig.~\ref{BgammagammaphidiagiiCCS}. Crucially, we see that for both diagrams, the oscillatory CC signals are well-approximated by the analytical cut results. Diagram (i) admits a $\mu^{-1}$ expansion, whose effect is captured very well with the leading order EFT background. In contrast, diagram (ii) has a non-zero background that is not covered by the cut result, as anticipated from the discussion above. It also becomes apparent that this background must be non-perturbative in $\mu^{-1}$, because the bispectrum amplitudes for $\mu=3$ and $\mu=5$ differ drastically by two orders of magnitude. Henceforth, in the equilateral shape limit, we can estimate the size of non-Gaussianity by
	\begin{align}
		\nonumber f_{NL,\lambda_1\lambda_2\zeta}&\equiv \frac{10}{9}\frac{\langle\gamma_{\lambda_1}(k)\gamma_{\lambda_1}(k)\zeta(k)\rangle'}{\langle\zeta(k)^2\rangle'^{2}}\\
		\nonumber&\simeq -1.3\times 10^{-4}\times|\alpha_{-2}|^2\left(\frac{\rho}{10}\right)^2\left(\frac{r}{0.05}\right)^2\left(\frac{m}{5H}\right)^{-2}\\
		&\quad+0.2\times\left(|\alpha_{\lambda_1}|^2|\alpha_{\lambda_2}^*\beta_{\lambda_2}|^2+|\alpha_{\lambda_1}^*\beta_{\lambda_1}|^2|\alpha_{\lambda_2}|^2\right)\tilde{\kappa}\left(\frac{\rho}{10}\right)^2\left(\frac{r}{0.05}\right)^2\left(\frac{m}{5H}\right)^{4}~,\label{fNLgammagammazeta}
	\end{align}
	where the two lines correspond to diagram (i) and (ii), respectively. Note that although the empirical mass power $m^4$ for diagram (ii) is positive, the implicit exponential suppression in the Bogoliubov coefficients ensures its size decreases fast with an increasing mass. The CC signals sourced by the enhanced $\lambda_1=-2$ mode take the form
	\begin{align}
		\nonumber \lim_{k_1\to 0}\langle\gamma_{-2}(\mathbf{k}_1)\gamma_{\pm2}(\mathbf{k}_2)\zeta(\mathbf{k}_3)\rangle_{(\text{i})}'=&\mp\frac{\pi\rho^2H^2\mathcal{B}(\mu,\tilde{\kappa})}{M_p^4} \frac{\cos^4\left(\frac{\theta_{13}}{2}+\frac{\pi}{4}\mp\frac{\pi}{4}\right)}{k_1^3 k_3^3}\left(\frac{k_1}{k_3}\right)^{3/2}\\
		&\times\left[|\alpha_{-2}^*\beta_{-2}|\sin\left(\mu\ln\frac{k_3}{k_1}+\varsigma(\mu,\tilde{\kappa})\right)+\mathcal{O}\left(|\beta_{-2}|^2\right)\right]~,\\
		\nonumber \lim_{k_1\to 0}\langle\gamma_{-2}(\mathbf{k}_1)\gamma_{\pm2}(\mathbf{k}_2)\zeta(\mathbf{k}_3)\rangle_{(\text{ii})}'=&\mp\frac{\pi\rho^2\dot{\phi} H^3\mathcal{C}(\mu,\tilde{\kappa})}{8\sqrt{2}M M_p^4\Lambda_{c}} \frac{\cos^4\left(\frac{\theta_{13}}{2}+\frac{\pi}{4}\mp\frac{\pi}{4}\right)}{k_1^3 k_3^3}\left(\frac{k_1}{k_3}\right)^{3/2}\\
		&\times\left[|\alpha_{-2}^*\beta_{-2}|\sin\left(\mu\ln\frac{k_3}{k_1}+\iota(\mu,\tilde{\kappa})\right)+\mathcal{O}\left(|\beta_{-2}|^2\right)\right]~.
	\end{align}
	Here
	\begin{equation}
		\mathcal{B}(\mu,\tilde{\kappa})\equiv \sqrt{\frac{\left(\mu ^2+\frac{1}{4}\right) \left(\mu ^2+\frac{9}{4}\right)}{\tilde\kappa  \mu }}~,\quad\mathcal{C}(\mu,\tilde{\kappa})\equiv \sqrt{\frac{\left(\mu ^2+\frac{1}{4}\right) \left(\mu ^2+\frac{9}{4}\right) \left(\mu ^2+\frac{25}{4}\right)}{\tilde\kappa  \mu }}~,
	\end{equation}
	and $\varsigma(\mu,\tilde{\kappa}),\iota(\mu,\tilde{\kappa})$ are parameter-dependent phases whose detailed form is irrelevant here. In principle, by observing the amplified oscillations as well as the angular dependence for different helicities, we are able to extract ample physical information of the massive spin-2 particle. However, (\ref{fNLgammagammazeta}) implies that the $\langle\gamma\gamma\zeta\rangle$ signal in our mode may be too small for a perturbative mixing diagram, and is inaccessible to current CMB observations such as Planck (which roughly requires $f_{NL,\gamma\gamma\zeta}\sim\mathcal{O}(100)$ \cite{Bartolo:2018elp}). Allowing a larger chemical potential can increase $\alpha_{-2},\beta_{-2}$ exponentially, yet the mixing between the graviton and the massive spin-2 particle will become non-perturbative. This scenario deserves a separate analysis and whether it can then be detected via future experiments is left for future works.

	\subsubsection{$\langle\zeta^4\rangle$}
\begin{figure}[h!]
		\centering
		\includegraphics[width=12cm]{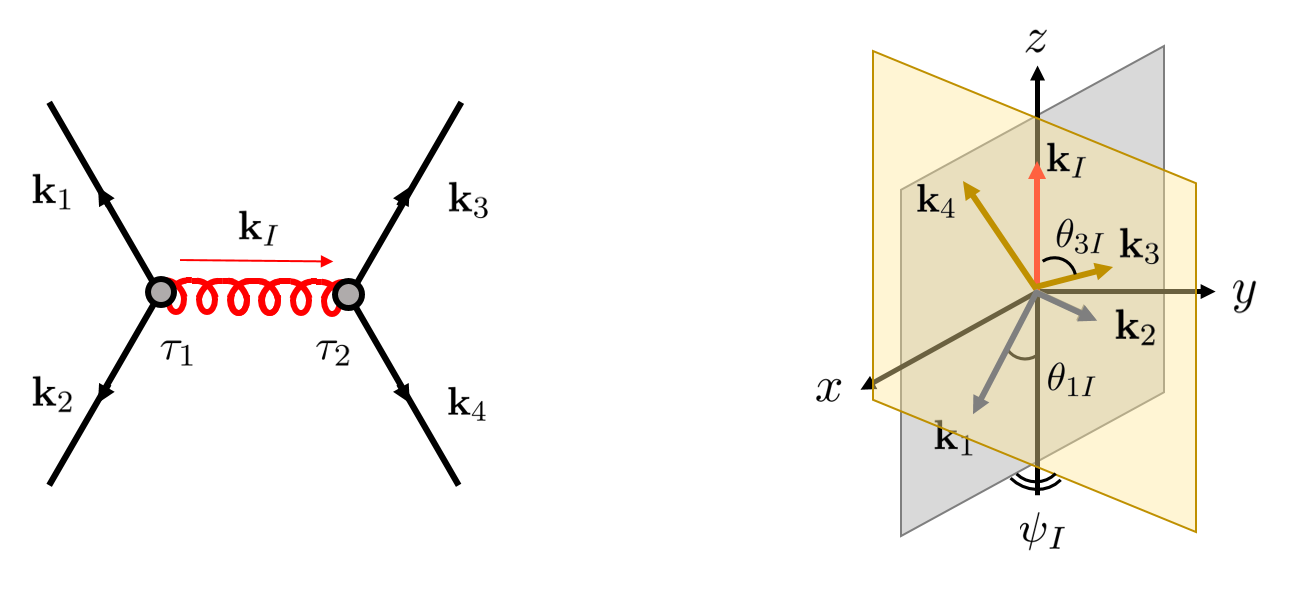}
		\caption{Left: The leading contribution of the amplified massive spin-2 field to $\langle\zeta^4\rangle$. Right: The kinematic configuration.}\label{phi4Diag}
	\end{figure}

\begin{figure}[h!]
		\centering
		\includegraphics[width=14cm]{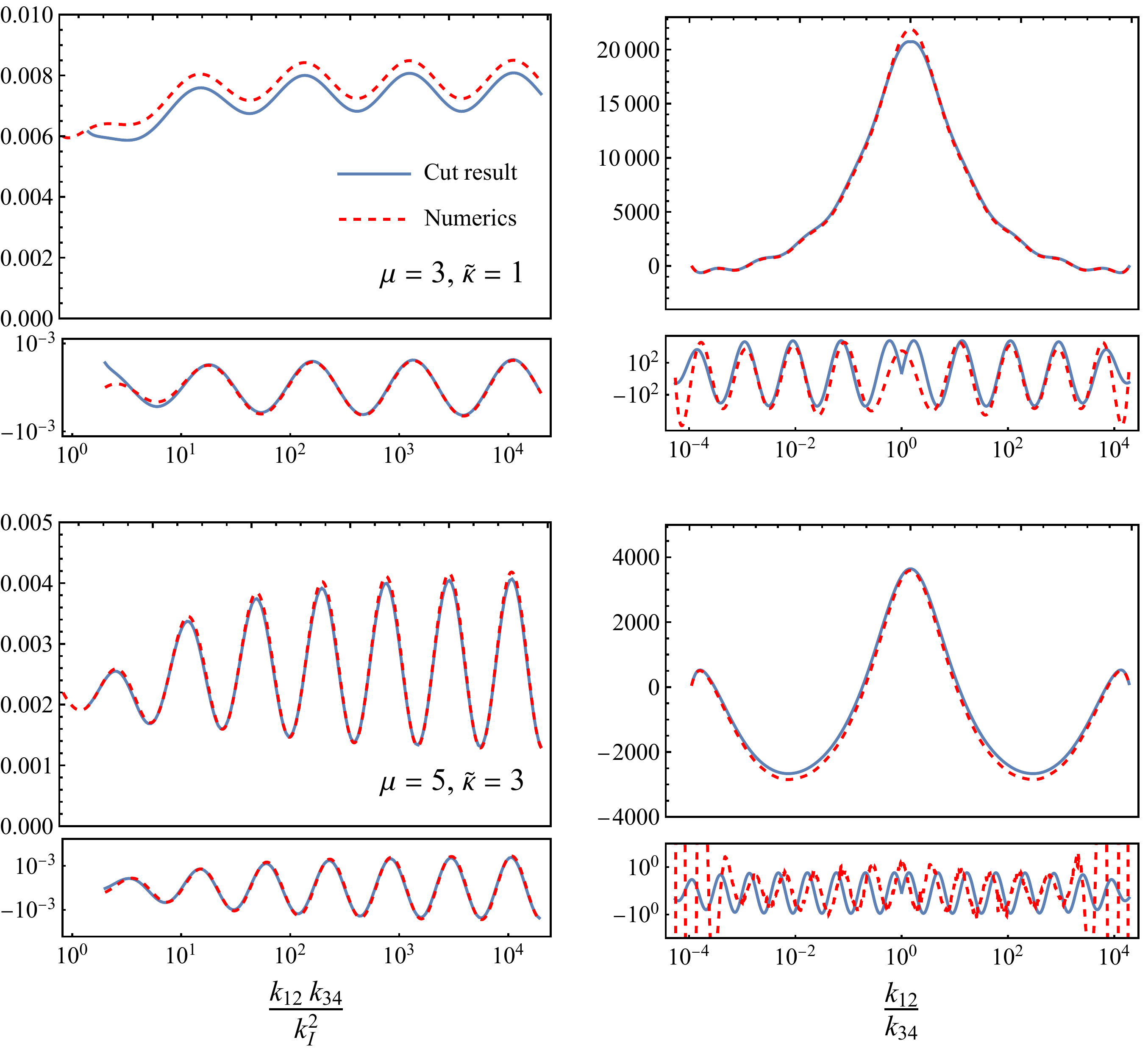}
		\caption{The CC signals in the $\langle\zeta^4\rangle$ for different combinations of mass $\mu$ and chemical potential $\tilde{\kappa}$ . Left column: The non-local type CC signal (oscillations in $k_{12}k_{34}/k_I^2$), with normalization $k_1^9\langle\zeta(\mathbf{k}_1)\zeta(\mathbf{k}_2)\zeta(\mathbf{k}_3)\zeta(\mathbf{k}_4)\rangle'_{s}/\frac{8H^{10}}{M^2\dot{\phi}_0^4}$ and momentum configuration $k_1=k_2=k_3=k_4$, $\psi_I=0$. Right column: The local type CC signal (oscillations in $k_{12}/k_{34}$), with normalization $k_1^6 k_3^6 k_I^{-3}\langle\zeta(\mathbf{k}_1)\zeta(\mathbf{k}_2)\zeta(\mathbf{k}_3)\zeta(\mathbf{k}_4)\rangle'_{s}/\frac{H^{10}}{M^2\dot{\phi}_0^4}$ and momentum configuration $k_1=k_2$, $k_3=k_4$, $\frac{k_{12}k_{34}}{k_I^2}=2\times10^{4}$, $\psi_I=0$. The upper panels show the full result while the lower panels show the oscillatory CC signal only, with the blue solid lines representing the analytical cut result and the red dashed line representing that from numerical results (the signals are filtered from the total numeric integration result using high-pass filters with a Blackman window function).}\label{Tphi4CCSignals}
	\end{figure}

	In addition to the mixed bispectra, the spin-2 particle also leaves characteristic imprints on the curvature trispectrum. As mentioned before, with a non-zero chemical potential, the massive spin-2 particle only propagates two tensor DoFs and a longitudinal scalar DoF. The trispectrum will thus receive contributions from only three helicity channels, as opposed to five in the ordinary case. Of these three helicities, the $\lambda=-2$ mode is enhanced, the $\lambda=0$ mode is unaffected, and the $\lambda=+2$ mode is suppressed. This special feature is again observable in the 4-pt correlator
	\begin{equation}
		\langle\zeta(\mathbf{k}_1)\zeta(\mathbf{k}_2)\zeta(\mathbf{k}_3)\zeta(\mathbf{k}_4)\rangle'=\langle\zeta(\mathbf{k}_1)\zeta(\mathbf{k}_2)\zeta(\mathbf{k}_3)\zeta(\mathbf{k}_4)\rangle'_{s}+(\text{$t,u$-channels})~,
	\end{equation}
	In the $s$-channel, the contributions from different helicities can be separated,
	\begin{equation}
		\langle\zeta(\mathbf{k}_1)\zeta(\mathbf{k}_2)\zeta(\mathbf{k}_3)\zeta(\mathbf{k}_4)\rangle'_{s}=\sum_{\lambda=0,\pm2}\Pi^\lambda \mathcal{J}_\lambda~,
	\end{equation}
	with the kinematic factor
	\begin{equation}
		\Pi^{\lambda}\equiv(\mathbf{k}_1)_i(\mathbf{k}_2)_j e^{\lambda}_{ij}(\mathbf{\hat{k}}_I)(\mathbf{k}_3)_m(\mathbf{k}_4)_n e^{\lambda}_{mn}(-\mathbf{\hat{k}}_I)~,
	\end{equation}
	and the dynamical factor $\mathcal{J}_\lambda$ whose explicit form is given by (\ref{zeta4J}). Choosing $\mathbf{k}_I$ to align along the $+z$-axis, the kinematic factor then reduces to that of a product of associated Legendre polynomials,
	\begin{align}
		\Pi^{\pm2}&=\frac{8k_1^2 k_3^2}{3}e^{\mp2i\psi_I} P_2^2(\cos\theta_{1I})P_2^{-2}(\cos\theta_{3I})~,\\
		\Pi^{0}&=k_1^2 k_3^2 P_2^0(\cos\theta_{1I})P_2^{0}(\cos\theta_{3I})+\mathcal{O}\left(k_I\right)~,
	\end{align}
	where $\psi_I$ is the dihedral angle between two planes spanned by $\mathbf{k}_1,\mathbf{k}_2$ and $\mathbf{k}_3,\mathbf{k}_4$ (see Fig.~\ref{phi4Diag}).
	Using the cutting rule, we can evaluate the 4-pt correlator analytically at the leading order as a combination of CC signals and an analytical background,
	\begin{equation}
		\langle\zeta(\mathbf{k}_1)\zeta(\mathbf{k}_2)\zeta(\mathbf{k}_3)\zeta(\mathbf{k}_4)\rangle'_{s}\simeq\sum_{\lambda=0,\pm2}\Pi^\lambda S_\lambda+B~.
	\end{equation}
	The signal part again consists combinations of Hypergeometric functions,
	\begin{align}
		S_{\pm 2}&=-\frac{H^{10}}{M^2\dot{\phi}^4}\frac{\pi ^2 e^{\mp\pi  \tilde{\kappa }} \text{sech}^2(\pi  \mu )}{16(k_1 k_2 k_3 k_4 )^3 k_I}\mathcal{K}^{\pm 2}_L(k_1,k_2,k_I)\mathcal{K}^{\pm 2}_R(k_3,k_4,k_I)+\text{c.c.}+\mathcal{O}\left(|\beta_{\pm2}|^2\right)~,\\
		S_{0}&=-\frac{H^{10}}{M^2\dot{\phi}^4}\frac{\pi^2 \text{sech}^2(\pi  \mu )}{96(k_1 k_2 k_3 k_4 )^3 k_I}\frac{\mu^2+9/4}{\mu^2+1/4}\mathcal{K}^{0}_L(k_1,k_2,k_I)\mathcal{K}^{0}_R(k_3,k_4,k_I)+\text{c.c.}+\mathcal{O}\left(|\beta_{0}|^2\right)~.
	\end{align}
	Here, we have abbreviated
	\begin{align}
		\nonumber\mathcal{K}^{\pm 2}_L(k_1,k_2,k_I)\equiv&\, {}_2\mathbf {F}_1\Bigg[\begin{array}{c} \frac{1}{2}-i \mu , \frac{1}{2}+i \mu\\[2pt] 1\mp i \tilde{\kappa }\end{array}\Bigg|\,\frac{k_{12I}}{2 k_I}\Bigg]\\
		\nonumber&-\frac{k_{12} }{2 k_I}\left(\mu ^2+\frac{1}{4}\right)\, {}_2\mathbf {F}_1\Bigg[\begin{array}{c} \frac{3}{2}-i \mu ,\frac{3}{2}+i \mu\\[2pt] 2\mp i \tilde{\kappa }\end{array}\Bigg|\,\frac{k_{12I}}{2 k_I}\Bigg]\\
		&+\frac{k_1 k_2}{4 k_I^2}\left(\mu ^2+\frac{1}{4}\right) \left(\mu ^2+\frac{9}{4}\right)\, {}_2\mathbf {F}_1\Bigg[\begin{array}{c} \frac{5}{2}-i \mu ,\frac{5}{2}+i \mu\\[2pt] 3\mp i \tilde{\kappa }\end{array}\Bigg|\,\frac{k_{12I}}{2 k_I}\Bigg]~,\\
		\nonumber\mathcal{K}^{0}_L(k_1,k_2,k_I)\equiv&\left(\mu -\frac{i}{2}\right) \left(\mu +\frac{9 i}{2}\right) \, {}_2\mathbf {F}_1\Bigg[\begin{array}{c} \frac{1}{2}-i \mu ,\frac{1}{2}+i \mu\\[2pt] 1\end{array}\Bigg|\,\frac{k_{12I}}{2k_I}\Bigg]\\
		&-4 i \left(\mu +\frac{i}{2}\right) \, {}_2\mathbf {F}_1\Bigg[\begin{array}{c} \frac{3}{2}-i \mu ,\frac{1}{2}+i \mu\\[2pt] 1\end{array}\Bigg|\,\frac{k_{12I}}{2k_I}\Bigg]+\mathcal{O}\left(\frac{k_{I}}{k_{12}},\frac{k_{I}}{k_{34}}\right)~,
	\end{align}
	and
	\begin{align}
		\mathcal{K}^{\pm 2}_R(k_3,k_4,k_I)&\equiv \mathcal{K}^{\pm 2}_L(k_1,k_2,k_I)|_{k_{1,2}\to {\scalebox{0.75}[1.0]{$-$}}k_{3,4},\tilde{\kappa}\to{\scalebox{0.75}[1.0]{$-$}}\tilde{\kappa}}~,\\
		\mathcal{K}^{0}_R(k_3,k_4,k_I)&\equiv \mathcal{K}^{0}_L(k_1,k_2,k_I)|_{k_{1,2}\to {\scalebox{0.75}[1.0]{$-$}}k_{3,4}}~.
	\end{align}
	Due to the lack of vector modes, the analytic background cannot be fully mimicked by local EFT operators. However, in special kinematic configurations, namely, when $\mathbf{k}_{1},\cdots, \mathbf{k}_{4}$ are approximately orthogonal to $\mathbf{k}_I$, it can be well-approximated by the effect of the operator
	\begin{equation}
		\int d\tau d^3x\frac{1}{m^2M^2}(\partial_i\varphi)^4~.
	\end{equation}
	In the $s$-channel, the analytic background reads
	\begin{align}
		\nonumber B=&\frac{H^{12}}{m^2M^2\dot{\phi}^4}\frac{(\mathbf{k}_1\cdot \mathbf{k}_3)(\mathbf{k}_2\cdot \mathbf{k}_4)+(\mathbf{k}_1\cdot \mathbf{k}_4)(\mathbf{k}_2\cdot \mathbf{k}_3)-(\mathbf{k}_1\cdot \mathbf{k}_2)(\mathbf{k}_3\cdot \mathbf{k}_4)}{2(k_1 k_2 k_3 k_4)^3}\\
		&\times\left(\frac{1}{k_{1234}}+\frac{k_1 k_2+k_3 k_4+k_{12} k_{34}}{k_{1234}^3}+\frac{3 \left(k_1 k_2 k_{34}+k_3 k_4 k_{12}\right)}{k_{1234}^4}+\frac{12 k_1 k_2 k_3 k_4}{k_{1234}^5}\right)~.
	\end{align}
	We plot the total trispectrum and the signals therein in Fig.~\ref{Tphi4CCSignals}. Again, the cut result agrees well with that from numerical results, since the vector component of the EFT background $B$ is subdominant in the chosen kinematic configurations. It becomes clear from the plot that the non-local type CC signal (for a explicit classification of signals, see Appendix~\ref{CRAppendix}) receive enhancement since it is interpreted as the dynamical phase of the particles produced by chemical potential. The local type signal, in contrast, comes from vertex particle production, for which dS symmetry is respected. For that reason, it is subjected to the original Boltzmann suppression, making its amplitude extremely small for large masses.

\begin{figure}[h!]
		\centering
		\includegraphics[width=15cm]{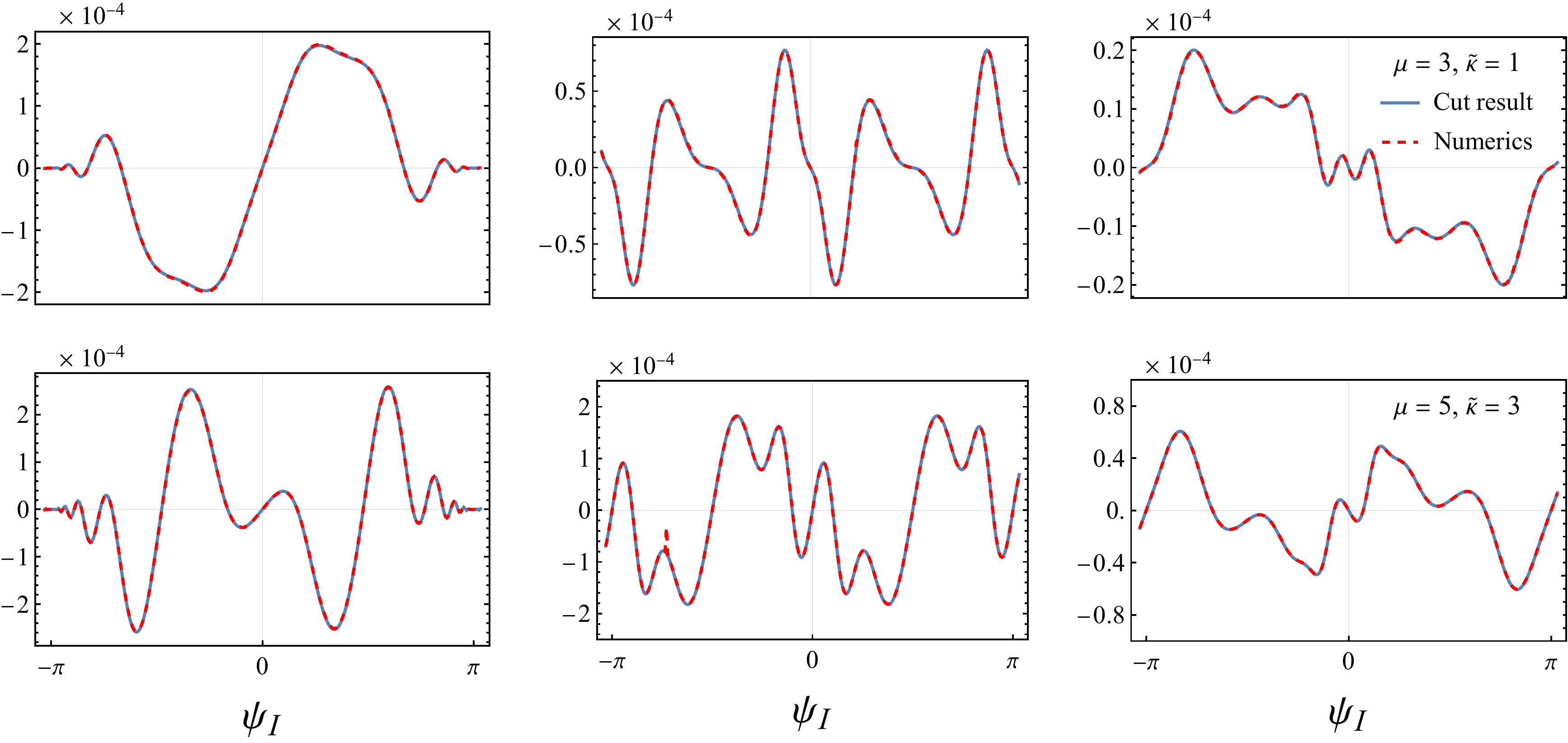}\\~\\
		~~~~\includegraphics[width=13cm]{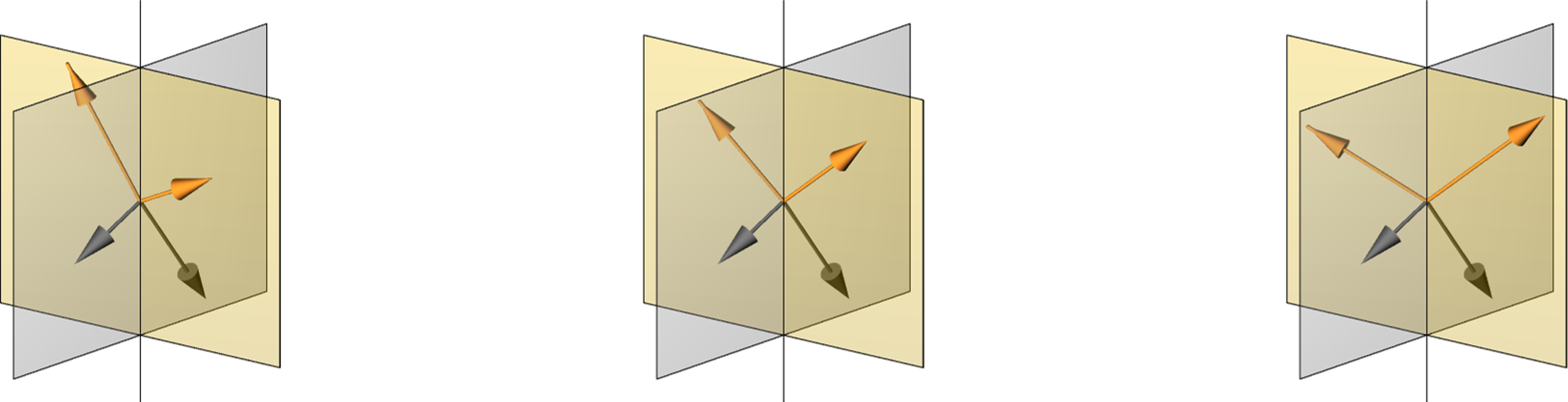}
		\caption{The imaginary part of the trispectrum $\Im\langle\zeta^4\rangle/\frac{H^{10}}{M^2\dot{\phi}_0^4}$ as a function of the dihedral angle $\psi_I$ for different momentum configurations and parameter choices. Left panel: $k_1=k_3=k_I/2,\theta_{1I}=\theta_{3I}=\pi/3$. Middle panel: $k_1=k_3/\sqrt{2}=k_I/2,\theta_{1I}=\pi/3,\theta_{3I}=\pi/4$. Right panel: $k_1=k_3/2=k_I/2,\theta_{1I}=\pi/3,\theta_{3I}=\pi/4$. The blue solid lines representing the analytical cut result and the red dashed line representing that from numerical results.}\label{Tphi4DihedralAngleDep}
	\end{figure}

	The trispectrum size estimator $g_{NL}$ is defined in the regular tetrahedron limit ($k_1=k_2=k_3=k_4=\frac{\sqrt{3}}{2}k_I$ with $\phi=\frac{\pi}{2}$) as,
	\begin{equation}
		g_{NL}\equiv\frac{25}{216}\frac{\langle\zeta(\mathbf{k}_1)\zeta(\mathbf{k}_2)\zeta(\mathbf{k}_3)\zeta(\mathbf{k}_4)\rangle'}{\langle\zeta(\mathbf{k}_1)^2\rangle'^3}\simeq 230\times|\alpha_{-2}|^2 \left(\frac{m}{5H}\right)^{-2}\left(\frac{M}{20H}\right)^{-2}~.
	\end{equation}
	In the collapsed limit, the trispectrum simplifies to the following characteristic structure,
	\begin{align}
		\nonumber&\lim_{k_I\to 0}\langle\zeta(\mathbf{k}_1)\zeta(\mathbf{k}_2)\zeta(\mathbf{k}_3)\zeta(\mathbf{k}_4)\rangle'_{s}\\
		\nonumber=&\frac{H^{10}}{48M^2\dot{\phi}^4} \frac{\pi\mathcal{D}(\mu)}{\left(k_1 k_3\right){}^{9/2}}\Bigg\{P_2^2(\cos\theta_{1I})P_2^{-2}(\cos\theta_{3I})\Bigg[e^{-2i\phi_I} |\alpha_{+2}^*\beta_{+2}|\sin \left(\mu  \ln
		\frac{k_{12} k_{34}}{k_{I}^2}+\xi(\mu,-\tilde{\kappa}) \right)\\
		\nonumber&\qquad\qquad\qquad\qquad\qquad\qquad\qquad\qquad\qquad\quad+e^{2i\phi_I} |\alpha_{-2}^*\beta_{-2}|\times\sin \left(\mu  \ln
		\frac{k_{12} k_{34}}{k_{I}^2}+\xi(\mu,\tilde{\kappa}) \right)\\
		\nonumber&\qquad\qquad\qquad\qquad\qquad\qquad\qquad\qquad\qquad\quad+2\cos(2\phi_I)e^{-\pi  \mu
		}\times\sin \left(\mu  \ln \frac{k_{12}}{k_{34}}\right)\Bigg]\\
		\nonumber&\qquad\qquad\qquad\qquad+P_2^0(\cos\theta_{1I})P_2^{0}(\cos\theta_{3I})\Bigg[|\alpha_{0}^*\beta_{0}|\sin\left(\mu\ln\frac{k_{12}k_{34}}{k_I^2}+\eta(\mu)\right)\\
		&\qquad\qquad\qquad\qquad\qquad\qquad\qquad\qquad\qquad\quad+e^{-\pi\mu}\sin\left(\mu\ln\frac{k_{12}}{k_{34}}\right)\Bigg]\Bigg\}+\mathcal{O}\left(|\beta_{0,\pm2}|^2\right)~,\label{trispectrumCCSignal}
	\end{align}
	where $\xi(\mu,\tilde{\kappa}),\eta(\mu)$ are parameter-dependent phases whose detailed form is of no importance here. And the overall factor is
	\begin{equation}
		\mathcal{D}(\mu)\equiv\frac{\left(\mu ^2+\frac{9}{4}\right) \left(\mu ^2+\frac{81}{4}\right)}{\mu}~.
	\end{equation}
	The various angular and shape dependence of the trispectrum is apparent from (\ref{trispectrumCCSignal}). The tensor modes $\lambda=\pm 2$ are separated from the scalar mode $\lambda=0$ in terms of their dependence on $\theta_{1I}\theta_{3I},\psi_i$, and each of them has their own amplitudes for the two types of CC signals. In the absence of the chemical potential, the first two rows in (\ref{trispectrumCCSignal}) combine into a \textit{real} number with a size comparable to the third row. However, introducing a non-zero chemical potential breaks this degeneracy. For $\tilde{\kappa}>0$, $|\beta_{-2}|>e^{-\pi\mu}>|\beta_{+2}|$, thus the dominating contribution comes from the second row, namely the non-local type CC signal contributed by the helicity $\lambda=-2$ mode. In addition, the trispectrum grows an \textit{imaginary} part because of the misalignment of the first two rows in (\ref{trispectrumCCSignal}). This is a unique signature of parity violation on the trispectrum, as suggested in \cite{Liu:2019fag}. The imaginary part is an odd function of the dihedral angle $\psi_I$ between two planes spanned by $\mathbf{k}_1,\mathbf{k}_2$ and $\mathbf{k}_3,\mathbf{k}_4$. Furthermore, due to a no-go theorem on parity violation in the single-field EFT, the parity-violating imaginary part must be proportional to a factor $e^{-\pi\mu}$, which is non-perturbative in $\mu^{-1}$ and is thence invisible at EFT level. This is indeed confirmed in (\ref{trispectrumCCSignal}):
	\begin{equation}
		\lim_{k_I\to 0}\Im \langle\zeta(\mathbf{k}_1)\zeta(\mathbf{k}_2)\zeta(\mathbf{k}_3)\zeta(\mathbf{k}_4)\rangle'|_{s}=e^{-\pi\mu}\sinh(\pi\tilde{\kappa})\sin(2\psi_I)\times (\text{real \& even function in $\psi_I$})~.
	\end{equation}
	This fact persists away from the collapsed limit even after taking into account the contributions from $t$-channel and $u$-channel. We verify this dependence on the dihedral angle in Fig.~\ref{Tphi4DihedralAngleDep}, where we have summed all three channels for different momentum configurations. The perfect agreement between the numerical results and the analytical cut result confirms that the imaginary part of the trispectrum comes purely from the non-perturbative CC signal $S_{\pm 2}$ cut from the transverse tensor modes.

	\subsection{Parameter space}
	Having exploited the rich phenomenology of the massive spin-2 chemical potential, we proceed to analyze its allowed parameter space and its feasibility for future observations. First, consistency requires the scale of the CC process to happen below the various cutoffs of the EFT, as illustrated in Fig.~\ref{PhysicalPicture}. Therefore, we impose the constraint (\ref{EFTConsistency}), with the cutoffs given by (\ref{StrongCouplingScale}) and(\ref{TimeDiffNonlinearScale}). In addition, to keep our perturbative computations reliable, we need several perturbativity constraints by requiring the self-energy corrections to be subdominant. First, the inflaton propagator receive a 1-loop correction from the enhanced $h$ field,
	\begin{align}
		\nonumber&\begin{gathered}
			\includegraphics[width=3cm]{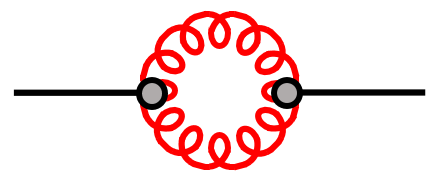}
		\end{gathered}\quad<\begin{gathered}
			\includegraphics[width=3.2cm]{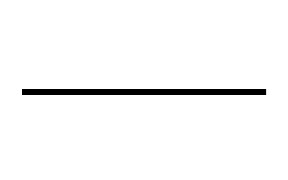}
		\end{gathered}\\
	&\Rightarrow\frac{|\alpha_{-2}|^4}{(4\pi)^2}\left(\frac{H}{2\Lambda_{c}}\right)^2\frac{H^4}{m^4}<1~.
	\end{align}
	The graviton propagator mixes into the massive spin-2 particle at tree level and receives contribution at 1-loop level,
	\begin{align}
		\nonumber&\max\Bigg\{\begin{gathered}
			\includegraphics[width=3cm]{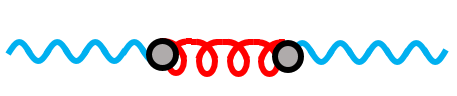}
		\end{gathered},~\begin{gathered}
			\includegraphics[width=3cm]{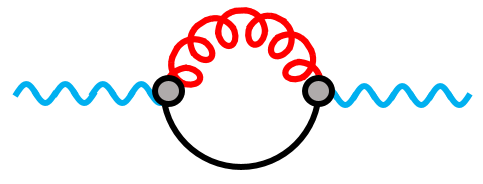}
		\end{gathered}\Bigg\}<\begin{gathered}
			\includegraphics[width=2.5cm]{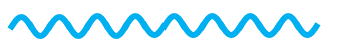}
		\end{gathered}\\
	&\Rightarrow \max\left\{\frac{|\alpha_{-2}|^2\rho^2\dot{\phi}_0^2}{M_p^2 m^2}~,~\frac{|\alpha_{-2}|^2}{(4\pi)^2}\frac{\rho^2 H^4}{M_p^2m^2}\right\}<1~.
	\end{align}
	The massive spin-2 propagator receives corrections from a tree-level mixing diagram and multiple 1-loop diagrams:
	\begin{align}
		\nonumber\max\Bigg\{&\begin{gathered}
			\includegraphics[width=3cm]{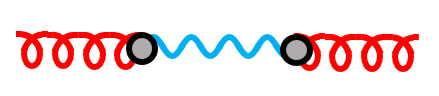}
		\end{gathered}~,\\
	\nonumber&\begin{gathered}
			\includegraphics[width=3cm]{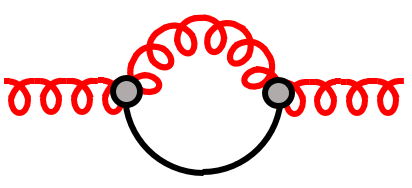}
		\end{gathered}~,~\begin{gathered}
			\includegraphics[width=3cm]{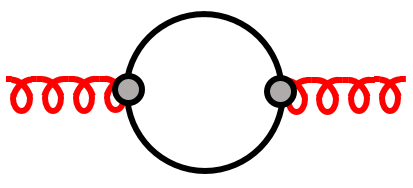}
		\end{gathered}~,~\begin{gathered}
			\includegraphics[width=3cm]{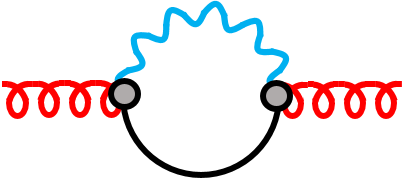}
		\end{gathered}\Bigg\}<\begin{gathered}
			\includegraphics[width=2.5cm]{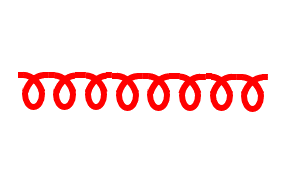}
		\end{gathered}\\
	\Rightarrow \max&\left\{\frac{|\alpha_{-2}|^2\rho^2\dot{\phi}_0^2}{M_p^2m^2}~,~\frac{|\alpha_{-2}|^4}{(4\pi)^2}\left(\frac{H}{2\Lambda_{c}}\right)^2\frac{H^4}{m^4}~,~\frac{|\alpha_{-2}|^2}{(4\pi)^2}\frac{H^4}{M^2m^2}~,~\frac{|\alpha_{-2}|^2}{(4\pi)^2}\frac{\rho^2H^4}{M_p^2 m^2}\right\}<1~.
	\end{align}
	One important aspect to notice here is that although (for simplicity) we have imposed perturbativity for the linear mixing diagram, there is nothing preventing it from taking a non-perturbatively large value. Unlike the cases with loops, we can still cope with this strongly coupled system by solving the linear EoM and performing diagonalization \cite{An:2017hlx,Bordin:2018pca}. This amounts to resumming all the chain diagrams. For the purpose of this work, however, we will take a simpler route and work with the weak-mixing case. At last, the current observational bounds set an upper limit of non-Gaussianities. The most stringent constraints come from Planck 2018, where the curvature bispectrum and trispectrum estimators are found to be $f_{NL}^{\text{equil}}=-26\pm47$ and $g_{NL}^{\text{loc}}=(-5.8\pm 6.5)\times 10^{4}$ (68\%CL). We need to make sure our signal strength do not get over-amplified by the chemical potential. Considering the leading diagrams, we impose
	\begin{align}
	\nonumber&\begin{gathered}
			\includegraphics[width=3cm]{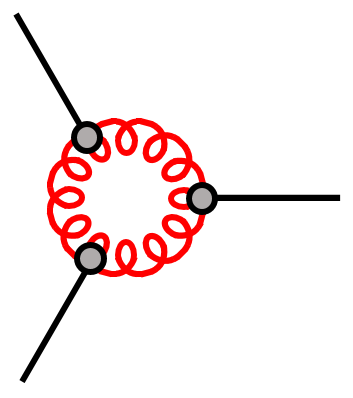}
		\end{gathered}\qquad\qquad\begin{gathered}
		\includegraphics[width=3cm]{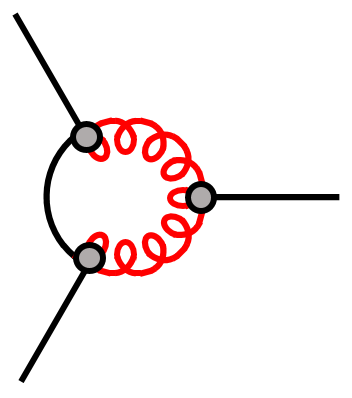}
		\end{gathered}\qquad\qquad\begin{gathered}
		\includegraphics[width=3cm]{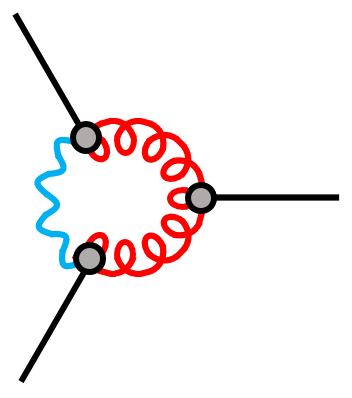}
		\end{gathered}\\
	&\Rightarrow~f_{NL}\sim \frac{\dot{\phi}}{H^2}\times\frac{1}{(4\pi)^2}\max\left\{|\alpha_{-2}|^6\frac{H^9}{\Lambda_{c}^3m^6}, |\alpha_{-2}|^4\frac{H^6}{M^2m^4},|\alpha_{-2}|^4\frac{\rho^2 H^6}{M_p^2m^4}\right\}<47~,\\
	&\begin{gathered}
		\includegraphics[width=3cm]{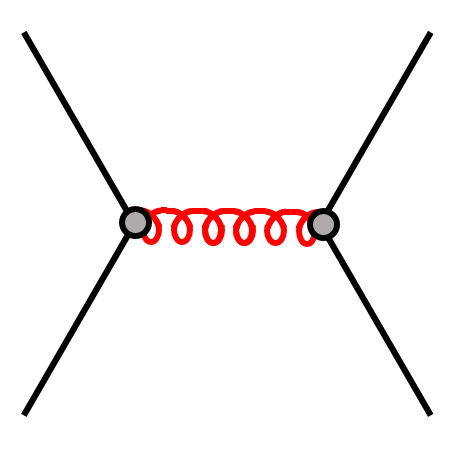}
	\end{gathered}\Rightarrow~g_{NL}\sim 230\times |\alpha_{-2}|^2\left(\frac{m}{5H}\right)^{-2}\left(\frac{M}{20H}\right)^{-2} <6.5\times 10^4~.
	\end{align}
	The total allowed parameter space is obtained by combining all these constraints. 
	
	Although no significant evidence for non-Gaussianities is found in any present-day observations, the next generation of CMB and Large Scale Structure (LSS) measurements are expected to improve the accuracy by at least one order of magnitude. For instance, LiteBIRD \cite{Matsumura:2013aja} is estimated to reduce the error bar on the mixed non-Gaussianity from the WMAP result ($\Delta f_{NL,\zeta\zeta\gamma}=\mathcal{O}(100)$) down to $\Delta f_{NL,\zeta\zeta\gamma}=\mathcal{O}(1)$ \cite{Shiraishi:2019yux}. Other next-generation CMB observations such as CMB-S4 \cite{CMB-S4:2016ple} or CORE \cite{CORE:2017oje} are expected to reach a comparable performance. Even more precision can be achieved via the futuristic 21-cm observations, which provide a tomographic map of the neutral Hydrogen atoms during the Dark Age \cite{Cooray:2006km,Pillepich:2006fj}. Interferometry based on radio telescopes on the far side of the moon \cite{Silk:2020bsr} potentially grants us access to all $10^{12}$ modes in the Dark Age, which is six orders of magnitude more than that observable with Planck, thereby greatly reducing the statistical errors. The forecast sensitivity for 21-cm interferometers with a $\mathcal{O}(10^3)$ km baseline can reach $\Delta g_{NL}^{\text{loc}}=\mathcal{O}(10)$ for the analytical background shape and $\Delta g_{NL}^{\text{clock}}=\mathcal{O}(100)$ for the oscillatory CC signals \cite{Floss:2022grj}. 
	
\begin{figure}[h!]
		\centering
		\includegraphics[width=\textwidth]{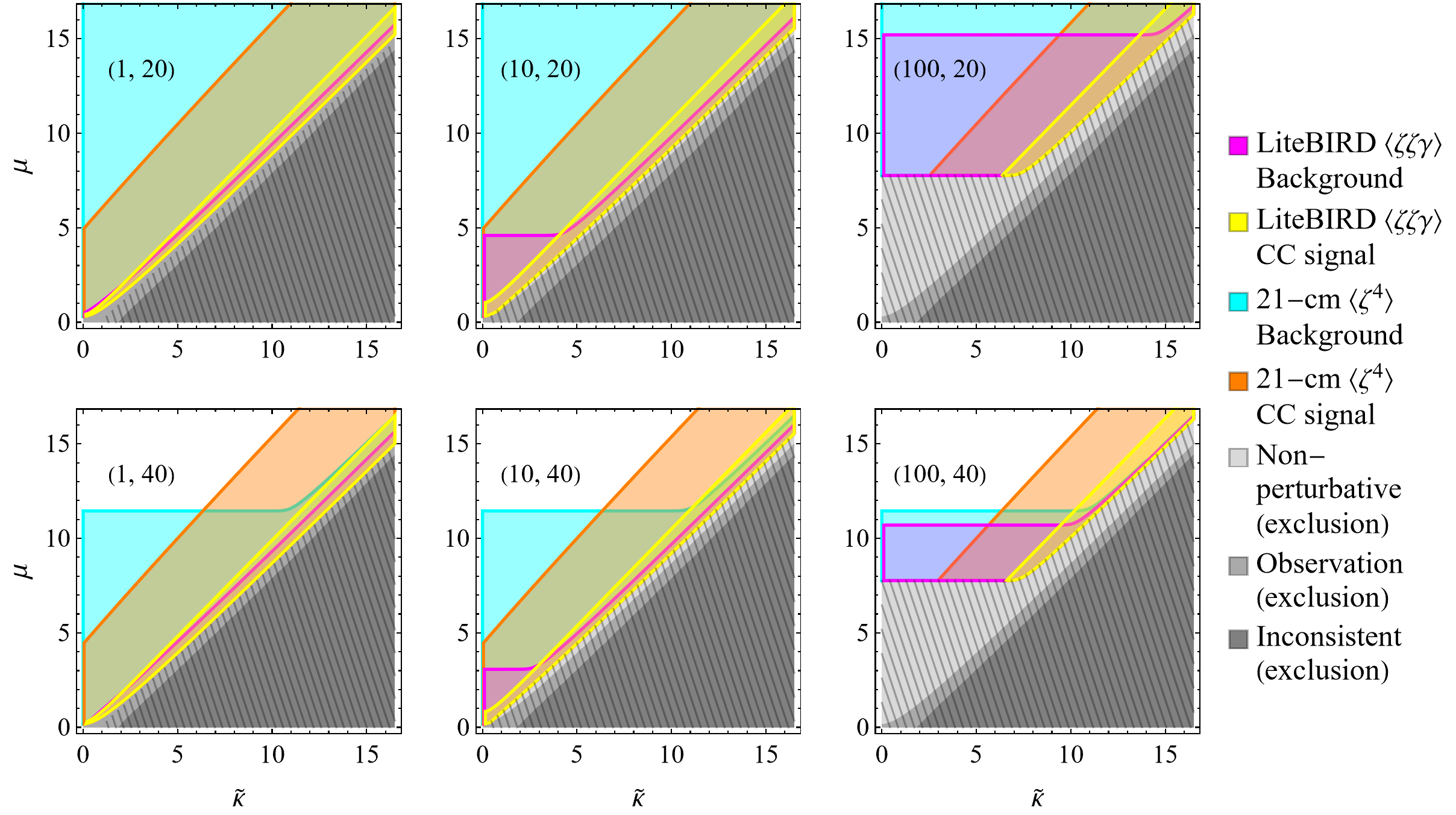}
		\caption{The feasible parameter regions in our model for future observations such as LiteBIRD and 21-cm interferometry. The various colored regions give detectably large non-Gaussian backgrounds/CC signals for the corresponding experiment. The gray dashed regions are excluded by the imposed constraints. The horizontal axis is the dimensionless chemical potential $\tilde{\kappa}=\frac{\dot{\phi}_0}{H\Lambda_{c}}$ and the vertical axis is the dS-invariant mass of the spin-2 particle $\mu=\sqrt{\frac{m^2}{H^2}-\frac{9}{4}}$. Here we have assumed a tensor-to-scalar ratio $r=0.05$, a value consistent with the Planck bounds. The mixing coupling and the coupling to the inflaton are given in the parenthesizes as $(\rho,M/H)$. The LiteBIRD forecast sensitivity is taken as $\Delta f_{NL,\zeta\zeta\gamma}=\mathcal{O}(1)$ for both the background and CC signals \cite{Shiraishi:2019yux}, while the 21-cm interferometer sensitivity is taken to be $\Delta g_{NL}^{\text{loc}}=\mathcal{O}(10)$ for the background shape and $\Delta g_{NL}^{\text{clock}}=\mathcal{O}(100)$ for CC signals \cite{Floss:2022grj}.}\label{figParameterSpace}
	\end{figure}

	We plot in Fig.~\ref{figParameterSpace} the detectability of our model for future observations, under the constraints from perturbativity, consistency and current non-Gaussianity bounds. The colored parameter regions give non-Gaussianities large enough to fall into the forecast sensitivities of LiteBIRD and 21-cm interferometry, while the gray, dashed regions are excluded by the constraints above. We see that our model enjoys considerably large feasible parameter space for the mixed bispectrum and curvature trispectrum. In particular, the introduction of the chemical potential makes it possible to detect the CC signals of the spin-2 particle even when its mass is much greater than the Hubble scale, $i.e.$, $\mu\gg 1$. It is also worthwhile to note that mixing perturbativity gives strongest constraint, as can be seen from Fig.~\ref{figParameterSpace}. However, as stated above, the mixing vertex needs not to be perturbative. In principle, we can give it a non-perturbative treatment by resumming the chain diagrams. This would further expand the feasible parameter space, possibly even to the extent of unlocking the $\langle\gamma\gamma\zeta\rangle$ channel for detection. We leave the detailed investigation of this interesting possibility to future studies.

	\section{Conclusion and outlooks}\label{ConclusionsSect}
	From massive gravity to extra dimensions and to string theory, massive spin-2 particles play an important role in different aspects of modern physics. In particular, if they exist during inflation, their interactions with the inflaton and graviton leave distinctive signatures on primordial non-Gaussianities, providing us invaluable insight into the particle spectrum of the early universe. Yet such spin-2 particles may be equipped with a large mass. Thus their pure dS-invariant gravitational production is exponentially suppressed by Boltzmann factors, resulting in cosmological collider signals potentially too small to detect.

	In this work, we have explored one possibility to enhance the signal strength by introducing a symmetry-breaking chemical potential. We first showed that the chemical potential operator is uniquely fixed at dimension-5 by the requirement of linearity, symmetry and consistency. Analysis of the linear theory shows that the spin-2 field propagates two tensor DoFs and a scalar DoF, with the vector modes suppressed by the chemical potential. The assistance of a positive chemical potential leads to the exponential enhancement in the production rate of the $\lambda=-2$ mode. After coupling the enhanced spin-2 particle to the visible sector, we found large non-Gaussianities of the type $\langle\zeta\zeta\gamma\rangle$, $\langle\zeta^4\rangle$ and large CC signals therein. Under the constraints from EFT consistency, perturbativity and current observations, our model encompass wide parameter regions feasible for future CMB observations and 21-cm tomography.
	
	Despite the interesting dynamics and rich phenomenology of our model, there are several questions untouched and we conclude by mentioning a few of them as prospects in the future.
	\begin{enumerate}
		\item[$\bullet$] First, the mixing between the massive spin-2 particle and graviton is taken to be perturbative in this work. It is interesting and straightforward to generalize the analysis to the non-perturbative case. This will broaden the parameter space and allow for large helical primordial gravitational wave power spectrum $\langle\gamma^2\rangle$, and possible observation of the $\langle\gamma\gamma\zeta\rangle$ type non-Gaussianity in addition. 
		\item[$\bullet$] Second, the disappearance of the vector modes even in the small chemical potential limit seems a bit mysterious to us, as it suggests a somewhat unphysical discontinuity in the $\tilde{\kappa}\to 0$ limit. It may be possible that this discontinuity is an artifact due to the linear approximation and will vanish in the nonlinear theory. Something like a Vainshtein mechanism may then be responsible for the suppressed vector modes and it is worthwhile to explore in more detail.
		\item[$\bullet$] Third, we have worked in a heavily bottom-up fashion in this paper, using Stueckelberg trick and NDA counting methods to keep unknown UV physics under control. It is certainly important to have a more UV complete understanding of the chemical potential operator, especially in light of the several lines of work mentioned in Sect.~\ref{UVcompletionSect}. A better understanding of the UV picture should also shed light on the previous point.
		\item[$\bullet$] Fourth, going beyond spin-2, chemical potential of higher-spin fields may aid probing the spin-mass relation of the inflationary particle spectrum. Experience from hadron resonances and string theory suggests that particles with a higher spin tend to have a larger mass. Assuming a naive Regge behavior for the massive spin-$S$ particles ($S \sim \alpha'm_S^2$, where $\alpha'$ is the Regge slope), the CC signal strength then decays quickly as $e^{-\pi\sqrt{S/\alpha'}/H}$. This deficiency motivates a generalization of chemical potential enhancement to higher spins. The generalization seems straightforward for bosons. Following the logic of Sect.~\ref{IntroducingSin2ChemPtl}, we can write down the following operator,
		\begin{equation}
			\int d^4x \epsilon^{\mu\nu\rho\sigma}\frac{\phi}{2\Lambda_{c,S}} \nabla_\mu\Sigma_{\nu\lambda_1\cdots\lambda_{S-1}}\nabla_\rho \Sigma_\sigma^{~\lambda_1\cdots\lambda_{S-1}}~,
		\end{equation}
		where $\Sigma_{\mu_1\cdots\mu_S}=\Sigma_{(\mu_1\cdots\mu_S)}$ is a totally symmetric rank-$S$ tensor. Then the massive spin-$S$ particle may also benefit from a chemical potential enhancement, with a boosted CC signal strength $\sim \exp\left[{-\frac{\pi}{H}\left(\sqrt{\frac{S}{\alpha'}}-\frac{S\dot{\phi}_0}{\Lambda_{c,S}}\right)}\right]$. It is even conceivable to detect a uniform family of signals from the whole high spin tower if $\Lambda_{c,S}\propto \sqrt{S}$. We leave the detailed study of the corresponding dynamics and interesting phenomenology to future studies. 
	\end{enumerate}

	\acknowledgments
	XT would like to thank Yuhang Zhu and Yi Wang for helpful discussions.
	XT and ZX are supported in part by the National Key R\&D Program of China (2021YFC2203100). ZX is supported in part by an Open Research Fund of the Key Laboratory of Particle Astrophysics and Cosmology, Ministry of Education of China, and a Tsinghua University Initiative Scientific Research Program.

	\appendix
	\section{The Feynman rules}\label{FeynmanRuleAppendix}
	We begin by noting the mode functions of inflaton and graviton,
	\begin{align}
		\varphi(\tau,k)&=\frac{H}{\sqrt{2k^3}}(1+ik\tau)e^{-ik\tau}~,\\
		\gamma_{\lambda}(\tau,k)&=\frac{H}{2\sqrt{k^3}}(1+ik\tau)e^{-ik\tau}~.
	\end{align}
	The mode functions for the massive spin-2 particle are given in Sect.~\ref{ModeFunSect}. The Schwinger-Keldysh (SK) propagators of $\varphi,\gamma_{ij},h_{ij}$ are labeled by $F,G,H$. Diagrammatically, we denote them by
	\begin{align}
		&~~~~~\begin{gathered}
			\includegraphics[width=3.2cm]{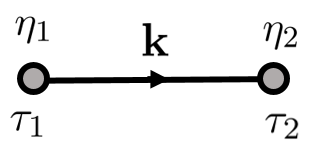}
		\end{gathered}=~F^{\eta_1\eta_2}(\tau_1,\tau_2,k)~,\\
		&\begin{gathered}
			\includegraphics[width=4cm]{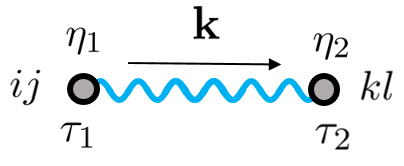}
		\end{gathered}=~G_{ij,kl}^{\eta_1\eta_2}(\tau_1,\tau_2,\mathbf{k})~,\\
		&\begin{gathered}
			\includegraphics[width=4cm]{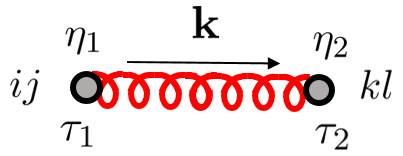}
		\end{gathered}=~H_{ij,kl}^{\eta_1\eta_2}(\tau_1,\tau_2,\mathbf{k})~.\\
	\end{align}
	The four SK propagators are all built from the basic Wightman functions with non-time ordering $\eta_1=-,\eta_2=+$,
	\begin{align}
		F^{-+}(\tau_1,\tau_2,k)&=\varphi(\tau_1,k)\varphi^*(\tau_2,k)~,\\
		G_{ij,kl}^{-+}(\tau_1,\tau_2,\mathbf{k})&=\frac{2}{M_p^2}\sum_{\lambda=\pm 2}\gamma_\lambda(\tau_1,k)e_{ij}^\lambda(\mathbf{\hat{k}})\left[\gamma_\lambda(\tau_2,k)e_{kl}^\lambda(\mathbf{\hat{k}})\right]^*~,\\
		\nonumber H_{ij,kl}^{-+}(\tau_1,\tau_2,\mathbf{k})&=\sum_{\lambda=\pm 2}h_\lambda(\tau_1,k)e_{ij}^\lambda(\mathbf{\hat{k}})\left[h_\lambda(\tau_2,k)e_{kl}^\lambda(\mathbf{\hat{k}})\right]^*\\
		&+\left(g_0(\tau_1,k)e_{ij}^0(\hat{\mathbf{k}})+\frac{1}{3}h_0(\tau_1,k)\delta_{ij}\right)\left(g_0(\tau_2,k)e_{kl}^0(\hat{\mathbf{k}})+\frac{1}{3}h_0(\tau_2,k)\delta_{kl}\right)^*~.
	\end{align}
	More explicitly, denoting $P\in\{F,G,H\}$ as a general propagator, then the other three SK propagators are
	\begin{align}
		P^{+-}(\tau_1,\tau_2,\mathbf{k})&=P^{-+}(\tau_1,\tau_2,-\mathbf{k})^*~,\\
		P^{++}(\tau_1,\tau_2,\mathbf{k})&=\Theta(\tau_1-\tau_2)P^{-+}(\tau_1,\tau_2,\mathbf{k})+\Theta(\tau_2-\tau_1)P^{+-}(\tau_1,\tau_2,\mathbf{k})~,\\
		P^{--}(\tau_1,\tau_2,\mathbf{k})&=P^{--}(\tau_1,\tau_2,-\mathbf{k})^*~.
	\end{align}
	It is often convenient to extract the kinematic factors from the spinning propagators as
	\begin{equation}
		P_{ij,kl}^{\eta_1\eta_2}(\tau_1,\tau_2,\mathbf{k})=\sum_{\lambda}e_{ij}^\lambda(\mathbf{\hat{k}})e_{kl}^{\lambda*}(\mathbf{\hat{k}})P_{\lambda}^{\eta_1\eta_2}(\tau_1,\tau_2,k)~.
	\end{equation}
	Note that although the longitudinal part of the $H$ propagator cannot be simply factored as above, only the $g_0 e_{ij}^0$ part dominate the IR limit. Hence when computing the CC signals, we can neglect the $h_0 \delta_{ij}$ contribution and factor out the kinematic structure as above.
	
	The vertex rules of the EFT operators (\ref{ChemPtlNonCovariantForm}), (\ref{MixingVertex}) (\ref{hphiphiVertex}) are given by
	\begin{align}
		\begin{gathered}
			\includegraphics[width=5cm]{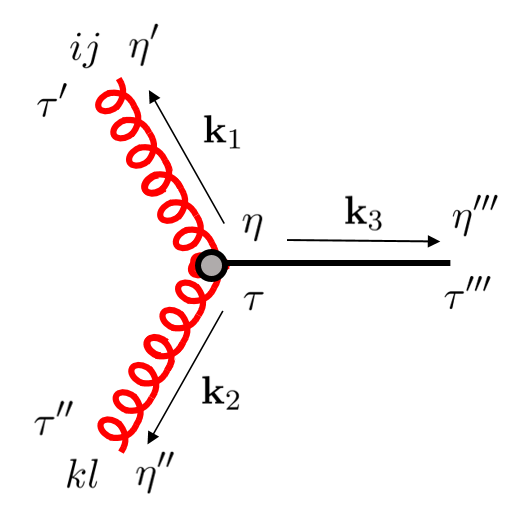}
		\end{gathered}
		&\qquad\begin{gathered}
			\frac{\eta}{2\Lambda_{c}}\epsilon_{mrs}(\mathbf{k}_1-\mathbf{k}_2)_r\int d\tau a^{-2}(\tau)\partial_\tau  F^{\eta'''\eta}(\tau''',\tau,k_3)\\
			\qquad\qquad\qquad\qquad\qquad\times H_{mn,ij}^{\eta\eta'}(\tau,\tau',\mathbf{k}_1)H_{sn,kl}^{\eta\eta''}(\tau,\tau'',\mathbf{k}_2)~,
		\end{gathered}
	\end{align}
	\begin{align}
		\begin{gathered}
			\includegraphics[width=5cm]{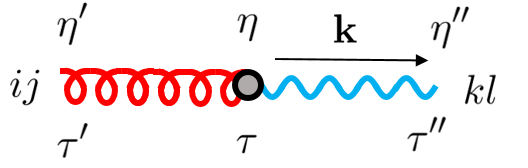}
		\end{gathered}
		&\qquad\begin{gathered}
			i\eta\rho\dot{\phi}_0\int d\tau a(\tau) H_{ij,mn}^{\eta'\eta}(\tau',\tau,\mathbf{k})\partial_\tau G_{mn,kl}^{\eta\eta''}(\tau,\tau'',\mathbf{k})~,
		\end{gathered}
	\end{align}
	\begin{align}
		\begin{gathered}
			\includegraphics[width=5cm]{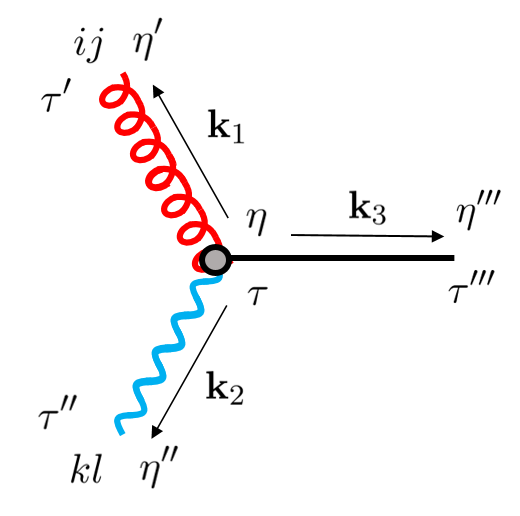}
		\end{gathered}
		&\qquad\begin{gathered}
			i \eta \rho\int d\tau H_{mn,ij}^{\eta\eta'}(\tau,\tau',\mathbf{k}_1)\partial_\tau G_{mn,kl}^{\eta\eta''}(\tau,\tau'',\mathbf{k}_2)\partial_\tau F^{\eta\eta'''}(\tau,\tau''',k_3)~,
		\end{gathered}
	\end{align}
	\begin{align}
		\begin{gathered}
			\includegraphics[width=5cm]{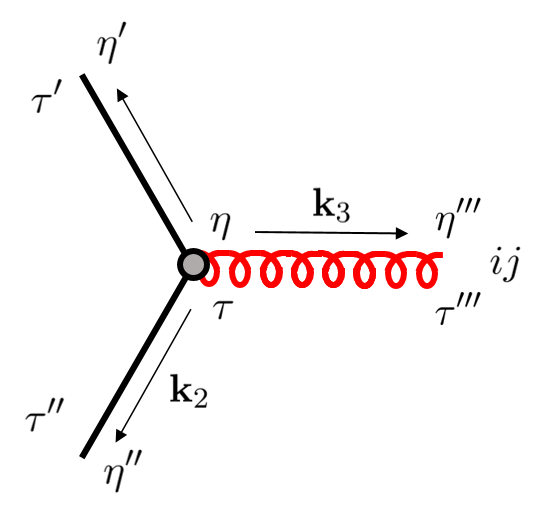}
		\end{gathered}
		&\qquad\begin{gathered}
			\frac{2i\eta}{M}(\mathbf{k}_1)_m(\mathbf{k}_2)_n\int d\tau F^{\eta'\eta}(\tau',\tau,k_1)F^{\eta''\eta}(\tau'',\tau,k_2)H_{mn,ij}^{\eta\eta'''}(\tau,\tau''',\mathbf{k}_3)~.
		\end{gathered}
	\end{align}

	The dynamical factor of the diagram contributing to $\langle\zeta\zeta\gamma\rangle$ reads
	\begin{align}
		\nonumber\mathcal{I}_\lambda&=-\frac{16\rho H^2}{M M_p^2\dot{\phi}_0}\sum_{\eta_1,\eta_2=\pm 1}\eta_1\eta_2\int d\tau_1 d\tau_2 a(\tau_2)F^{\eta_1-}(\tau_1,0,k_1)F^{\eta_1-}(\tau_1,0,k_2)\\ &~~~~~~~~~~~~~~~~~~~~~~~~~~~~~~~~~~~~~~~~~~~~~~~\times H_\lambda^{\eta_1\eta_2}(\tau_1,\tau_2,k_3)\partial_{\tau_2}G_\lambda^{\eta_2-}(\tau_2,0,k_3)~.\label{zetazetagammaI}
	\end{align}
	The dynamical factors of the two diagrams contributing to $\langle\zeta\zeta\gamma\rangle$ are given by
	\begin{align}
		\nonumber\mathcal{I}^{(\text{i})}_{\lambda_1\lambda_2}\equiv& \frac{16\rho^2 H}{M_p^4}\sum_{\eta_1,\eta_2}\eta_1\eta_2\int d\tau_1 d\tau_2a(\tau_1)\partial_{\tau_2}F^{\eta_2-}(\tau_2,0,k_3)\partial_{\tau_2}G^{\eta_2-}_{\lambda_2}(\tau_2,0,k_2)\\
		\nonumber&\qquad\qquad\qquad\qquad\qquad\qquad\qquad\times H^{\eta_2\eta_1}_{\lambda_1}(\tau_2,\tau_1,k_1)\partial_{\tau_1}G^{\eta_1-}_{\lambda_1}(\tau_1,0,k_1)\\
		&+(k_1\leftrightarrow k_2)~,\label{gammagammazetaIi}\\
		\nonumber\mathcal{I}^{(\text{ii})}_{\lambda_1\lambda_2}\equiv& -\frac{32\rho^2\dot{\phi}_0Hk_3}{M_p^4\Lambda_{c}}\sum_{\eta_1,\eta_2,\eta_3}\eta_1\eta_2\eta_3\int d\tau_1 d\tau_2 d\tau_3a(\tau_1)a(\tau_2)a(\tau_3)^{-2}\partial_{\tau_3}F^{\eta_3-}(\tau_3,0,k_3)\\
		\nonumber&\qquad\qquad\qquad\qquad\qquad\qquad\qquad\qquad\times H^{\eta_3\eta_1}_{\lambda_1}(\tau_3,\tau_1,k_1)\partial_{\tau_1}G^{\eta_1-}_{\lambda_1}(\tau_1,0,k_1)\\
		&\qquad\qquad\qquad\qquad\qquad\qquad\qquad\qquad\times H^{\eta_3\eta_2}_{\lambda_2}(\tau_3,\tau_2,k_2)\partial_{\tau_2}G^{\eta_2-}_{\lambda_2}(\tau_2,0,k_2)~,\label{gammagammazetaIii}
	\end{align}
	respectively. The dynamical factor of the diagram contributing to $\langle\zeta^4\rangle$ reads
	\begin{align}
		\nonumber\mathcal{J}_\lambda=&-\frac{4H^4}{M^2\dot{\phi}_0^4}\sum_{\eta_1,\eta_2=\pm 1}\eta_1\eta_2\int d\tau_1 d\tau_2 F^{\eta_1-}(\tau_1,0,k_1)F^{\eta_1-}(\tau_1,0,k_2)\\
		&~~~~~~~~~~~~~~~~~~~~~~~~~~~~~~~~~~~~~~\times F^{\eta_2-}(\tau_2,0,k_3)F^{\eta_2-}(\tau_2,0,k_4) H_\lambda^{\eta_1\eta_2}(\tau_1,\tau_2,k_I)~.\label{zeta4J}
	\end{align}

	\section{The cutting rule for CC signals}\label{CRAppendix}
	In this appendix, we give a lighting review of the cutting rule for CC signals. At tree-level, the analytical expression of the oscillatory CC signals can be conveniently extracted via the recently proposed cutting rule. The key insight is that the CC signals originate from two distinct physical processes during inflation, and can be classified according to their different oscillation patterns,
	\begin{align}
		\textbf{I. }\textbf{Non-local Type:}&\quad \sin\left(\mu\ln\frac{k_L k_R}{k_I^2}+\xi\right)~,\\
		\textbf{II. }\textbf{Local Type:}&\quad \sin\left(\mu\ln\frac{k_L}{k_R}\right)~.
	\end{align}
	Here $k_{L,R}$ are the corresponding injection frequencies at the left/right blobs to which the massive propagator is attached.
	\begin{figure}[h!]
		\centering
		\includegraphics[width=13cm]{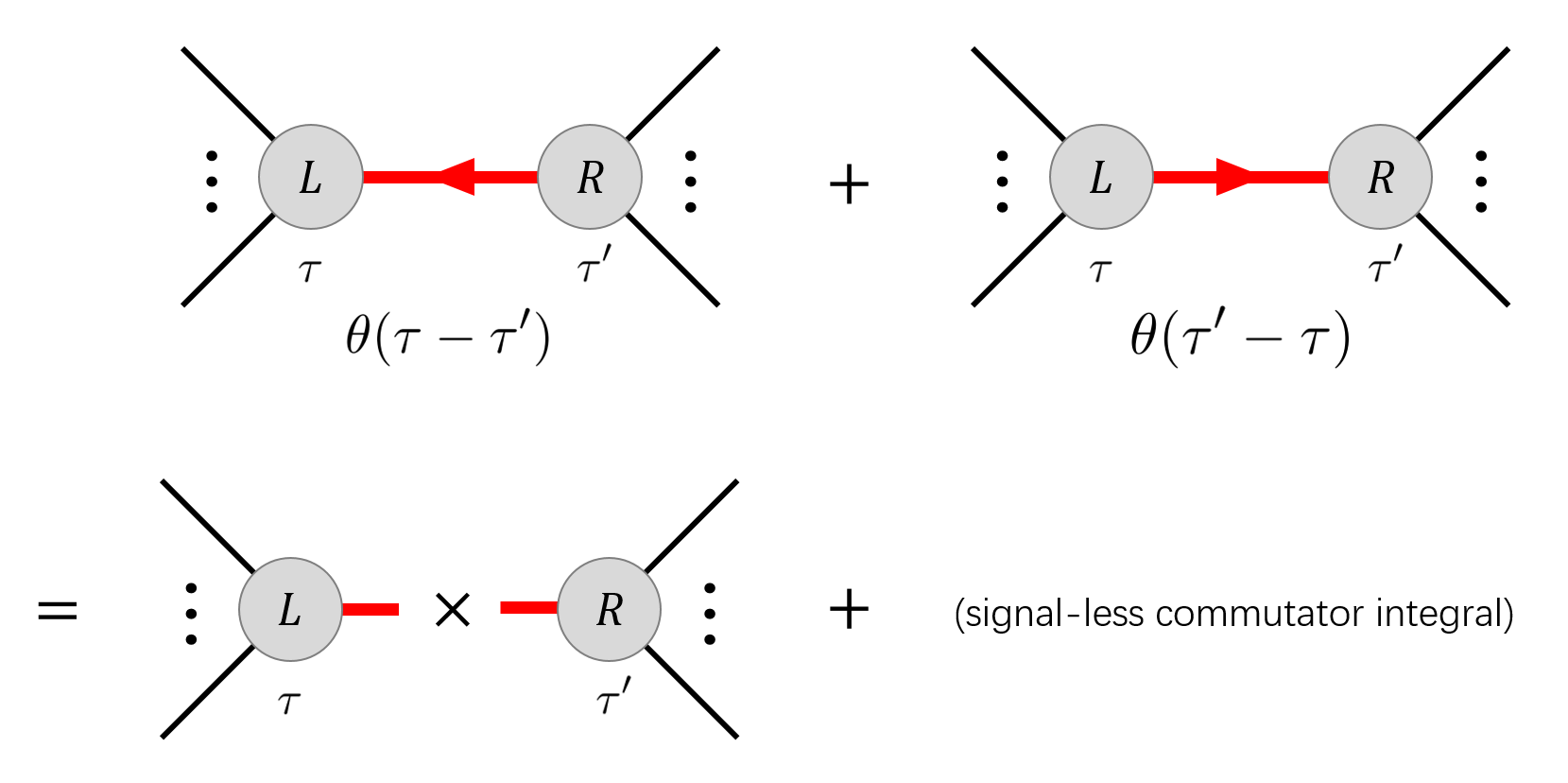}
		\caption{A cartoon illustration of the cutting rule for CC signals.}\label{CRillustration}
	\end{figure}
	The non-local type signal represents non-local bulk pair production and resonance decay events, while the local type signal represents local vertex production and resonance decay\footnote{Notice that in 3-point correlators, the two types of CC signals take the same form, since $k_I=k_R$. However, they still represent different contributions and can be distinguished from the external line helicities.}. Since all the events happen in the form of resonances localized in time, one can utilize their time ordering to discard certain part of the Feynman diagram that is free of any signals. In particular, when $k_L>k_R$, the complicated commutator integral from flipping one of the Heaviside theta function can be thrown away, leaving a simple factorized integral that is often easy to compute. The diagram is then cut into simple product of two Laplace transformations. This procedure is illustrated in Fig.~\ref{CRillustration}. Based on this observation, one can approximate any tree diagrams by a combination of CC signals and EFT backgrounds. We briefly summarize the general algorithm below, and refer interested readers to \cite{Tong:2021wai} for more details.
	
	\begin{framed}
		{\noindent {\it Cutting algorithm for CC signals}
			\begin{enumerate}
				\item Consider only completely time-ordered or totally anti-time-ordered diagrams.
				\item For all massive propagators in the given diagram, work with one at a time and mimic other massive propagators by the corresponding large-mass EFT operators.
				\item Work out the left/right blobs and reduce them to a sum of products of exponentials and polynomials.
				\item Flip the time-ordering Heaviside step function and discard the commutator integral, obtaining two factorized integrals as Laplace transformations.
				\item Perform a left $\leftrightarrow$ right symmetrization.
				\item Repeat Step 2-5 for each massive propagator, with their union giving the total signal $S$.
				\item Finally, dress the total signal $S$ by an EFT background $B$ where all massive fields are integrated out.
			\end{enumerate}
		}
	\end{framed}

	\section{Some helpful identities and checklists of dimensionality}\label{HelpfulIds}
	The dS curvature tensors are made of metric tensors alone, since it is a maximally symmetric spacetime:
	\begin{subequations}
		\begin{align}
			R_{\mu\nu\rho\sigma}&=H^2(g_{\mu\rho}g_{\nu\sigma}-g_{\mu\sigma}g_{\nu\rho})~,\\
			R_{\mu\nu}&=3H^2 g_{\mu\nu}~,\\
			R&=12H^2~.
		\end{align}
	\end{subequations}
	Then using the commutator of covariant derivatives,
	\begin{equation}
		[\nabla_\mu,\nabla_\nu] T^{\rho_1\cdots\rho_m}_{~~~~~~~~\sigma_1\cdots\sigma_n}=R^{\rho_1}_{~~\lambda\mu\nu}T^{\lambda\cdots\rho_m}_{~~~~~~~~\sigma_1\cdots\sigma_n}+\cdots-R^{\lambda}_{~~\sigma_1\mu\nu}T^{\rho_1\cdots\rho_m}_{~~~~~~~~\lambda\cdots\sigma_n}-\cdots~,
	\end{equation}
	one can prove the following quantity vanishes in dS:
	\begin{align}
			\nonumber\mathcal{E}^{\mu\nu\rho\sigma}[\nabla_\mu,\nabla_\nu]T_{\rho\alpha_1\cdots\alpha_n}&=-\mathcal{E}^{\mu\nu\rho\sigma}R^\lambda_{~\rho\mu\nu}T_{\lambda\alpha_1\cdots\alpha_n}-\mathcal{E}^{\mu\nu\rho\sigma}R^\lambda_{~\alpha_1\mu\nu}T_{\rho\lambda\cdots\alpha_n}-\cdots\\
			\nonumber&=0-H^2\mathcal{E}^{\mu\nu\rho\sigma}(\delta^\lambda_\mu g_{\alpha_1\nu}-\delta^\lambda_\nu g_{\alpha_1\mu})T_{\rho\lambda\cdots\alpha_n}-\cdots\\
			&=0~,\label{epsilonIdentity}
		\end{align}
	for any totally symmetric tensor $T_{\lambda\alpha_1\cdots\alpha_n}=T_{(\lambda\alpha_1\cdots\alpha_n)}$.
	
	In Table.~\ref{massDimTable} and Table.~\ref{conformalDimTable}, we list the mass and conformal dimensions of various quantities that frequently appear in this work. They are often convenient in consistency checks, for instance, when counting powers of couplings and scale factors in EFT operators.
	
	\begin{table*}[hbt!]
		\centering 
		\begin{tabular}{cc}
			\toprule[1.2pt] 
			Mass dimension & Quantities   \\
			\midrule[1pt]  
			$4$	&	$\mathcal{L},~ \Delta \mathcal{L}$\\
			\specialrule{0em}{1.5pt}{1.5pt}
			$3$	&	$\delta^3(\mathbf{x})$\\
			\specialrule{0em}{1.5pt}{1.5pt}
			$2$	&	$\dot{\phi}$\\
			\specialrule{0em}{1.5pt}{1.5pt}
			$1$ &	$m,~\kappa,~H,~M_p,~M,~\mathcal{M}_{*,\pm2},~\Lambda_{c},~\partial_\tau,~\partial_i,~k,~\varphi(x),~h_{\mu\nu}(x)$\\ \specialrule{0em}{1.5pt}{1.5pt}
			$0$	&	$\mu,~\tilde{\kappa},~\rho,~a(\tau),~f_{NL},~g_{NL},~\gamma_{\mu\nu}(x),~\zeta(x)$\\
			\specialrule{0em}{1.5pt}{1.5pt}
			$-1$	&	$\tau,~\mathbf{x},~\langle\varphi^2\rangle'$\\
			\specialrule{0em}{1.5pt}{1.5pt}
			$-2$	&	$h_{ij}(\mathbf{k}),~\varphi(\mathbf{k})$\\
			\specialrule{0em}{1.5pt}{1.5pt}
			$-3$	&	$\gamma_{ij}(\mathbf{k}),~\zeta(\mathbf{k}),~\delta^3(\mathbf{k}),~\langle\zeta^2\rangle',~\langle\gamma^2\rangle'$\\
			\specialrule{0em}{1.5pt}{1.5pt}
			$-6$	&	$\langle\zeta^2\gamma\rangle',~\langle\gamma^2\zeta\rangle'$\\
			\specialrule{0em}{1.5pt}{1.5pt}
			$-9$	&	$\langle\zeta^4\rangle'$\\
			\bottomrule[1.2pt]
		\end{tabular}
		\caption{A list of the mass dimensions of various quantities.}\label{massDimTable}
	\end{table*}

	\begin{table*}[hbt!]
		\centering 
		\begin{tabular}{cc}
			\toprule[1.2pt] 
			Conformal dimension & Quantities   \\
			\midrule[1pt]  
			$4$	&	$\mathcal{L},~ \Delta \mathcal{L}$\\
			\specialrule{0em}{1.5pt}{1.5pt}
			$3$	&	$\delta^3(\mathbf{x})$\\
			\specialrule{0em}{1.5pt}{1.5pt}
			$2$	&	$h_{\mu\nu}(x)$\\
			\specialrule{0em}{1.5pt}{1.5pt}
			$1$ &	$a(\tau),~\partial_\tau,~\partial_i,~k$\\ \specialrule{0em}{1.5pt}{1.5pt}
			$0$	&	$H,~M_p,~m,~M,~\mathcal{M}_{*,\pm2},~\Lambda_{c},~\kappa,~\rho,~\mu,~\tilde{\kappa},~ \dot{\phi},~f_{NL},~g_{NL},~\gamma_{\mu\nu}(x),~\zeta(x)$\\
			\specialrule{0em}{1.5pt}{1.5pt}
			$-1$	&	$\tau,~\mathbf{x},~h_{ij}(\mathbf{k})$\\
			\specialrule{0em}{1.5pt}{1.5pt}
			$-2$	&	 \\
			\specialrule{0em}{1.5pt}{1.5pt}
			$-3$	&	$\varphi(\mathbf{k}),~\gamma_{ij}(\mathbf{k}),~\zeta(\mathbf{k}),~\delta^3(\mathbf{k}),~\langle\varphi^2\rangle',~\langle\zeta^2\rangle',~\langle\gamma^2\rangle'$\\
			\specialrule{0em}{1.5pt}{1.5pt}
			$-6$	&	$\langle\zeta^2\gamma\rangle',~\langle\gamma^2\zeta\rangle'$\\
			\specialrule{0em}{1.5pt}{1.5pt}
			$-9$	&	$\langle\zeta^4\rangle'$\\
			\bottomrule[1.2pt]
		\end{tabular}
		\caption{A list of the conformal dimensions of various quantities.}\label{conformalDimTable}
	\end{table*}
% BIBLIOGRAPHY
% use BIBTEX if you want
\bibliographystyle{JHEP}
\bibliography{reference}

\end{document}